\documentclass[smallextended]{svjour3} 
\usepackage[colorlinks,
linkcolor=blue,
anchorcolor=blue,
citecolor=blue]{hyperref}      
\usepackage{amsmath}
\usepackage{subfigure}
\usepackage{amssymb}
\usepackage[pagewise]{lineno}
\smartqed  
\usepackage{graphicx}
\usepackage{epstopdf}
\usepackage{tabularx}
\usepackage{ctable}
\usepackage{booktabs}
\usepackage{indentfirst}
\usepackage{color}
\usepackage[titletoc,title]{appendix}

\begin{document}

\title{Impacts of seasonality and parasitism on honey bee population dynamics \thanks{} 
}

\titlerunning{Seasonality and Parasitism Impacts on Honey Bee}        

\author{Jun Chen, Jordy O Rodriguez Rincon, Gloria DeGrandi-Hoffman, Jennifer Fewell, Jon Harrison \and Yun Kang}

\authorrunning{J. Chen, J. Rodriguez R, G. DeGrandi-Hoffman, J. Fewell, J. Harrison, Y. Kang} 

\institute{J. Chen \at
             Simon A. Levin Mathematical and Computational Modeling Sciences Center, Arizona State University, Tempe, AZ 85281, USA.\\
             \email{jchen152@asu.edu}    
                  \and   
        J. Rodriguez Rincon \at
             Simon A. Levin Mathematical and Computational Modeling Sciences Center, Arizona State University, Tempe, AZ 85281, USA.\\
             \email{jordyrodriguez@asu.edu} 
          \and
         G. DeGrandi-Hoffman \at
             Carl Hayden Bee Research Center, United States Department of Agriculture-Agricultural Research Service, Tucson, AZ  85719, USA.\\
             \email{gloria.hoffman@ars.usda.gov} 
        \and
        J. Fewell \at
             Life Sciences, Arizona State University, Tempe, AZ 85287, USA. \\
              \email{ j.fewell@asu.edu}
              \and
              J. Harrison \at
             Life Sciences, Arizona State University, Tempe, AZ 85287, USA. \\
              \email{j.harrison@asu.edu}
              \and
          Y. Kang \at
             Sciences and Mathematics Faculty, College of Integrative Sciences and Arts, Arizona State University, Mesa, AZ 85212, USA\\
              \email{yun.kang@asu.edu}
}

\date{Received: date / Accepted: date}

\maketitle

\begin{abstract}

The honeybee plays an extremely important role in ecosystem stability and diversity and in the production of bee pollinated crops. Honey bees and other pollinators are under threat from the combined effects of nutritional stress, parasitism, pesticides, and climate change that impact the timing, duration, and variability of seasonal events. To understand how parasitism and seasonality influence honey bee colonies separately and interactively, we developed a non-autonomous nonlinear honeybee-parasite interaction differential equation model that incorporates seasonality into the egg-laying rate of the queen. Our theoretical results show that parasitism negatively impacts the honey bee population either by decreasing colony size or destabilizing population dynamics through supercritical or subcritical Hopf-bifurcations depending on conditions. Our bifurcation analysis and simulations suggest that seasonality alone may have positive or negative impacts on the survival of honey bee colonies. More specifically, our study indicates that (1) the timing of the maximum egg-laying rate seems to determine when seasonality has positive or negative impacts; and (2) when the period of seasonality is large it can lead to the colony collapsing. Our study further suggests that the synergistic influences of parasitism and seasonality can lead to complicated dynamics that may positively and negatively impact the honey bee colony's survival. Our work partially uncovers the intrinsic effects of climate change and parasites, which potentially provide essential insights into how best to maintain or improve a honey bee colony's health.
 
\end{abstract}

\section{Introduction}
\indent Honey bee, \emph{Apis mellifera}, the colony is not only an excellent example of a complex adaptive system \cite{wilson2000sociobiology}, but also has great value to our ecosystem and economic development. Per USDA statistics, 80\% of crops benefit from pollination by honey bees, including more than 130 types of fruits and vegetables \cite{USDApollinate}, worth \$215 billion annually worldwide \cite{smith2013pathogens}. Additionally, honey bees produce  honey and other hive products that are beneficial to human health. For example, the average American consumed 1.0 pounds of honey per person in 2019, which has increased from 0.5 pounds in 1990 \cite{USDA}. Unfortunately, honey bee colonies are collapsing at an alarming rate, especially during winter \cite{neumann2010micro} causing unsustainable losses to commercial beekeepers and colony shortages to growers.\\

Research \cite{perry2015rapid,oldroyd2007s,smith2013pathogens} suggests that there are many factors contributing to the global decline of the honey bee population. Those factors include nutritional stress from lack of flowering plants, environmental stressors such as global warming, lack of genetic variation, and vitality, parasites such as Varroa mites and Nosema, and diseases such as acute bee paralysis virus and deformed wing virus. Most notably, Varroa mites pose a huge threat to the health of honey bees  \cite{Peng:1987aa,Vetharaniam:2006aa,degrandi2004mathematical,messan2017migration,messan2021population,kang2016disease}. They can parasitize honey bees, transmit viruses, and also make honey bees more susceptible to viral outbreaks \cite{koleoglu2017effect}. Mites parasitize workers and drones (male bees), larvae and adults, but not the queen \cite{degrandi2017dispersal}. Parasitized honey bees have shortened lifespans, lower weight, and weakened immune systems \cite{Peng:1987aa}. Foragers that have been parasitized during development are more easily disoriented during foraging as adults \cite{koleoglu2017effect}. Infected colonies also are more prone to viral diseases and struggle to survive in the winter \cite{degrandi2019economics,degrandi2004mathematical,martin2012global,chen2007honey}.\\

Seasonality has important effects on honey bee foraging behaviors. For example, in temperate areas during in fall and winter, food can become unavailable as temperatures drop below freezing. During this time, honey bees remain in their hives and form a thermoregulated cluster of bees  \cite{stabentheiner2003endothermic}, but if the bees fail to maintain cluster warmth, the colony will perish \cite{simpson1961nest}. Moreover, the queen bee stops or reduces egg laying \cite{seeley1985survival,degrandi1989beepop,seasonality} in preparation for overwintering \cite{martin2001role}. Overwintering is stressful to colonies and losses may exceed 30\% \cite{doeke2015overwintering}. \\

Both experimental and simulated bee population data show seasonal patterns in colony population dynamics \cite{degrandi1989beepop,harris1980population}.
Seasonality also plays a role in the dynamics of parasites and viruses in colonies  \cite{degrandi2004mathematical,martin2001role,smolinski2021raised}. Thus, there is increased attention to including seasonality in honey bee population models. For example, \cite{ratti2015mathematical,eberl2010importance,sumpter2004dynamics} adding seasonality equations using four sets of parameter values to differentiate seasons revealed that  seasonal dynamics can lead to colonies with persistant Varroa infestations to suddenly collapse in late fall or spring because of the compounding effects of parasitism and viruses transmitted by Varroa \cite{ratti2015mathematical}. The seasonal models also generated recommendations that controls for Varroa should occur in summer to reduce the colony losses \cite{sumpter2004dynamics}.  The work of \cite{betti2016age,betti2014effects} directly used two sets of models to represent the dynamics of non-winter and winter, respectively. The model \cite{betti2014effects} has no egg laying in the winter system and considering the age structure of the colony during its yearly cycle. The model \cite{betti2016age} added 21-day transition equations for colonies to wake-up between the end-of-winter and a new active season. This model captured the sharp decline in colony size often seen in the spring (spring dwindling) and showed that the timing of the onset of disease in a colony can impact its severity and persistence in the population.\\


Here, motivated by the experimental work shown in \cite{degrandi1989beepop,harris1980population}, we describe a model where seasonality has been incorporated into the queen's egg-laying rate through cosine functions.
An age-structure model of honey bees’ population dynamics Chen et al. (2020) \cite{chen2020model} showed that seasonality may reduce colony survival but may also prevent colony collapse. Messan et al. (2021) \cite{messan2021population} focused on the colonies with parasites, and found seasonality can help colonies recover under certain conditions. Messan et al. (2018) \cite{messan2018effects} focused on the nutrition of colonies, and found that seasonality can effects from stress and cause  colony death.  \\

Based on the data 
\cite{kang2016disease,degrandi1989beepop} and previously reported models 
\cite{chen2020model,messan2018effects,messan2021population}, we formulate a mathematical modeling framework describing honeybee-mite interactions with seasonality in the queen's egg-laying rate to address the following questions:
\begin{itemize}
\item How may seasonality impact honey bee populations in the absence of parasitism?
\item How may parasitism impact the honey bee population?
\item What are the synergistic impacts of seasonality and parasitism on the honey bee population?
\end{itemize}

The remaining parts of this article are structured as follows: In Section 2, we provide details of how we modeled seasonality in the egg-laying rate and a general modeling framework for the interactions of parasitism and honey bees. In Section 3, we address how seasonality impacts the survival of honey bee colonies and their population dynamics. We theoretically demonstrate the impacts of parasitism on the honey bee populations without seasonality. In Section 4, we explore how parasites and seasonality might influence colony survival and population dynamics. In the last section, we conclude our theoretical and bifurcation analysis results regarding the effects of seasonality and parasites on colony dynamics and propose future studies needed for understanding how climate-related factors may threaten honey bee colonies.\\

\section{Model Derivation}

In this section, we focus on modeling the honeybee-parasite colony dynamics with seasonality. Let $H(t)$ be the population of the honey bee and $M(t)$ be the population of the mites in a given colony at time $t$. We assume that:

\begin{itemize}
\item[\textbf{A1:}]The term $\frac{H^2}{K+H^2} $ reflects the cooperative brood care from adult bees that perform nursing and collecting food for brood \cite{chen2020model,messan2021population,messan2018effects,schmickl2007hopomo,kang2016disease,eischen1984some}, where $\sqrt{K}$ indicates the colony size at which brood survival rate is half maximum.\\

\item[\textbf{A2:}] We assume that the queen egg-laying rate is seasonal ($r(t)$) due to resource constraints. The literature work suggests that food, temperature, weather, and oviposition place would affect the queen \cite{bodenheimer1937studies,khoury2011quantitative,degrandi1989beepop}. Motivated by literature \cite{chen2020model,messan2021population,chen2021review} and analysis of recent experimental data \cite{fisher2021colony}, we model the egg-laying rate with seasonality as follows:\\
	\begin{equation}
	    r(t)=r_0(1+\epsilon \cos(\frac{2\pi(t -\psi)}{\gamma}))
	\end{equation}
with $\epsilon\in(0,1)$ measuring the intensity of seasonal impacts, $r_0$ representing the average of egg-laying rate,  $\gamma$ representing the length of seasonality, and $\psi$ being the time of the maximum laying rate.\\

\item[\textbf{A3:}] Female mites breed offspring in the cell, and complete the mating in the cell. In the phoretic phase, female mites feed on adult bees and immigrate to other colonies \cite{Vetharaniam:2006aa}. In the reproductive phase, mites attach to foraging bees and then reproduce offspring in the cell \cite{ramsey2019varroa}. Based on the biological background and literature work \cite{messan2021population,betti2014effects,sumpter2004dynamics}, we model the honeybee-parasite interaction as follows: $$\frac{a H}{b + c H}$$ where $a$ is the mite parasitism rate to the honey bee, $c$ is parasite attachment effects, and $b$ is the size of honey bee population at which rate of attachment is half maximal. \\

\item[\textbf{A4:}] Female mites need nutrition from honey bees to produce the next generation. The parameter $\sigma$ indicates conversion rate of nutrient consumption obtained from bees into nutrients needed by mites to reproduce.
\end{itemize}

The four assumptions above lead to the following nonautonomous and non-linear ordinary differential equations of the honeybee-parasite interaction model with seasonality (Model \eqref{Honeybee-mite}):
\begin{equation}\label{Honeybee-mite}
\begin{array}{lcl}
H' &=& \frac{r(t) H^2}{K+H^2}-d_hH-\frac{a H}{b + c H} M,\\
M' &=& \frac{\sigma a H}{b+c H} M - d_m M,
\end{array}
\end{equation} 
with $r(t)=r_0(1+\epsilon \cos(\frac{2\pi(t -\psi)}{\gamma}))$.\\

\textbf{Note:} If $b=1$ and $c=0$, Model \eqref{Honeybee-mite} reduces to the previous work of Kang et al. (2016) \cite{kang2016disease} disease free model; and if $c=1$, Model \eqref{Honeybee-mite} reduces to our previous works of Messan et al. (2017 \& 2021) \cite{messan2017migration,messan2021population}. Thus the current model \eqref{Honeybee-mite} processes the general interaction properties of honey bees and parasitism. \\

\begin{table}[]
\footnotesize
\begin{tabular}{|l|l|l|l|}
\hline
Parameter                 & Definition    (Units)                                               & Parameter             & Definition        (Units)                                                                                          \\ \hline
H                         & Honey bee population     (bees)                                    & M                     & Parasite (mites) population   (bees)                                                                             \\ \hline
a                         & The parasitism rate to honey bee    (per day)                          & b                     & \begin{tabular}[c]{@{}l@{}}The size of honey bee population \\ at which the rate of attachment is \\ half-maximal (bees)  \end{tabular}                        \\ \hline
c                         & Parasite attachment effects                                  & $r_0$                     & The average of egg-laying rate (bees/day)                                                                                         \\ \hline
$\sqrt{K}$ & \begin{tabular}[c]{@{}l@{}}The colony size at which \\ brood survival rate is\\  half-maximum (bees)  \end{tabular} & $\sigma$ &\begin{tabular}[c]{@{}l@{}}The conversion rate of nutrient \\ consumption obtained from bees \\ to sustenance for mites' reproduction\end{tabular} \\ \hline
$d_h$    \&       $d_m $           & \begin{tabular}[c]{@{}l@{}}The death rate of honey bee \\ and parasite (mites) (per day) \end{tabular}                                  &      $\gamma$            &       the length of seasonality     (days)                                        
\\ \hline
$\psi$ & \begin{tabular}[c]{@{}l@{}}The time of the maximum\\ laying rate (days)\end{tabular}  & $\epsilon$ & the strength of  seasonality\\ \hline
\end{tabular}
\end{table}

In the following two sections, we will provide our detailed study to obtain insights regarding how may seasonality and/or parasitism alone or combined impact honey bee population dynamics.

\section{Mathematics Analysis}
To facilitate our analysis of the proposed system, we start with re-scaling our system \eqref{Honeybee-mite}. Assume that $b\neq0$, $c\neq0$ and $\sigma \neq 0$, let $u=\frac{c}{b}H$, $v = \frac{c}{b\sigma}M$, $\hat{K}=\frac{K c^2}{b^2}$, $\omega=\frac{a \sigma}{c}$, $\bar{r}(t)=\frac{r(t) c}{b}$, $\bar{d}_h=d_h$ and $\bar{d}_m=d_m$, then system \eqref{Honeybee-mite} can scaled by following:

\begin{equation}\label{Honeybee-mite-scaled}
\begin{array}{lcl}
u' &=& \frac{\bar{r}(t)u^2}{\hat{K}+u^2}-\bar{d}_h u-\frac{\omega u}{1+u} v\\
v'&=& \frac{\omega u}{1+u} v-\bar{d}_m v
\end{array}
\end{equation}

{
We first show that the proposed model \eqref{Honeybee-mite-scaled} is positive invariant and bounded in $\mathbb{R}^{2}_{+}$ as the following theorem:}\\

\begin{theorem}
\label{th:positive}
Assume that all parameters are non-negative. Model \eqref{Honeybee-mite-scaled} with initial value $u(0)=u_0$, $v(0)=v_0$, and $(u_0,v_0) \in \mathbb  X$ possesses a unique solution, and  the space $\mathbb  X$ is positively invariant and bounded in $\mathbb{R}^{2}_{+}$. 
\end{theorem}

\begin{remark} Theorem \ref{th:positive} provides us reassurances that the proposed model \eqref{Honeybee-mite-scaled} is well defined biologically, provides bases for our careful designed numerical studies.
\end{remark}

\subsection{Impact of Seasonality on Honeybee-Only Population Dynamics} \label{sec:honeybee-only}
If there is no mites, i.e., $v(0)=0$, 
the model \eqref{Honeybee-mite} reduces to the following bee-only population model with seasonality:
\begin{equation}\label{honeybee}
u' = \frac{\bar{r}(t)u^2}{\hat{K}+u^2}-\bar{d}_h u
\end{equation} with $\bar{r}=r_0(1+\epsilon \cos(\frac{2\pi(t -\psi)}{\gamma}))$ which satisfies a Lipschitz condition for all $u\geq0$. Thus according to Theorem \ref{th:positive},  the initial value problem with $u(0)\geq0$ has a unique non-negative and bounded solution.\\

In order to study the effects of the strength of seasonality ($\epsilon$) and the length of seasonality ($\gamma$) on bee populations, we start with the dynamics of the Honeybee-only model \eqref{honeybee} when  $\bar{r}(t)=r_0$ is a constant. The honeybee-only system without seasonality \eqref{honeybee} has two equilibria $u_i^{*}, i=1,2$ shown as below provided $r_0>2\bar{d}_h\sqrt{\hat{K}}$: $$u_1^{*}=\frac{r_0-\sqrt{r_0^2-4\bar{d}_h^2\hat{K}}}{2\bar{d}_h}, \quad u_2^{*}=\frac{r_0+\sqrt{r_0^2-4\bar{d}_h^2\hat{K}}}{2\bar{d}_h}.$$ 

\begin{figure}[htbp]
		\centering
		\subfigure[Seasonality leads to the collapsing of the colony when $u_0=1$]{
			\includegraphics[width=5cm]{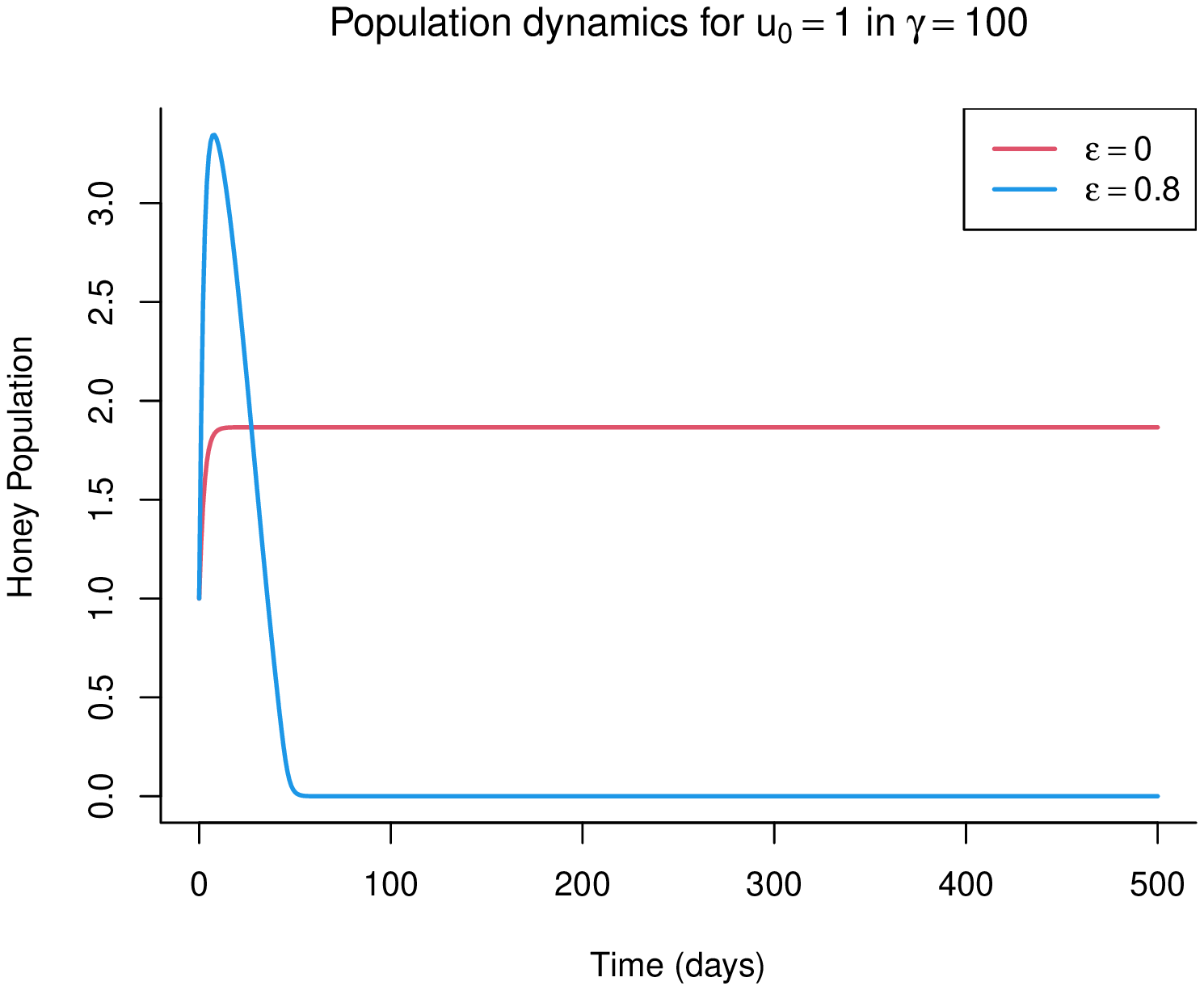}\label{fig:pop_suv}
		}
		\subfigure[Seasonality can promote the survival of the colony when $u_0=0.1$]{
			\includegraphics[width=5cm]{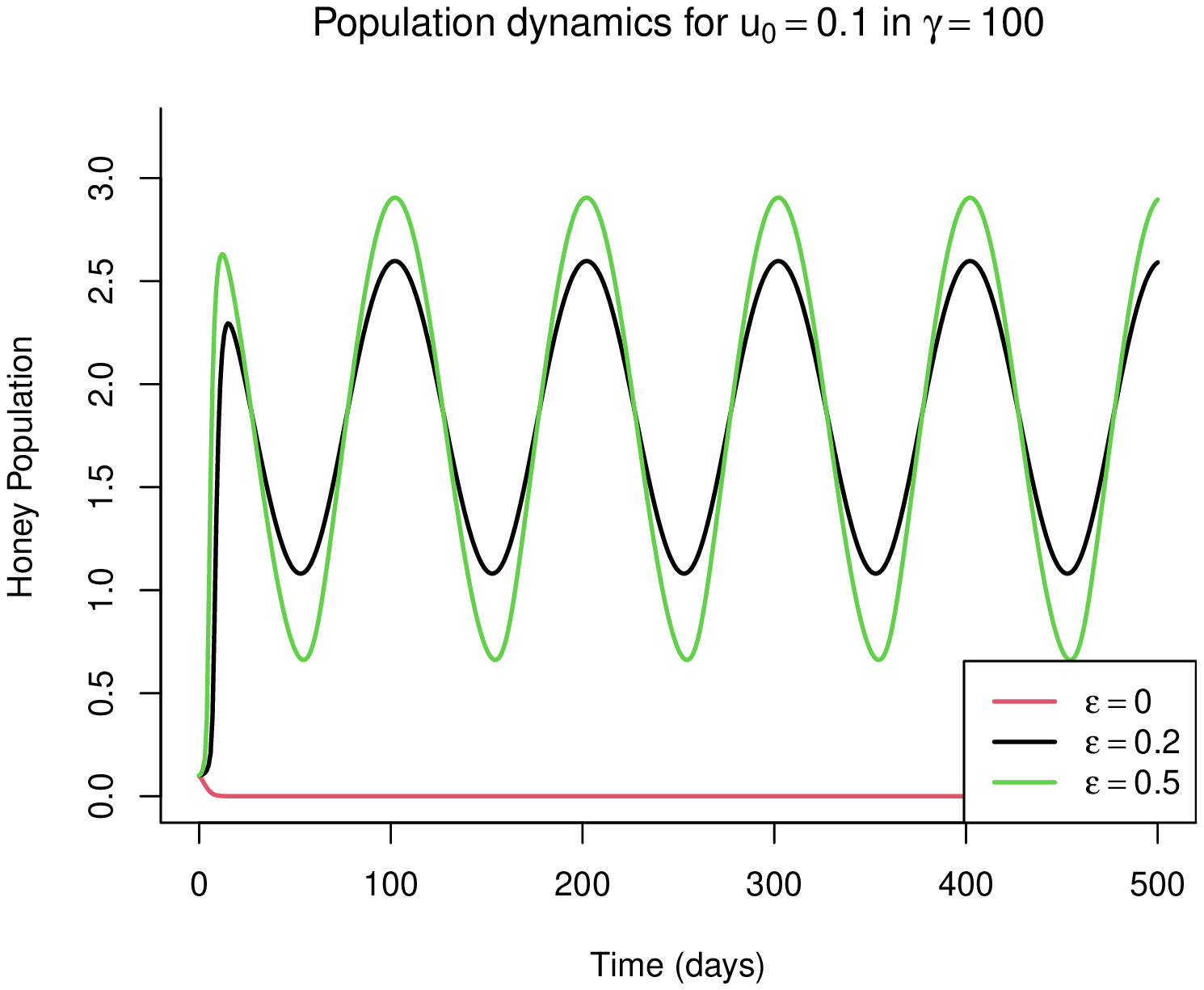}\label{fig:pop_clps}
		}
			\caption{Population dynamics of honeybee-only model \eqref{honeybee} with or without seasonality by setting $r_0=1$, $\bar{d}_h=0.5$, $\hat{K}=1/4$, $\psi=0$ and $\gamma=100$ with $u_0=0.1$ or $1$ as its initial population. }
			\label{fig:1D-climate}
\end{figure}

The global dynamics of \eqref{honeybee} when $\bar{r}(t)=r_0$ can be summaries as the following proposition:
\begin{proposition}\label{p4honeybee}
If $r_0<2\bar{d}_h\sqrt{\hat{K}}$, then the population of $u(t)$ converges to 0 for any initial condition $u(0)\geq 0$. In the case that $r_0>2\bar{d}_h\sqrt{\hat{K}}$, $u(t)$ converges to 0 for any initial condition $u(0)<u_1^* $ while $u(t)$ converges to $u_2^*$ for any initial condition $u(0)>u_1^* $.
\end{proposition}
\noindent\textbf{Notes:} Proposition \ref{p4honeybee} indicates that the relationship among the constant egg-laying rate $r_0$, the honey bee mortality, and the half-maximum rate $\hat{K}$, as well as initial conditions, determine whether the honey bee colony can survive. With the larger egg-laying rate $r_0$ with the larger initial condition $u_0$, the honey bee colony is more likely to survive. 
In the case that the egg-laying rate is seasonal, $\bar{r}(t)=r_0(1+\epsilon \cos(\frac{2\pi(t -\psi)}{\gamma}))$ with  its average value over each seasonal length $\gamma$ being $r_0$, the consequence of honey bee population dynamics can be complicated. Examples shown in Figure \ref{fig:1D-climate} suggest that the seasonality in the egg-laying rate can promote the survival of honey bees when the intensity of seasonality is not too high, and it can also make the honey bee colony prone to collapsing when the intensity of seasonality is high. \\


In Figure \ref{fig:1D-climate}, without seasonality $\epsilon=0$, the honey bee colony with $r_0=1$, $\bar{d}_h=0.5$, $\hat{K}=1/4$, $\psi=0$ and $\gamma=100$ can survive under its initial condition $u(0)=1$ (red curve in Figure \ref{fig:pop_suv}) while it collapses under its initial condition $u(0)=0.1$ (red curve in Figure \ref{fig:pop_clps}). When the intensity of seasonality is not too high, i.e., $\epsilon=0.2$ or $0.5$, the honey bee colony can survive under its initial condition $u(0)=0.1$ (black and green curves in Figure \ref{fig:pop_clps}). This is an example showing that seasonality can promote the survival of a honey bee colony. On the other hand, When the intensity of seasonality is high, i.e., $\epsilon=0.8$ (blue curve in Figure \ref{fig:pop_suv}), the honey bee colony collapses with the initial condition of $u(0)=1$ when the honey bee colony can survive without seasonality. This is an example showing that seasonality can make honey bee colony collapse under certain conditions. \\

In order to  explore the impact of the intensity of seasonality  $\epsilon$, 
we first define the minimum and maximum value of the egg-laying rate function: $r_m=\min{\bar{r}(t)}=r_0(1-\epsilon)$ and $r_M=\max{\bar{r}(t)}=r_0(1+\epsilon)$. Motivated by Proposition \ref{p4honeybee}, the intensity of seasonality can be classified into the following three cases:
\begin{enumerate}
    \item The low egg-laying rate if  $r_M=r_0(1+\epsilon)\leq 2\bar{d}_h\sqrt{\hat{K}}$. This case is equivalent to 
    $$0 \leq \epsilon \leq 1-\frac{2\bar{d}_h\sqrt{\hat{K}}}{r_0}$$
    \item The high egg-laying rate if $r_m =r_0(1-\epsilon)\geq 2\bar{d}_h\sqrt{\hat{K}}$. This case 
    is equivalent to 
    $$0 \leq \epsilon \leq \frac{2\bar{d}_h\sqrt{\hat{K}}}{r_0}-1$$
    \item The intermediate egg-laying rate if $r_m=\min{\bar{r}(t)}=r_0(1-\epsilon) < 2\bar{d}_h\sqrt{\hat{K}} \leq r_M=\max{\bar{r}(t)}=r_0(1+\epsilon)$. This is the case when $$\max\{1-\frac{2\bar{d}_h\sqrt{\hat{K}}}{r_0}, \frac{2\bar{d}_h\sqrt{\hat{K}}}{r_0}-1\} \leq \epsilon \leq 1.$$
\end{enumerate}
Now we have the following theorem:
\begin{theorem}\label{th:honeybee-rt}Let $r_M=r_0(1+\epsilon)$ and $r_m=r_0(1-\epsilon)$.
If the egg-laying rate $\bar{r}(t)=r_0(1+\epsilon \cos(\frac{2\pi(t -\psi)}{\gamma}))$ is low, i.e., $r_M=r_0(1+\epsilon)\leq 2\bar{d}_h\sqrt{\hat{K}}$, the honey bee population $u(t)$ converges to zero for any initial condition $u(0)\geq 0$. In the case that the egg-laying rate $\bar{r}(t)$ is high, i.e., $r_m =r_0(1-\epsilon)\geq 2\bar{d}_h\sqrt{\hat{K}}$, honey bee population $u(t)$ can survive if the initial condition $u(0)>\frac{r_m-\sqrt{r_m^2-4\bar{d}_h^2\hat{K}}}{2\bar{d}_h}$. More specifically, we have
$$ \frac{r_m-\sqrt{r_m^2-4\bar{d}_h^2\hat{K}}}{2\bar{d}_h}<\liminf_{t\rightarrow\infty} u(t)\leq \limsup_{t\rightarrow\infty} u(t) <\frac{r_M+\sqrt{r_M^2-4\bar{d}_h^2\hat{K}}}{2\bar{d}_h} $$ if 
$r_m\geq 2\bar{d}_h\sqrt{\hat{K}}$ and $u(0)>\frac{r_m-\sqrt{r_m^2-4\bar{d}_h^2\hat{K}}}{2\bar{d}_h}$.
\end{theorem}
\noindent\textbf{Notes:} Theorem \ref{th:honeybee-rt} implies that we can focus on how seasonality impacts honey bee population when the egg-laying rate $\bar{r}(t)$ is not low, i.e.,$r_M=r_0(1+\epsilon)\geq 2\bar{d}_h\sqrt{\hat{K}}$ which includes the case 2 and 3. Because the low egg-laying rate  leads the colony to collapse. Thus, we can reduce the three cases above to the following two cases by introducing the critical intensity of seasonality $\epsilon_c=\frac{2\bar{d}_h\sqrt{\hat{K}}}{r_0}-1$
\begin{enumerate}
 \item The low intensity of seasonality, i.e.,
    $$0 \leq \epsilon \leq \epsilon_c$$
    \item The high intensity of seasonality, i.e.,  $$0<\epsilon_c\leq \epsilon \leq 1.$$
\end{enumerate}

By applying Proposition 3.1 and the method used in Ratti et al.(2015) \cite{ratti2015mathematical}, we obtain the stability condition when Model \ref{honeybee} processes a periodic solution $u^*$ as the following theorem:
\begin{theorem}\label{th:honeybee-stability}
    Suppose $u(t) = u^*$ are periodic solutions of the Model \ref{honeybee}, and $f(u)=\frac{u^2}{\hat{K}+u^2}$. Then $u(t) = u^*$ is stable if $\lambda=\int_0^t\left[\bar{r}(z)*f'(u^*)-\bar{d}_h\right]dz < 0$, or is unstable if $\lambda > 0$, where $f'(u^*)=\frac{2 \hat{K} u^*}{\left(\hat{K}+(u^*)^2\right)^2}$.
\end{theorem}

\noindent\textbf{Notes:} Theorem \ref{th:honeybee-stability} shows that the stability of the periodic solution of Model \ref{honeybee}  requires $\int_0^t\left[\bar{r}(z)*f'(u^*)-\bar{d}_h\right]dz < 0$, thus $u=0$ is always locally stable as the case without seasonality.  

\begin{figure}[ht]
		\centering
		\subfigure[$\gamma=4$]{
			\includegraphics[width=3cm]{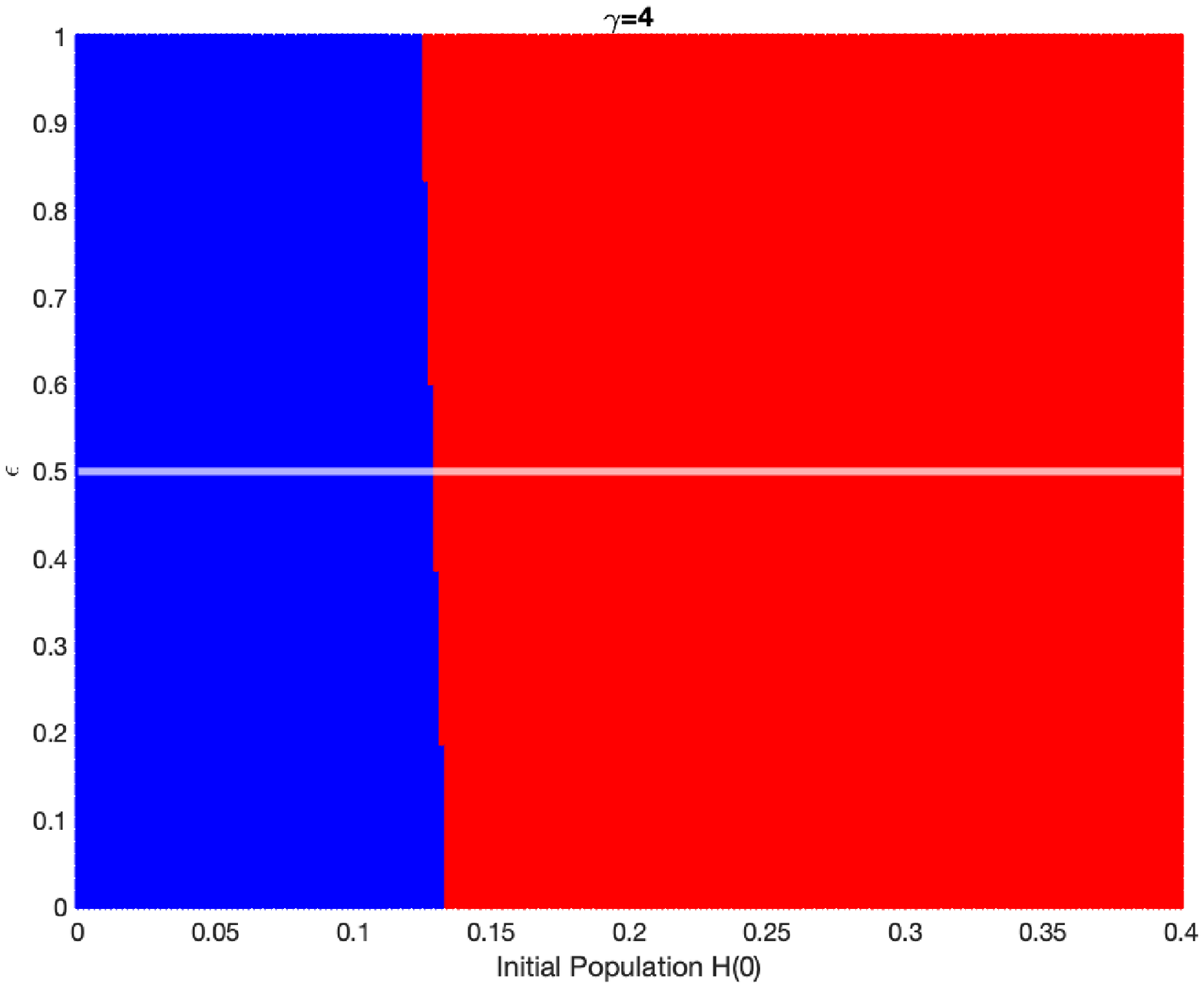}
		}
		\subfigure[$\gamma=40$]{
			\includegraphics[width=3cm]{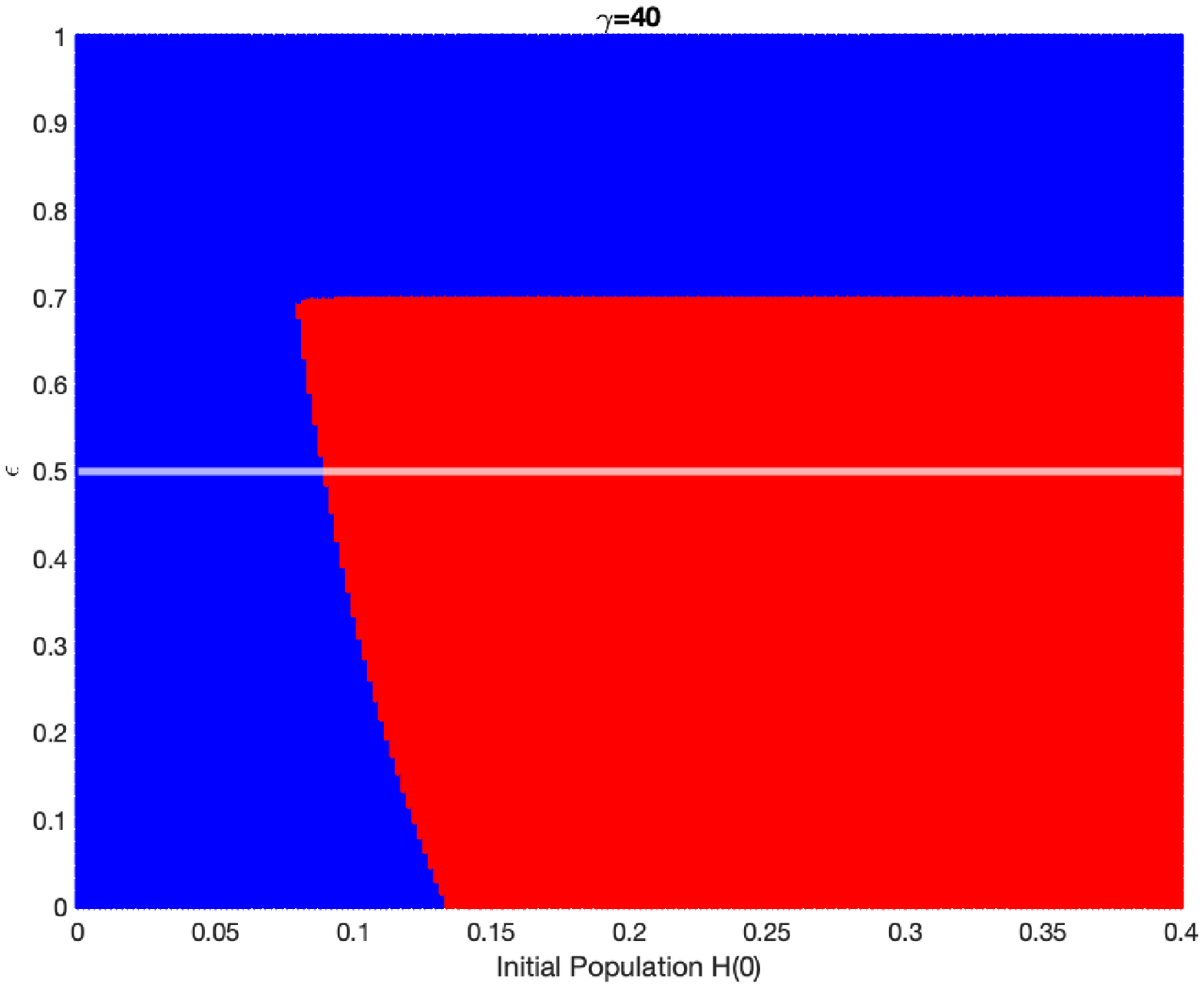}
		}
		\quad
		\subfigure[$\gamma=400$]{
			\includegraphics[width=3cm]{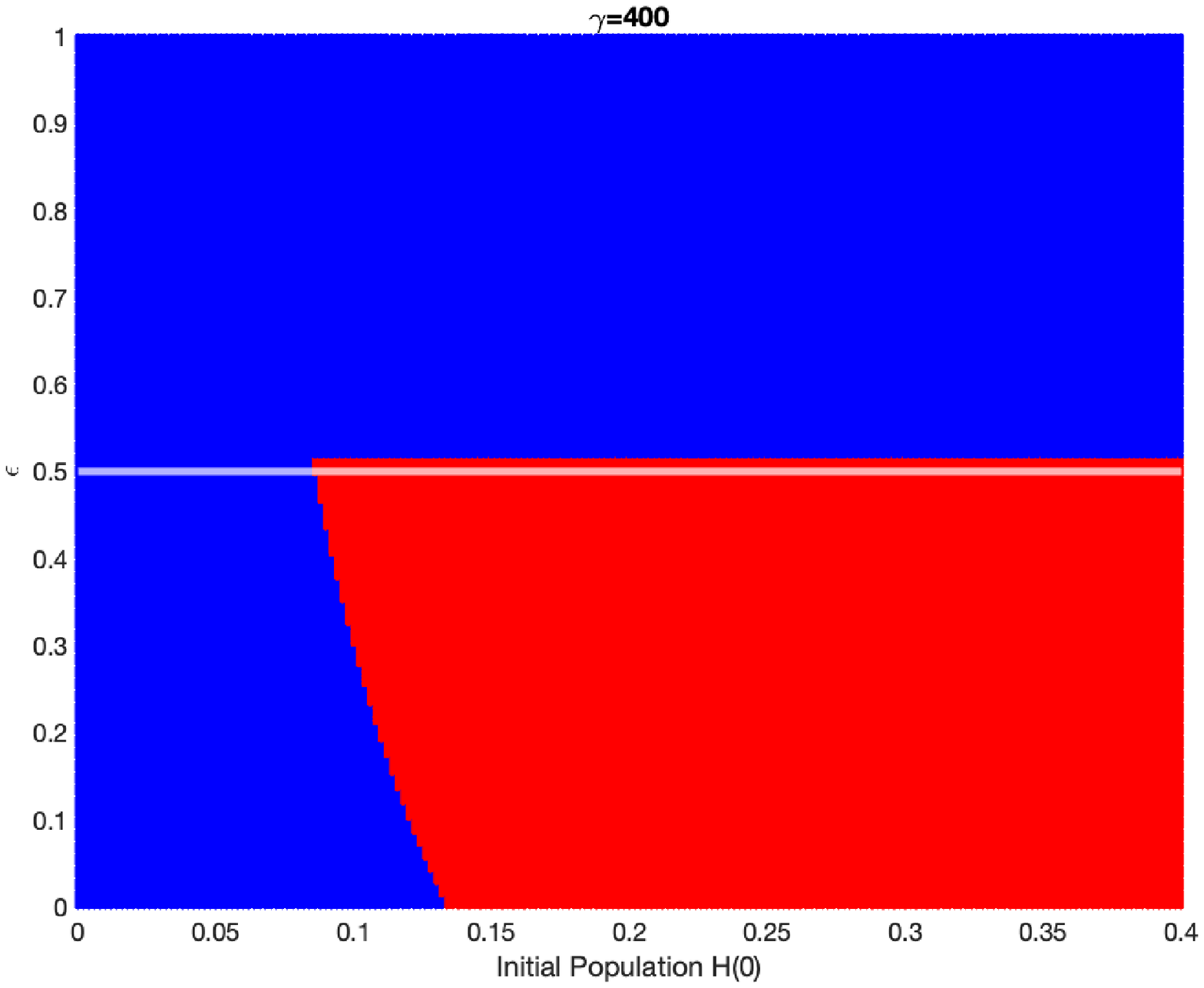}
		}
			\caption{Impacts of the strength of seasonality ($\epsilon$) and the length of seasonality ($\gamma$). The blue area is colony collapse and red area is colony survive.  $r_0=1$, $\bar{d}_h=0.5$, $\hat{K}=1/4$ and $\psi=0$. Honey bee initial population is $u_0 \in [0, 0.4]$ }
			\label{fig:1D-gamma}
\end{figure}

\begin{figure}[ht]
		\centering
		\subfigure[$\gamma=4, \psi=0$]{
			\includegraphics[width=2.5cm]{Figure/gamma4.eps}
		}
		\subfigure[$\gamma=4, \psi=1$]{
			\includegraphics[width=2.5cm]{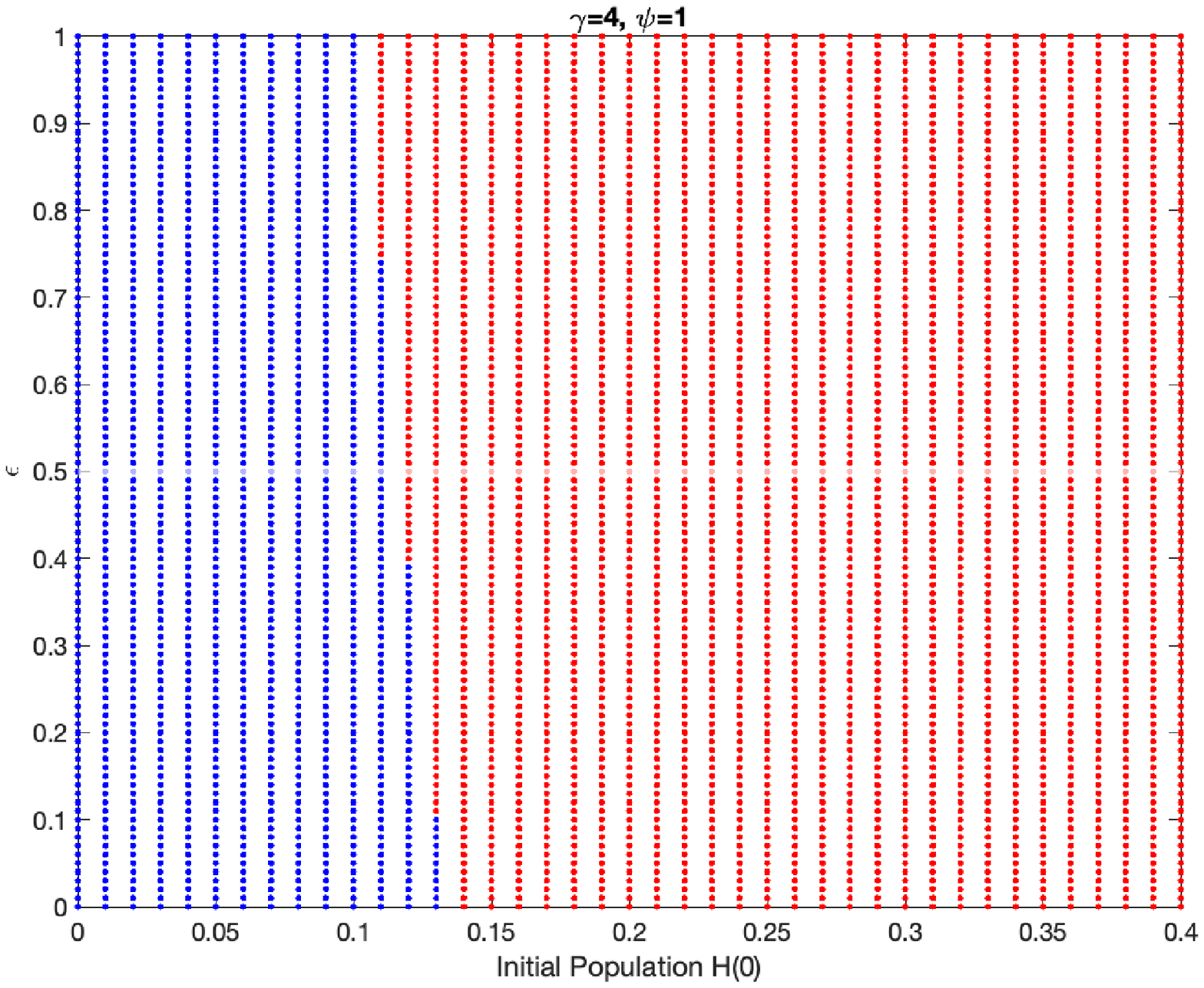}
		}
		\subfigure[$\gamma=4, \psi=2$]{
			\includegraphics[width=2.5cm]{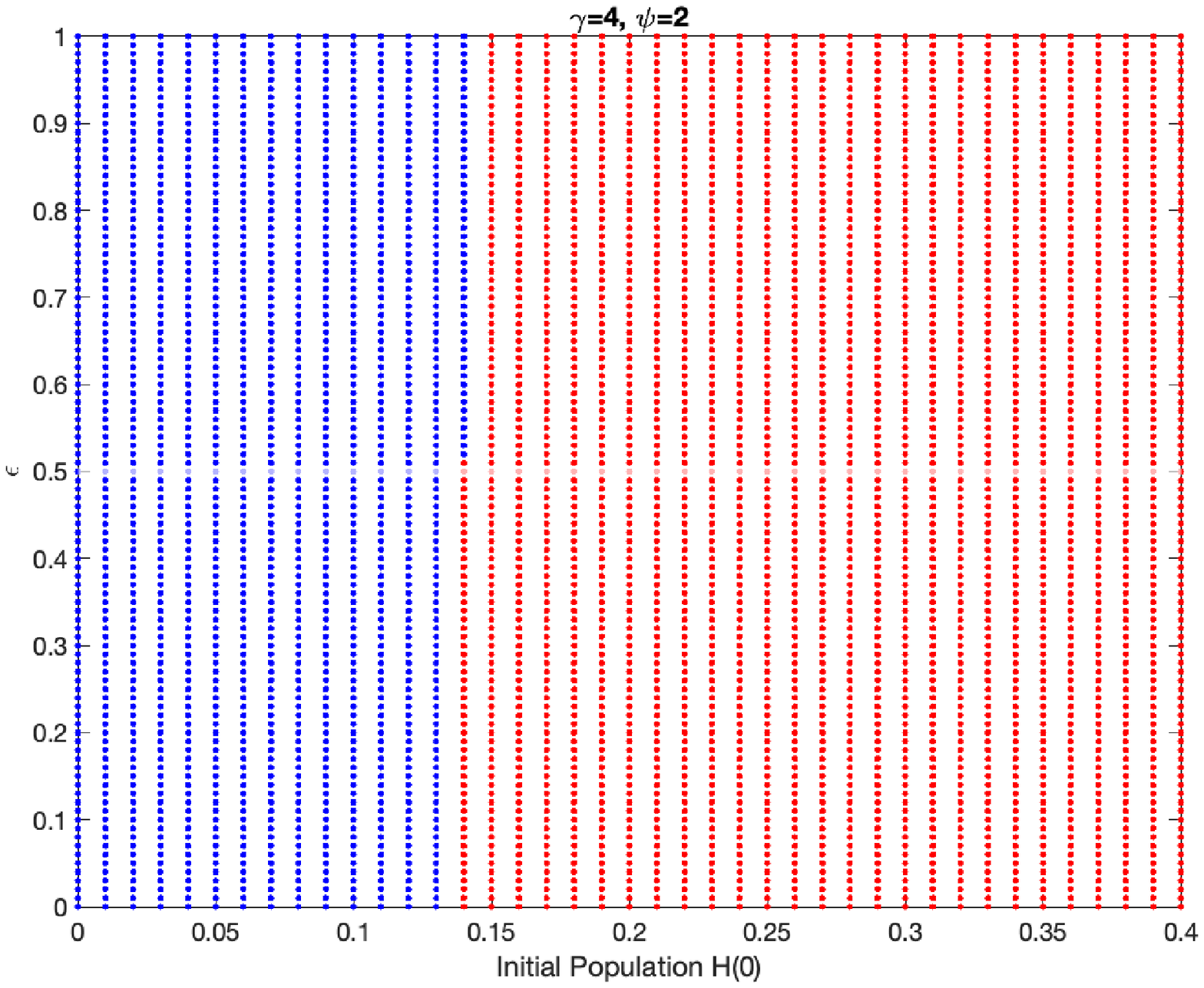}
		}
			\subfigure[$\gamma=4, \psi=3$]{
			\includegraphics[width=2.5cm]{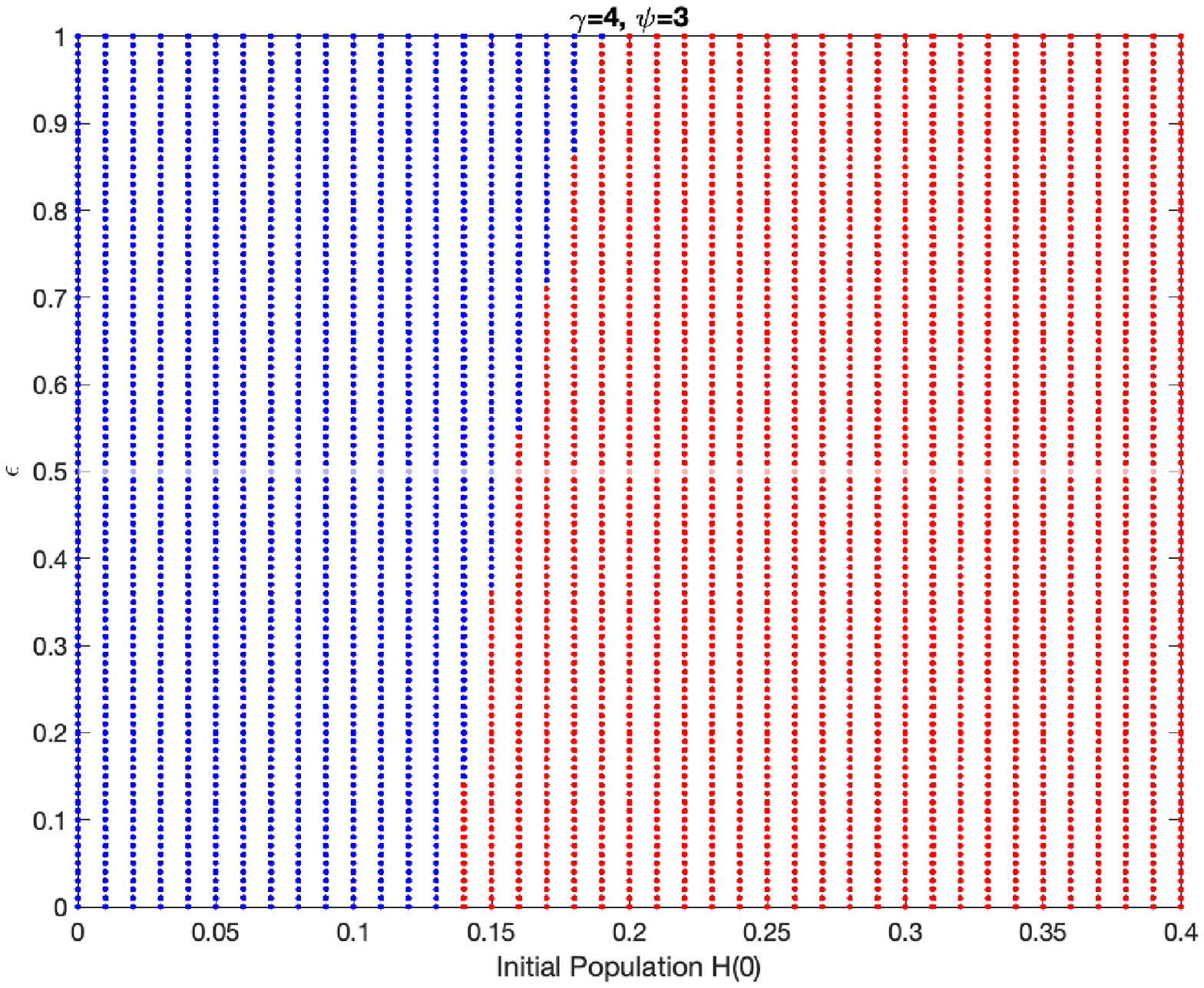}
		}
		\quad
		\subfigure[$\gamma=40, \psi=0$]{
			\includegraphics[width=2.5cm]{Figure/gamma40.eps}
		}
		\subfigure[$\gamma=40, \psi=10$]{
			\includegraphics[width=2.5cm]{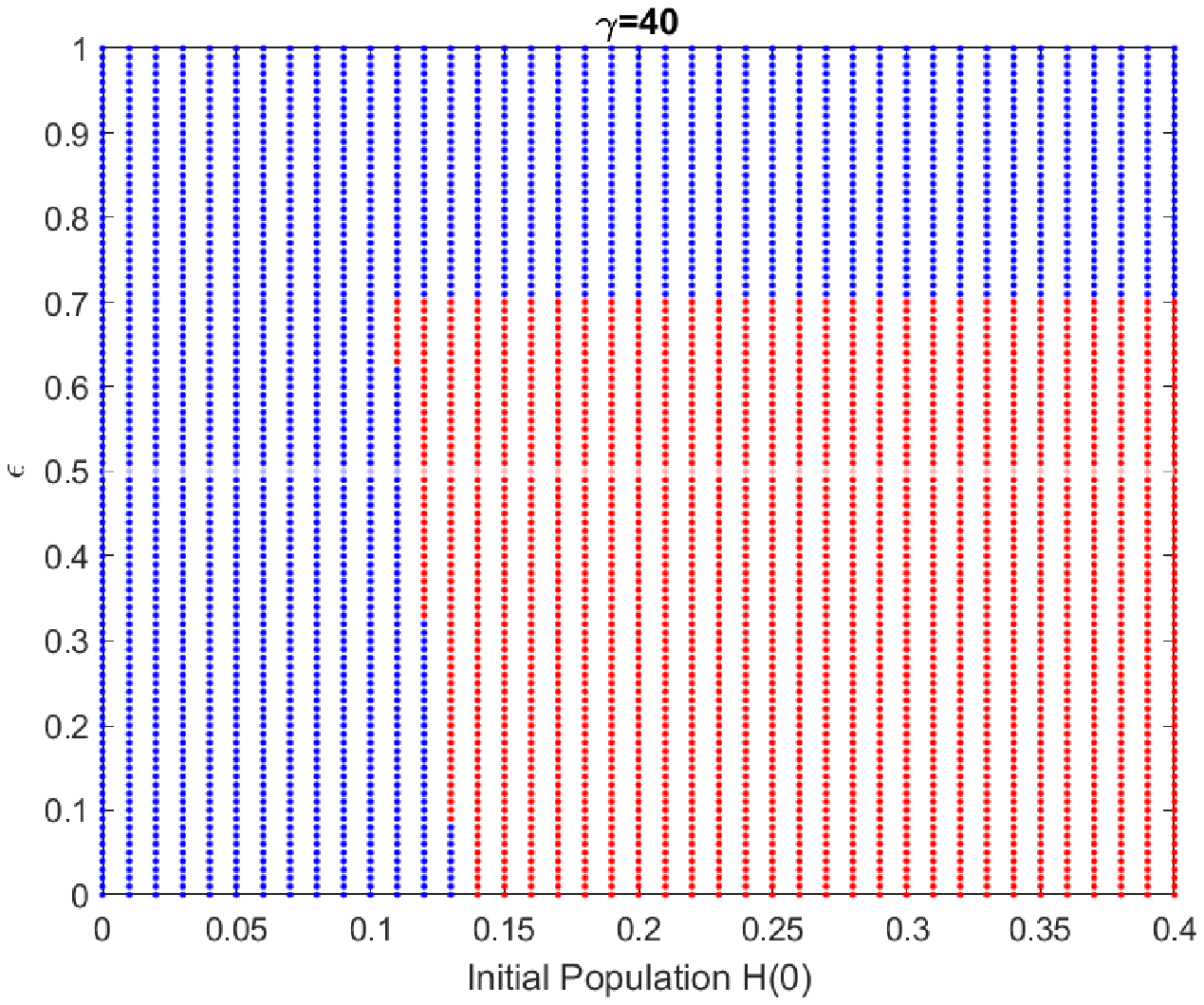}
		}
		\subfigure[$\gamma=40, \psi=20$]{
			\includegraphics[width=2.5cm]{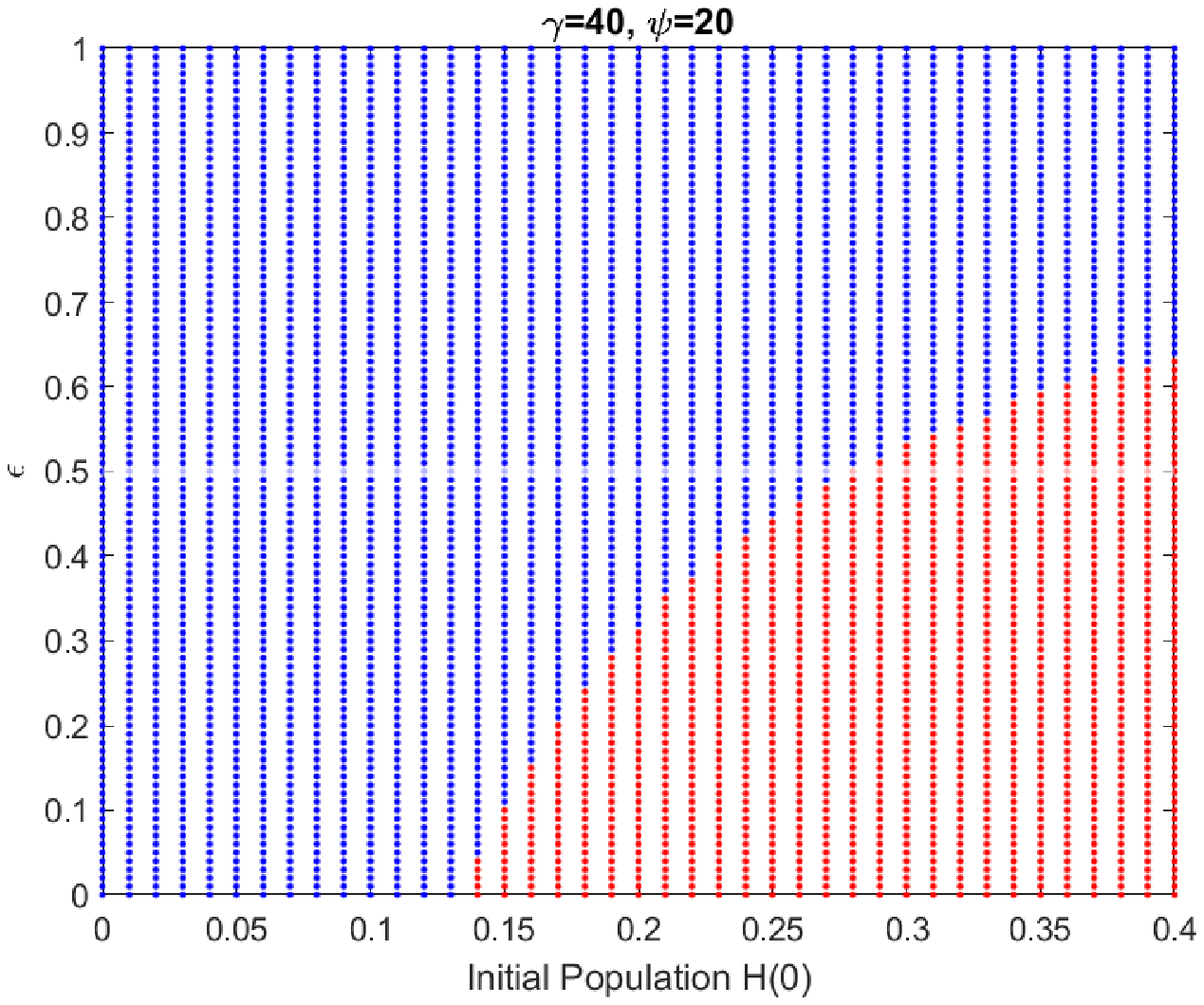}
		}
			\subfigure[$\gamma=40, \psi=30$]{
			\includegraphics[width=2.5cm]{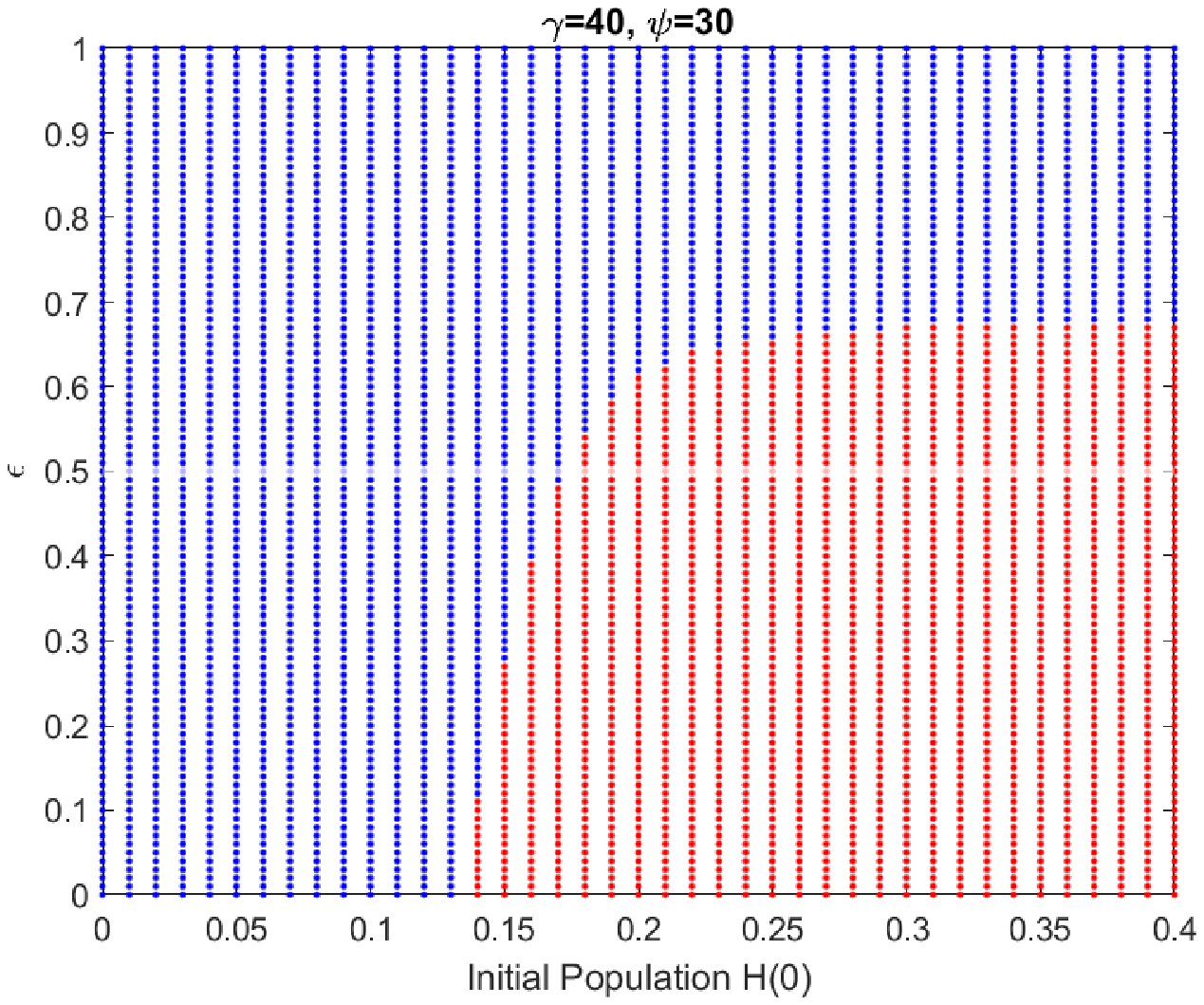}
		}
			\caption{Impacts of the maximum laying rate ($\psi$). The blue area is colony collapse and the red area is colony survival. The horizontal line is the dividing line between $\epsilon$ in results 1 and 3. $r_0=1$, $\bar{d}_h=0.5$ and $\hat{K}=1/4$. Honey bee initial population is $u_0 \in [0, 0.4]$ }
			\label{fig:1D-psi}
\end{figure}
To further address the impacts of seasonality on honey bee population dynamics, 
we provide basins of attractions for Model \eqref{honeybee} in Figure \ref{fig:1D-gamma} and Figure \ref{fig:1D-psi} by setting $\bar{d}_h=0.5, \hat{K}=1/4,	r_0=1$.
We set $\psi=0$ in Figure \ref{fig:1D-gamma}. The x-axis is the initial honey bee population $u(0)$, and the y-axis is the intensity of seasonality measured by $\epsilon$.  Those parameter values gives $\epsilon_c=0.5$ which is a white horizontal line in Figure \ref{fig:1D-gamma} and Figure \ref{fig:1D-psi}.
The blue region in Figures is the  value of the strength of seasonality ($\epsilon$) and the corresponding initial conditions that lead the colony to collapse, while the red region is the value of $\epsilon$ and $u(0)$ that lead to the colony survival.\\

Figure \ref{fig:1D-gamma} and Figure \ref{fig:1D-psi} suggest that the strength of seasonality ($\epsilon$),  the length of seasonality ($\gamma$), and the time of the maximum laying rate ($\psi$) impact the survival of honey bee colony in the synergistic ways:
\begin{enumerate}
    \item The length of seasonality ($\gamma$) is small, e.g., $\gamma=4$: 
    \begin{itemize}
        \item If the time of the maximum laying rate ($\psi$) is less than the half period $\gamma$, the seasonality seems to promote the survival of the colony in the sense that the initial bee population that originally leads to collapsing but it leads to colony survival with seasonality. 
        \item If the time of the maximum laying rate ($\psi$) is larger than the half period $\gamma$, the seasonality seems to suppress the survival of the colony in the sense that the initial bee population that originally leads to survival but it leads to colony collapsing with seasonality. 
        
    \end{itemize}
    \item When the length of seasonality ($\gamma$) is larger, e.g., $\gamma=40, 400$, the large intensity of seasonality $\epsilon$ can lead to the collapsing of the colony while the impacts of the smaller intensity of seasonality $\epsilon$ depends on the timing of the maximum laying rate ($\psi$) as follows: 
    \begin{itemize}
        \item If the time of the maximum laying rate ($\psi$) is less than the half period $\gamma$, the seasonality seems to promote the survival of the colony.
        \item If the time of the maximum laying rate ($\psi$) is larger than the half period $\gamma$, the seasonality seems to suppress the survival of the colony. 
        
    \end{itemize}
\end{enumerate}

\subsection{Impact of Parasitism on honey bee Population without Seasonality} \label{sec:bee-parasite}

In this subsection, we focus on dynamics of the honeybee-parasite interaction model \eqref{Honeybee-mite-scaled} in the absence of seasonality, i.e., $\bar{r}(t)=\bar{r}$. Thus, we have the following rescaled model \eqref{honeybee-mite-constant}:
\begin{equation}\label{honeybee-mite-constant}
    \begin{array}{lcl}
u' &=& \frac{\bar{r}u^2}{\hat{K}+u^2}-\bar{d}_h u-\frac{\omega u}{1+u} v\\
v'&=& \frac{\omega u}{1+u} v-\bar{d}_m v
\end{array}
\end{equation} that would allow us to obtain biological insights on how parasitism impacts the honey bee population by comparing the dynamics of $v(0)=0$ versus $v(0)>0$. In the case that $v(0)=0$, the model \eqref{Honeybee-mite-scaled} reduces to the honey bee only model in the constant environment \eqref{honeybee} whose dynamics are summarized in Proposition \ref{p4honeybee}. \\

Let $(u^*,v^*)$ be an equilibrium of Model \eqref{Honeybee-mite-scaled}, then it satisfies the following equations:

\begin{equation}\label{M1}
  \frac{\bar{r}(u^*)^2}{\hat{K}+(u^*)^2}-\bar{d}_h u^*-\frac{\omega u^*}{1+u^*} v^*=0, 
\end{equation}
\begin{equation}\label{M2}
 \frac{\omega u^*}{1+u^*} v^*-\bar{d}_m v^*=0 \Rightarrow (\frac{\omega u^*}{1+u^*}-\bar{d}_m) v^*=0
\end{equation}

Solving Eqt.\ref{M2} gives $v^*=0$ or $u^*=\frac{ \bar{d}_m}{\omega  -\bar{d}_m}$. And if $v^*=0$, then Eqt.\ref{M1} is $$  \frac{\bar{r}(u^*)^2}{\hat{K}+(u^*)^2}-\bar{d}_hu^*=0,$$ which gives the following two positive solutions provided that $\bar{r}>2\bar{d}_h\sqrt{\bar{K}}$, $$u^*_1=\frac{\bar{r}-\sqrt{\bar{r}^2-4 \hat{K} \bar{d}_h^2}}{2 \bar{d}_h}$$ or $$u^*_2=\frac{\bar{r}+\sqrt{\bar{r}^2-4 \hat{K} \bar{d}_h^2}}{2 \bar{d}_h}.$$\\

In the case that $\omega  >\bar{d}_m$, we have $u^*=\frac{ \bar{d}_m}{\omega  -\bar{d}_m}$ and $v^*=\frac{\left[\bar{r}u^*-\bar{d}_h \left((u^*)^2+\hat{K}\right)\right](1 +u^*)}{ \omega((u^*)^2+\hat{K})}$ as the unique interior equilibrium of Model \ref{Honeybee-mite-scaled}. The stability of the equilibrium point can be evaluated through the following Jacobean matrix of Model \ref{Honeybee-mite-scaled} is
$$J = \begin{Bmatrix}
-\bar{d}_h+\frac{2 \hat{K} \bar{r} u}{\left(\hat{K}+u^2\right)^2}-\frac{  \omega v}{(1 +u)^2} & -\frac{  \omega u}{1 +u}\\
\frac{\omega v }{(u+1)^2} & \frac{ \omega u}{u+1}-\bar{d}_m
\end{Bmatrix}$$
Now we are the following on the dynamics of the Honeybee-Parasite system \eqref{Honeybee-mite-scaled}:

\begin{theorem}\label{th:Ee}
[\textit{Dynamics of Honeybee-Parasite system \eqref{Honeybee-mite-scaled}}] 
The system \eqref{Honeybee-mite-scaled} can have one, three, or four equilibria whose existence and stability conditions are listed in Table \ref{t1:DF}. The global dynamics of Model \eqref{Honeybee-mite-scaled} can be summarized as follows:
\begin{enumerate}
    \item The system \eqref{Honeybee-mite-scaled} converges to extinction $(0, 0)$ for almost all initial conditions if one the three conditions holds (1)$\frac{\bar{r}}{2\sqrt{\hat{K}}}<d_h$; (2) $\omega>\bar{d}_m$; or (3)$\bar{N}^c_h>u^*$.
    \item If $\omega<\bar{d}_m$ or $\bar{N}^*_h<u^*$, depending on initial condition, the trajectory of system \eqref{Honeybee-mite-scaled} converges to either $(0, 0)$ or $(\bar{N}^*_h, 0)$.
    \item If $\bar{N}^c_h<u^*<\bar{N}^*_h$, then system \eqref{Honeybee-mite-scaled} has a unique interior equilibrium $(u^*, v^*)$ which is locally asymptotically stable when $\hat{K} < \hat{K}_1 $ and is a source when $\hat{K} > \hat{K}_1 $.
\end{enumerate}
\end{theorem}

\begin{table}[ht]
\begin{tabular}{|l|l|l|l|l|l|}
\hline
Equilibria             & Existence condition                                      & Stability condition \\ \hline
  $(0, 0)$           &    Always exists                                      & Always locally stable   \\ \hline
$(\bar{N}^c_h, 0)$ & $\frac{\bar{r}}{2\sqrt{\hat{K}}}>d_h$                & Saddle if $\bar{N}^c_h<u^*$; Source if $\omega<\bar{d}_m$ or $\bar{N}^c_h>u^*$  \\ \hline
 $(\bar{N}^*_h, 0)$& $\frac{\bar{r}}{2\sqrt{\hat{K}}}>\bar{d}_h$ &  Sink if $\bar{N}^*_h<u^*$ or $\omega<\bar{d}_m$; Saddle if $\bar{N}^*_h>u^*$ \\ \hline
$(u^*, v^*)$      & $\omega>\bar{d}_m$ \& $\frac{\bar{r} u^*}{\hat{K}+(u^*)^2}>\bar{d}_h$ & Sink if $\hat{K} < \hat{K}_1 $; Source if $\hat{K} > \hat{K}_1$ \\ \hline
\end{tabular}
\caption{The existence and stability of equilibrium for Model \ref{Honeybee-mite-scaled}, where 
$\bar{N}^c_h=\frac{\bar{r}-\sqrt{\bar{r}^2-4 \hat{K} \bar{d}_h^2}}{2 \bar{d}_h},\,\,\bar{N}^*_h=\frac{\sqrt{\bar{r}^2-4 \hat{K} \bar{d}_h^2}+\bar{r}}{2 \bar{d}_h}, u^*= \frac{  \bar{d}_m}{ \omega -\bar{d}_m}, v^*=\frac{\left[\bar{r}u^*-\bar{d}_h \left((u^*)^2+\hat{K}\right)\right](1 +u^*)}{\omega( (u^*)^2+\hat{K})}, \hat{K}_1=\frac{-\sqrt{\bar{r}} \sqrt{\bar{r} (2 u^*+1)^2-8 \bar{d}_h  (u^*)^2 (u^*+1)}+2 \bar{r} u^*+\bar{r}-2 \bar{d}_h  (u^*)^2}{2 \bar{d}_h }$.}
\label{t1:DF}
\end{table}

\noindent\textbf{Notes:} Theorem \ref{th:Ee} provides us a global picture of the dynamics of the system \eqref{Honeybee-mite-scaled} and the related biological implications of the impact of parasitism on honey bee population dynamics in constant conditions. Theorem \ref{th:Ee} suggests that parasitism can have negative impacts on the honey bee population in three ways: (1) May lead to the collapsing of the colony; (2) May lead to the coexistence of both honey bee and parasitism but the honey bee population decreases compared to the case without parasitism, or (3) May destabilize the honey bee population. \\

Item (3) needs further theoretical exploration regarding how may parasitism destabilize the colony dynamics. For example, the colony destabilizes to show fluctuating dynamics through supercritical Hopf-bifurcation; or to collapse supercritical Hopf-bifurcation.\\

By applying the results in \cite{wang2011predator}, our system \ref{Honeybee-mite-scaled} undergoes a Hopf-bifurcation. To study further, we re-scaled the system \ref{Honeybee-mite-scaled} to the following model:
\begin{equation}\label{Honeybee-mite-scaled3}
\begin{array}{lcl}
u'&=& g(u)(f(u)-v)\\
v'&=& v(g(u)-\bar{d}_m),
\end{array}
\end{equation}where $g(u)=\frac{\omega u}{1+u}$ and $f(u)=\frac{\bar{r}}{g(u)}\cdot\frac{u^2}{\hat{K}+u^2}-\frac{\bar{d}_h}{g(u)}\cdot u$. We can verify that our system \ref{Honeybee-mite-scaled3} satisfies the following conditions:\\

(a1) $f \in C^1(\bar{\mathbb{R}}), f(a)=f(b)=0$, where $0<a<b$; $f(u)$ is positive for $a<u<b$, and $f(u)$ is negative otherwise; there exists $\bar{\lambda} \in (a,b)$ such that $f'(u)>0$ on $[a,\bar{\lambda})$, $f'(u)<0$ on $(\bar{\lambda},b]$;\\

(a2) $g \in C^1(\bar{\mathbb{R}}), g(0)=0$; $g(u)>0$ for $u>0$ and $g'(u) > 0$ for $u>0$, and there exists $\lambda>0$ such that $g(\lambda)=d$.\\

(a3) $f(u)$ and $g(u)$ are $C^3$ near $\lambda=\bar{\lambda}$ and $f''(\bar{\lambda})<0$.\\

Then according to Theorem 3.1 in Wei et al. (2011) \cite{wang2011predator}, we can conclude that our system \ref{Honeybee-mite-scaled3} exists the first Lyapunov coefficient

\begin{equation*}
\begin{array}{lcl}
a(\bar{\lambda
})&=&\frac{f'''(\bar{\lambda}) g(\bar{\lambda}) g'(\bar{\lambda})+2f''(\bar{\lambda})[g'(\bar{\lambda})]^2-f''(\bar{\lambda})g(\bar{\lambda})g''(\bar{\lambda})}{16g'(\bar{\lambda})}\\
&=&\frac{\omega}{16(1+\bar{\lambda})}(2 f''(\bar{\lambda})+\bar{\lambda} f'''(\bar{\lambda}))
\end{array}
\end{equation*}
where 

\begin{equation}\label{hopf_a}
\small{
2 f''(\bar{\lambda})+\bar{\lambda} f'''(\bar{\lambda})=\frac{2 \bar{r} \left(2 \hat{K}^3-\hat{K}^2 (2 \bar{\lambda} (2 \bar{\lambda}+9)+3)+2 \hat{K} (\bar{\lambda} (4-3 \bar{\lambda})+9) \bar{\lambda}^2+(2 \bar{\lambda}-3) \bar{\lambda}^4\right)}{\omega  \left(\hat{K}+\bar{\lambda}^2\right)^4}}
\end{equation}

Thus, we have the following results on Hopf-bifurcations:
\begin{theorem}\label{th:hopf}
The system \ref{Honeybee-mite-scaled} undergoes a supercritical Hopf-bifurcation at $\hat{K} = \hat{K}_1$ with $a(\bar{\lambda})<0$, and a subcritical Hopf-bifurcation at $\hat{K} = \hat{K}_1 $ with  $a(\bar{\lambda})>0$.
\end{theorem}

\noindent\textbf{Note:} Theorem \ref{th:hopf} implies that the system \ref{Honeybee-mite-scaled} can undergo a supercritical or subcritical Hopf-bifurcation depending on the relationship between $\hat{K}$ and $\bar{\lambda}$. If the system goes supercritical bifurcation at $\hat{K}_1$, then it has a stable limit cycle surrounding a source equilibrium when $\hat{K} > \hat{K}_1$. When the system \ref{Honeybee-mite-scaled} undergoes a subcritical Hopf-bifurcation, then both population of honey bees and the parasitic mites go to zero through the unstable limit cycle. Biologically, it implies that parasitism in the constant environment can destabilize the dynamics and even lead to colony collapse, thus parasitism has negative impacts on honey bee population dynamics.




\section{Synergistic Impacts of Parasitism and Seasonality}
In the previous two sections, we explore the impacts of seasonality on the honey bee population and the impacts of parasitism on the honey bee population in a constant environment, respectively. Our study shows that seasonality can have positive or negative effects on the survival of honey bee colonies depending on the values of the strength of seasonality $\epsilon$, the period $\gamma$, and the timing of the maximum egg-laying rate $\psi$. Our theoretical work shows that parasitism in general has negative impacts on the survival of honey bee colonies in a constant environment.\\

In this section, we will explore how seasonality combined with parasitism affects honey bee population dynamics.  We start with the following theorem regarding the stability condition when Model \ref{Honeybee-mite-scaled} processes a periodic solution of $(u^*,0)$ by applying Floquet theory theorem and the approach in Ratti et al.(2015) \cite{ratti2015mathematical}.

\begin{theorem} \label{th:non-2d}
    Suppose $u^*(t)$ is a periodic positive solution of the Model \ref{honeybee}, and $f(u)=\frac{u^2}{\hat{K}+u^2}$. Then, $(u^*,0)$ is a periodic solution of Model \ref{Honeybee-mite-scaled}, and $f'(u)=\frac{2 \hat{K} u}{\left(\hat{K}+u^2\right)^2}$. It is stable if $\int_0^T\left[\bar{r}(t)*f'(u^*)-\bar{d}_h\right]dt < 0$ and $\int_0^T\left[\frac{\omega u^*}{1+u^*}-\bar{d}_m\right] dt<0$.
\end{theorem}

\noindent \textbf{Note:} Theorem \ref{th:non-2d} implies that $(0,0)$ is always locally stable, thus initial conditions play important roles in the survival of honeybee colonies. \\

By comparing the results of Theorem \ref{th:Ee} and Theorem \ref{th:non-2d}, we can see that the impact of seasonality: the seasonality in the egg laying rate $r(t)$ generates the periodic solution $u^*(t)$ whose stability requires $\int_0^T\left[\bar{r}(t)*f'(u^*)-\bar{d}_h\right]dt<0 \quad \text{and} \quad \int_0^T\left[\frac{  \omega u^*}{1 +u^*}-\bar{d}_m\right]dt<0$. Those conditions reduce to $rf'(u^*)<\bar{d}_h$ \mbox{ and } $\frac{  \omega u^*}{1 +u^*}<\bar{d}_m$ when $r(t)=r$ being  a constant. \\

By comparing the results of Theorem \ref{th:honeybee-stability} and Theorem \ref{th:non-2d}, we can see the impact of parasitism. Specifically, the stability of nontrial periodic boundary solution $(u^*,0)$ requires $\int_0^T\left[\frac{\omega u^*}{1+u^*}-\bar{d}_m\right] dt<0$.\\

Note that our honeybee-parasite model \eqref{honeybee} exhibits strong Allee effects in honey bees due to collaborative behavior in the colony. There is limited theoretical work on exploring the impacts of both parasitism and seasonality. Ratti et al.(2015)\cite{ratti2015mathematical} developed a honeybee-mite-virus model with seasonality.  Their model also exhibits strong Allee effects in honey bees while their mite-free solution is always unstable due to their formulation of the mite population. They discussed the existence of periodic solution and its stability in the bee-only model and discussed the stability of the disease-free solution and mite-free solution through linearization and the method of Floquet theory in the bee-mite model and bee-mite-virus model respectively. The most recent work that can be related to our topic is the paper by Rebelo and Soresina (2020) \cite{rebelo2020coexistence}. Their paper proposed and studied a prey-predator model with weak or strong Allee effects in a periodic environment. They discussed the stability conditions of trivial, nontrivial solutions, and periodic solutions. They also showed that different initial conditions might lead to the extinction of both species or the coexistence of two species that  converges to a stable periodic orbit. \\

To further our understanding of the impacts of seasonality and parasitism, we perform simple time series simulations and observe the following by setting 
$$\bar{r}_0=1,\,\bar{d}_h=0.2,\, \bar{d}_m=0.21,\,\omega=0.3,\,\hat{K}=4.49,\,\psi=0$$
\begin{enumerate}
\item In the absence of seasonality and parasitism, a honey bee colony can establish its population when its initial condition is greater than $1.173$ otherwise, it collapses. 
\item With seasonality but without parasitism, Figure \ref{cycle_time_H1} suggests that seasonality can promote the survival of a honey bee colony when its initial condition is 1 ($<1.173$) and it can also make a honey bee colony prone to collapse when its initial condition is above $1.173$ (see the black curve in Figure \ref{cycle_time_H1.2}).
\item With parasitism but without seasonality, a honey bee colony can survive through the stable limit cycle around the interior equilibrium $(2.33, 0.3875)$ for the right initial conditions. For example, a honey bee colony survives when $u_0=1.2$ and $v_0=0.02$ (see the red curve in Figure \ref{Compare_no}) while it collapses when the initial parasites population grows up to $0.05$ (see the red curve in \ref{Compare_40}).
\begin{figure}[ht]
		\centering
\subfigure[Seasonality promotes honey bee survival]{
			\includegraphics[width=5cm]{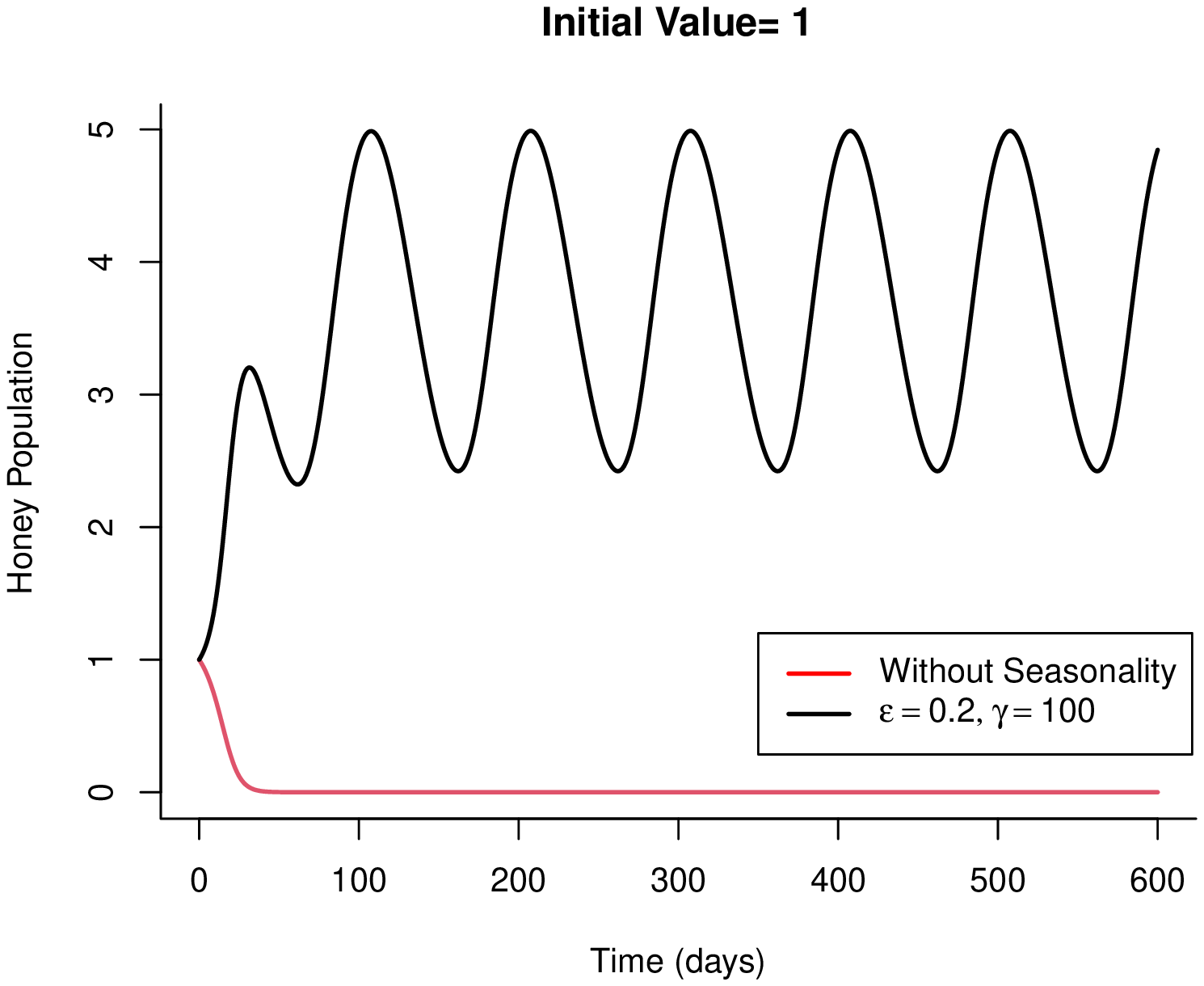}
			\label{cycle_time_H1}
		}	
		\subfigure[Seasonality leads to the colony collapsing]{
			\includegraphics[width=5cm]{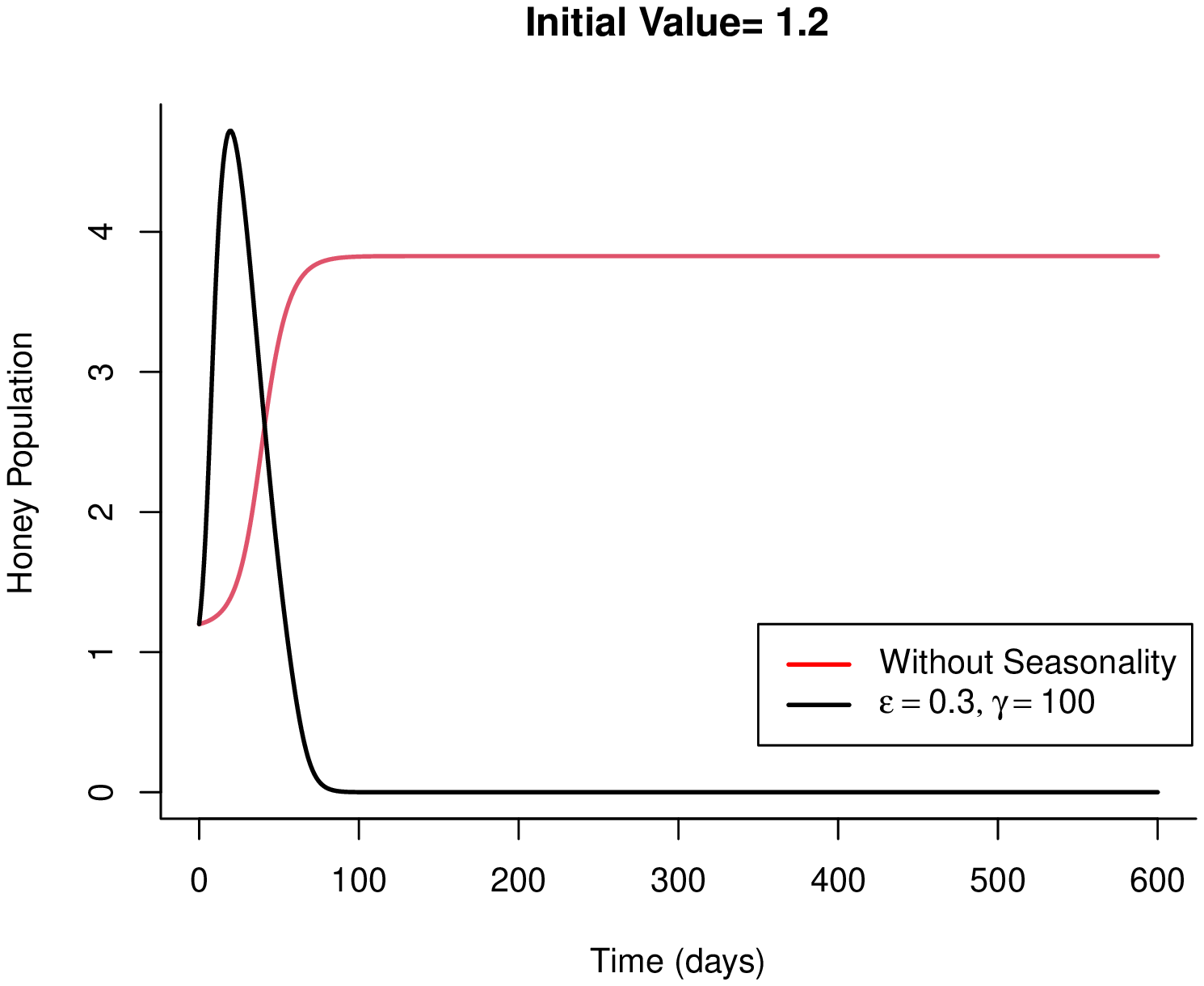}
			\label{cycle_time_H1.2}
		}
\caption{Comparison examples of seasonality having positive or negative effects in the honey bee colony survival without parasitism. Red curves are honey bee populations without seasonality and black curves are honey bee populations with seasonality.} 
			\label{fig:bee_only_time}
\end{figure}
\item With both seasonality and parasitism, Figure \ref{Compare_40} suggests that seasonality can promote the survival of a honey bee colony when the parasite's initial population is $0.05$ and the seasonality can also make the honey bee colony prone to collapse when the parasite initial population is $0.02$ (see Figure \ref{Compare_no}).

\begin{figure}[ht]
		\centering
\subfigure[Seasonality promotes honey bee survival]{
			\includegraphics[width=5cm]{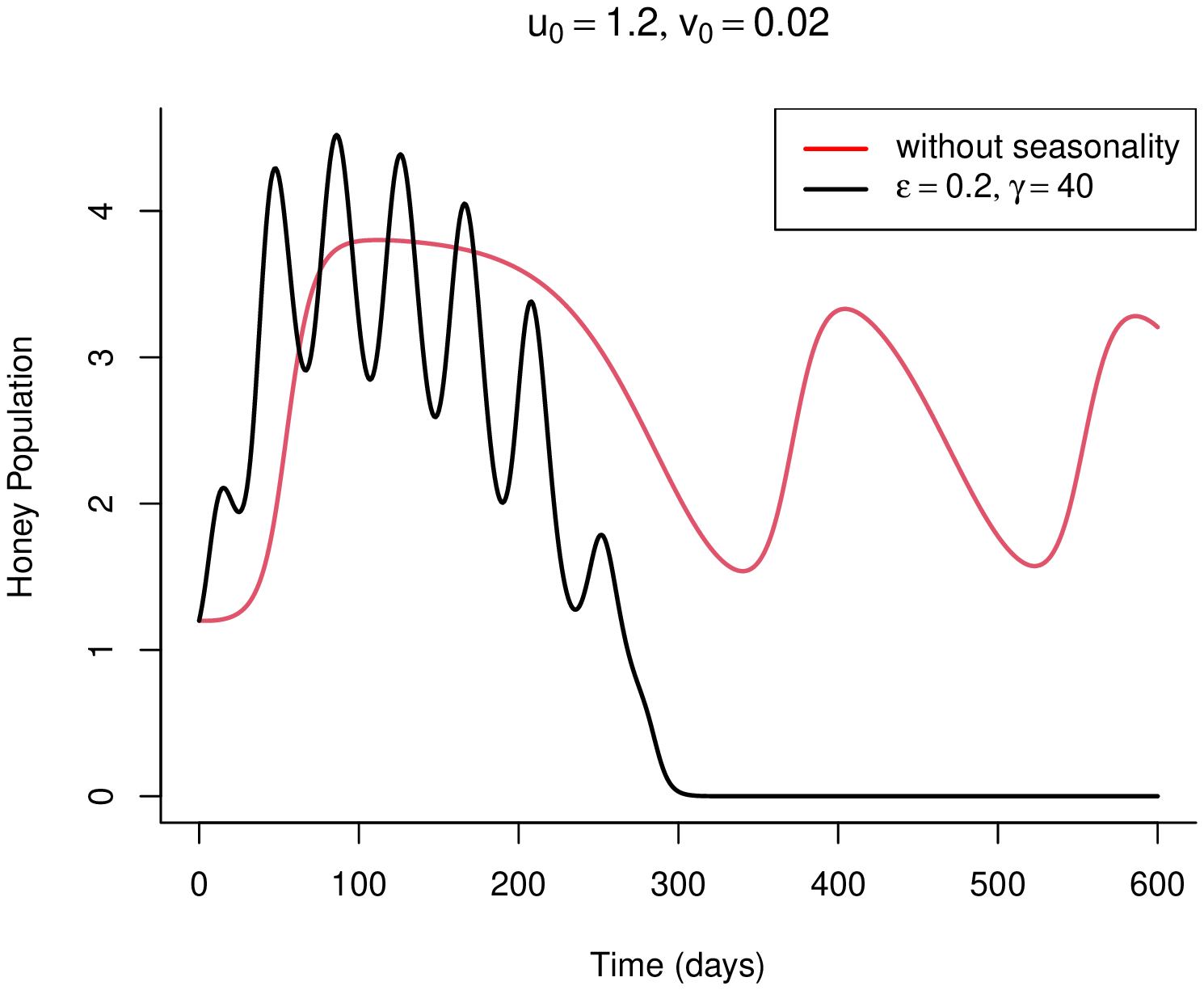}
			\label{Compare_no}
		}	
		\subfigure[Seasonality leads to the colony collapsing]{
			\includegraphics[width=5cm]{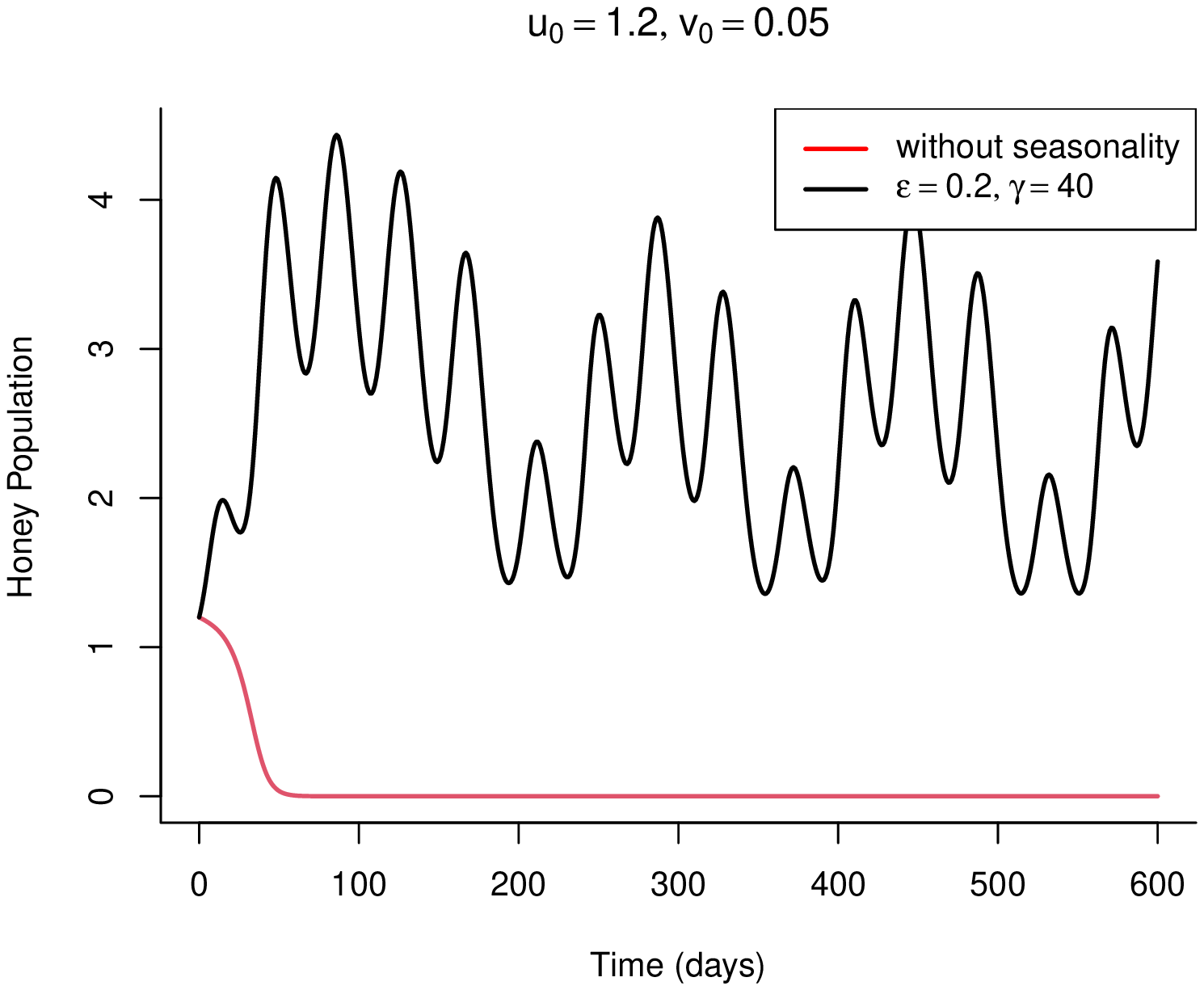}
			\label{Compare_40}
		}
\caption{Comparison examples of seasonality having positive or negative effects in the honey bee colony survival with parasitism. Red curves are the honey bee populations without seasonality and black curves are the honey bee population with seasonality.} 
			\label{fig:limit_cycle_time}
\end{figure}

\end{enumerate}

The observations above suggest that seasonality combined with parasitism may have positive or negative impacts on the honey bee colony survival depending on varied conditions. To explore further, we will perform a bifurcation analysis to understand how may the strength of seasonality $\epsilon$, the length of seasonal period $\gamma$, the timing of the maximum egg-laying rate $\psi$, and the severity of parasitism measured by $\omega$ in the following two scenarios of honeybee-parasitism dynamics in the absence of seasonality:
\begin{itemize}
    \item \textbf{Honey bee and parasitism Coexists at a stable equilibrium}
      \item \textbf{Honey bee and parasitism Coexists as a stable limit cycle}
\end{itemize}

\subsection{Impacts of seasonality on the stable equilibrium coexistence}
We choose a typical example of our honeybee-parasite interaction model \eqref{Honeybee-mite-scaled}  by setting 
$$\bar{r}_0=2.86, \bar{d}_h=\bar{d}_m=0.25, \omega=0.3, \hat{K}=2.04$$ which has a bistability between the colony collapsing state $(0,0)$ and the survival equilibrium at the locally stable equilibrium point $(5,5.5769)$ whose basins of attractions are red area shown in Figure \ref{before_no season1}.\\

To further explore the impacts of the seasonality strength $\epsilon$ and the period of seasonality $\gamma$ on the colony survival and population dynamics, without loss of generosity, we set the queen laying her maximum number of eggs at time $\psi=0$, and we perform the following simulations (Figure \ref{fig:2D_gammaS}, \ref{fig:2D_gammaM}, \ref{fig:2D_gammaL}) on basin's attractions of our honeybee-parasite model \eqref{Honeybee-mite-scaled}.
\begin{enumerate}
    \item When the period of the seasonality  $\gamma$ is small, e.g., $\gamma=4$, comparisons of areas of basin attractions for the colony survival among Figure \ref{before_no season1} (no seasonality), \ref{no_42} (the seasonality strength $\epsilon=0.2$), and \ref{no_48} (the seasonality strength $\epsilon=0.8$), suggest that seasonality strength $\epsilon$ may not impact the basin attractions of the colony survival but it impacts the population dynamics as shown in Figure \ref{pop-gammaS}. Simulations suggest that the larger value of the strength of seasonality $\epsilon$, the larger amplitude  of the population.\\

\begin{figure}[ht]
		\centering
	\subfigure[no seasonality]{
			\includegraphics[width=5cm]{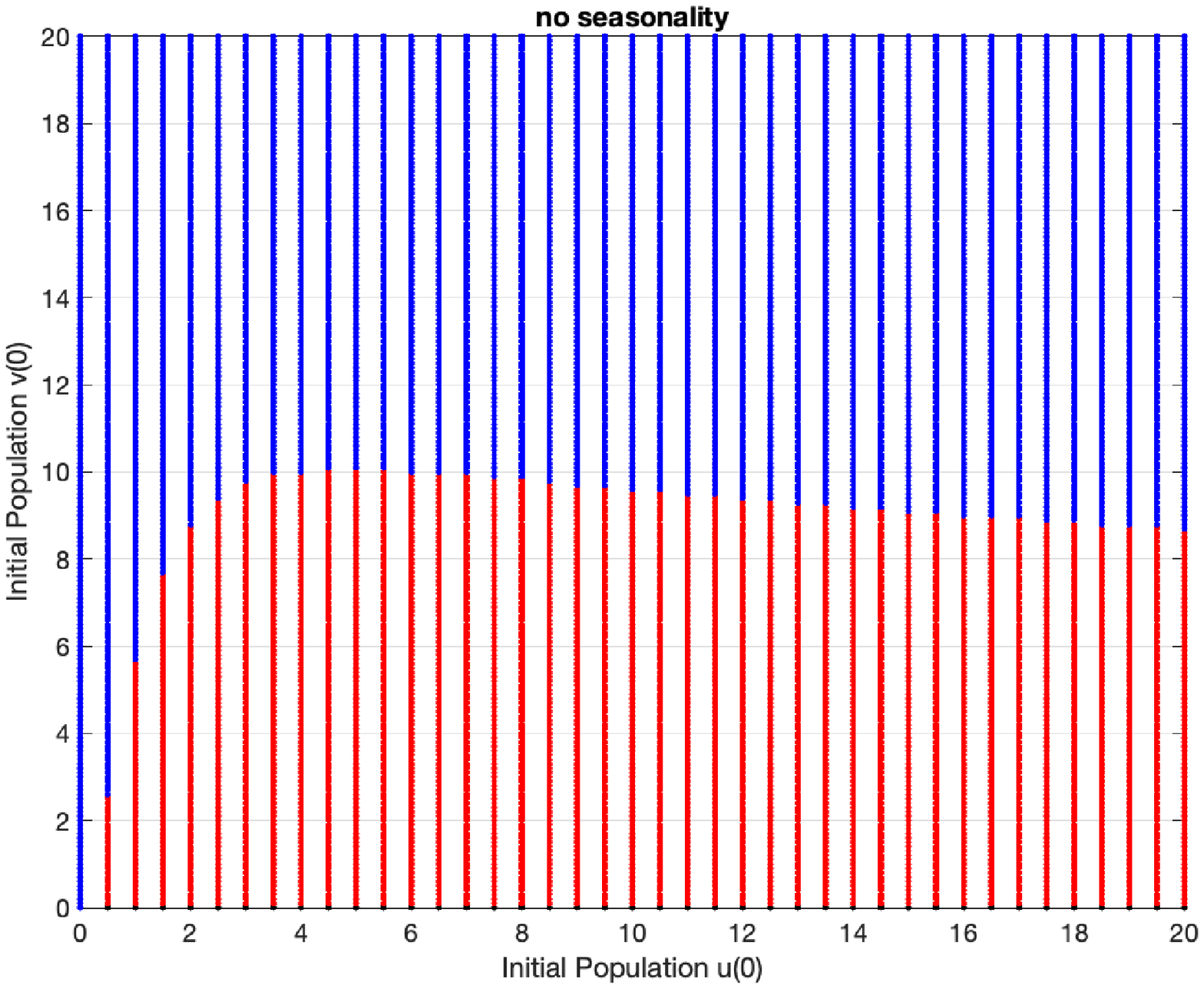}\label{before_no season1}
		}
		\subfigure[Honey bee Population for those three cases]{
			\includegraphics[width=5cm]{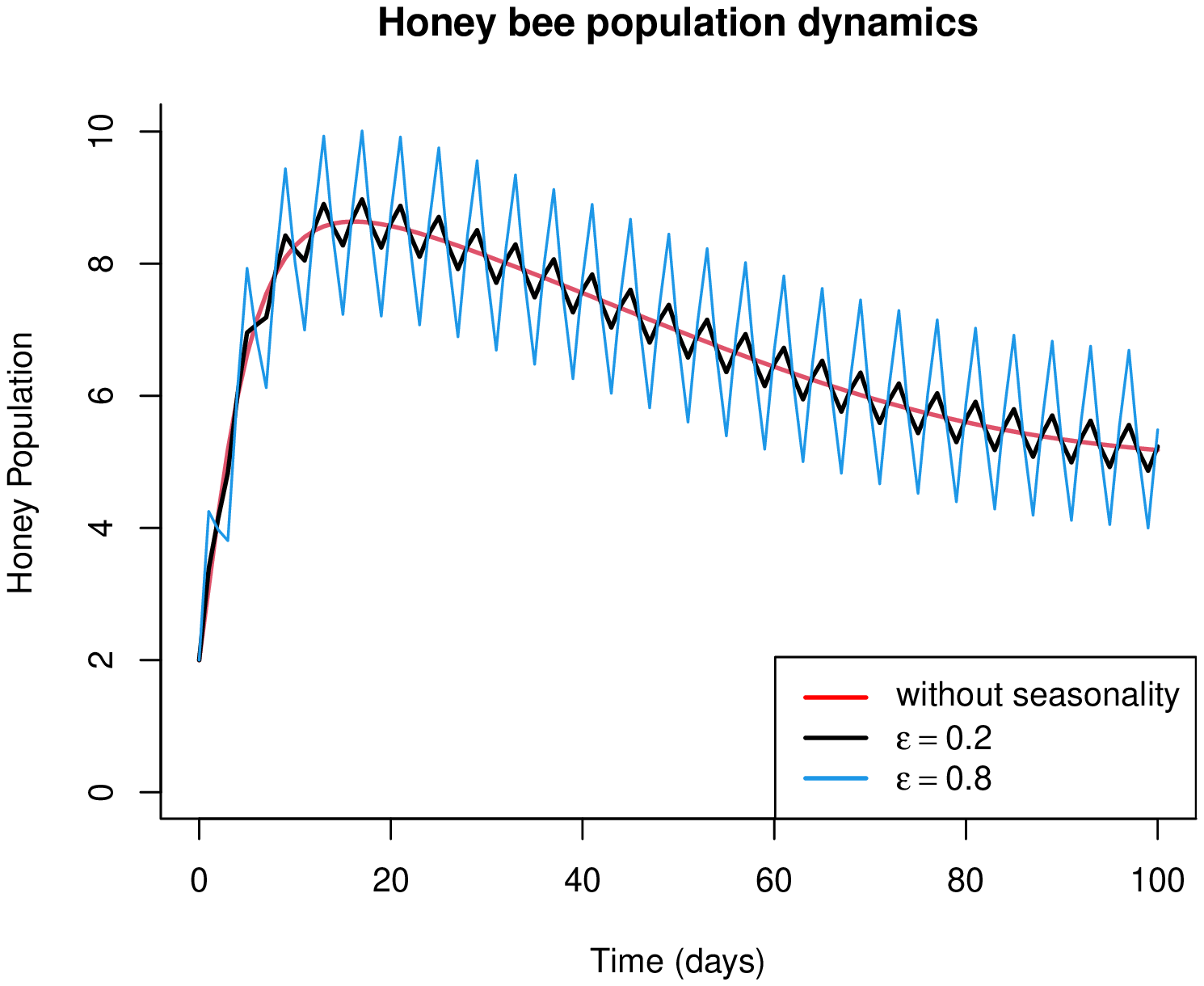}\label{pop-gammaS}
		}
		\subfigure[$\gamma=4, \epsilon=0.2$]{
			\includegraphics[width=5cm]{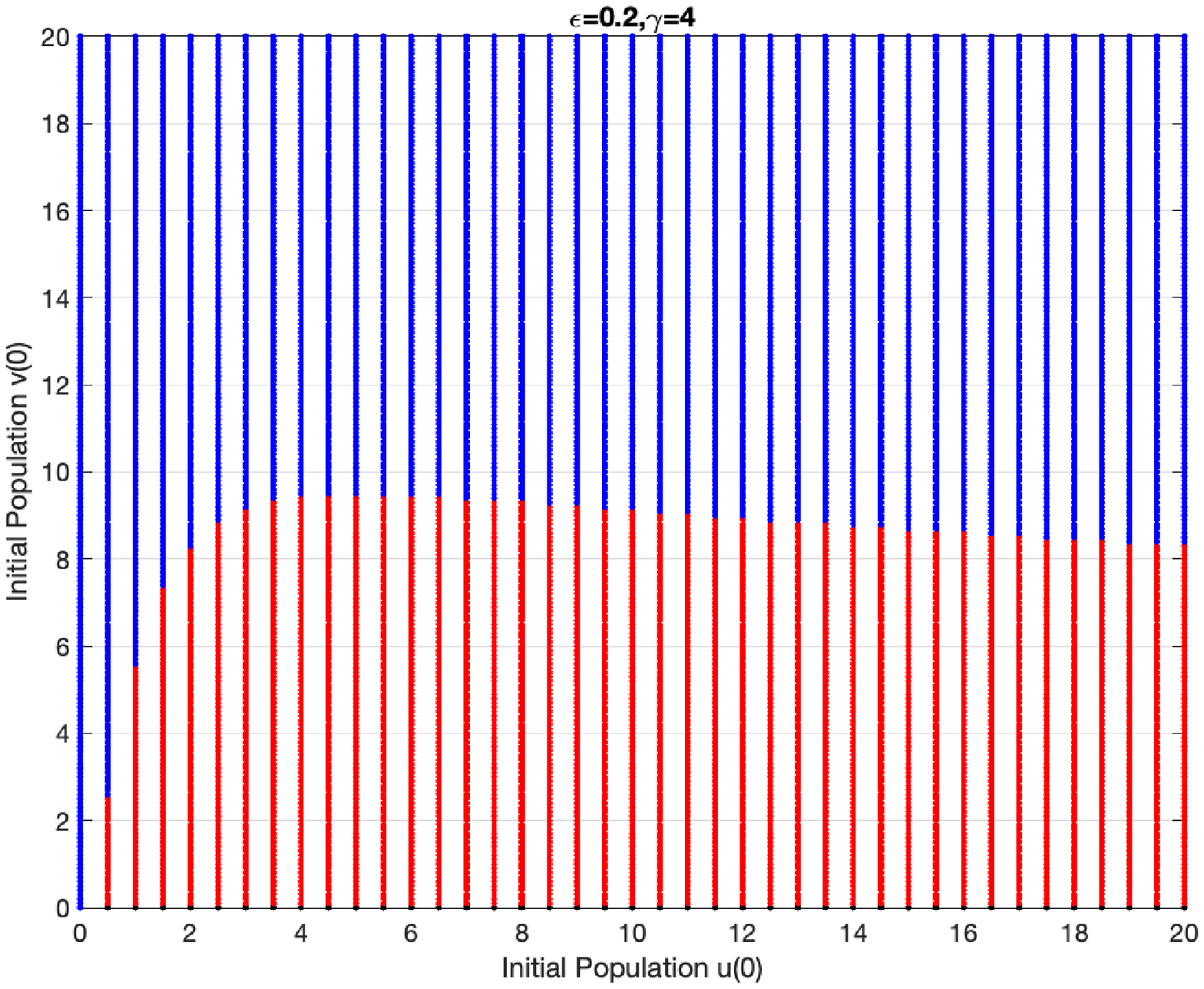}
			\label{no_42}
		}
		\subfigure[$\gamma=4, \epsilon=0.8$]{
			\includegraphics[width=5cm]{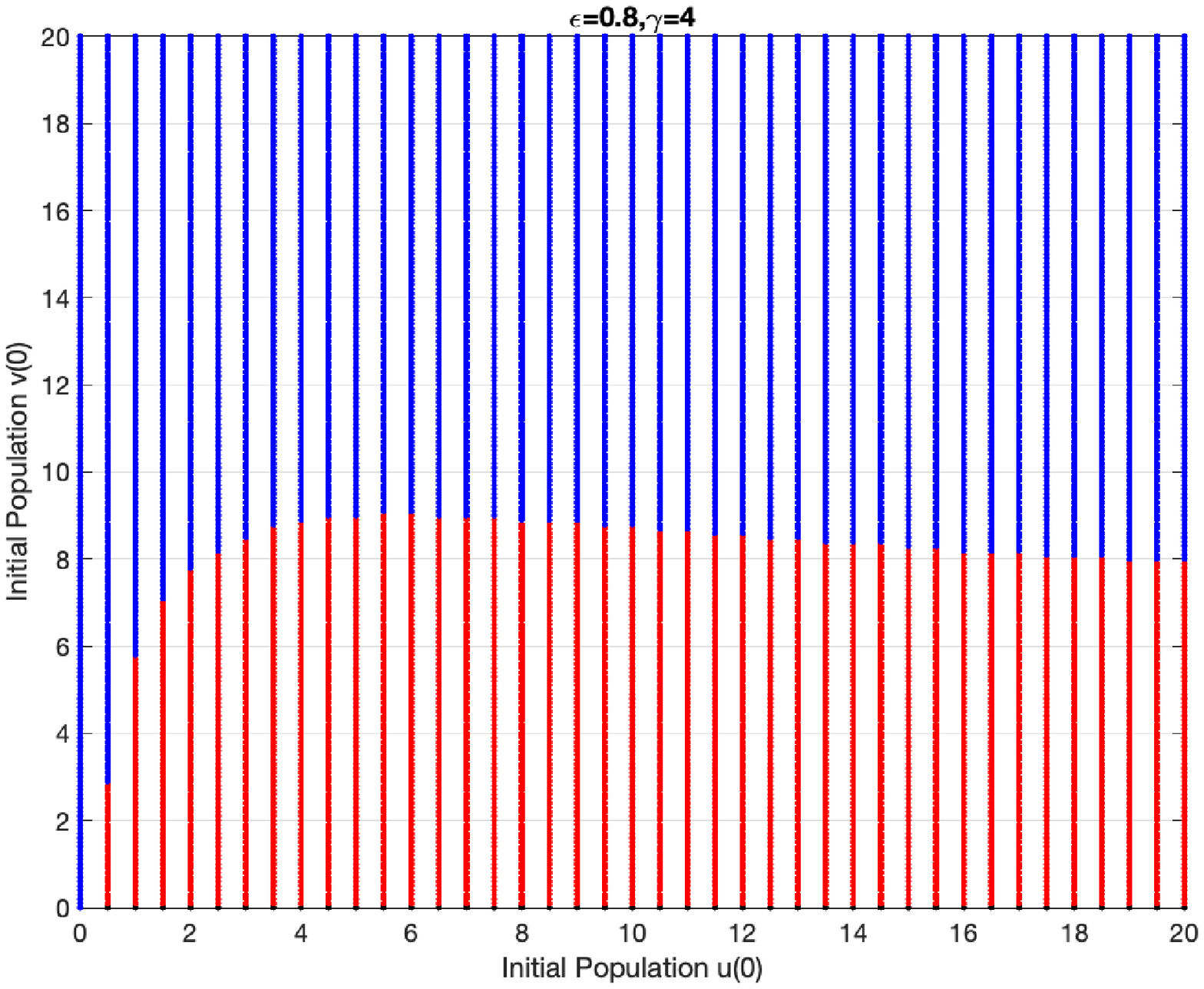}
			\label{no_48}
		}
		\caption{Impacts of seasonality on the honey bee colony survival when the period of seasonality $\gamma$ is large; and
		$\bar{r}_0=2.86$, $\bar{d}_h=\bar{d}_m=0.25$, $\omega=0.3$ and $\hat{K}=2.04$ and $\psi=0$. Initial population is $u_0 \in [0, 20]$, and  $v_0 \in [0, 20]$. The blue area is the basin attraction that leads to colony collapse, while the red area is the basin attraction the colony can survive.}
			\label{fig:2D_gammaS}
		\end{figure}
		
		 \item When the period of the seasonality  $\gamma$ is in the intermediate range, e.g., $\gamma=80$, the impacts from the strength  seasonality  $\epsilon$ can be very complicated. For example, Figure \ref{before_8035} shows that basins of attractions for the colony survival are splitted into two red areas, and Figure \ref{pop-gammaM} shows larger $\epsilon$ gives larger population amplitude.\\

	\begin{figure}[ht]
		\centering
	\subfigure[no seasonality]{
			\includegraphics[width=5cm]{Figure/scale_no_season.eps}\label{before_no season2}
		}
		\subfigure[Population Dynamics]{
			\includegraphics[width=5cm]{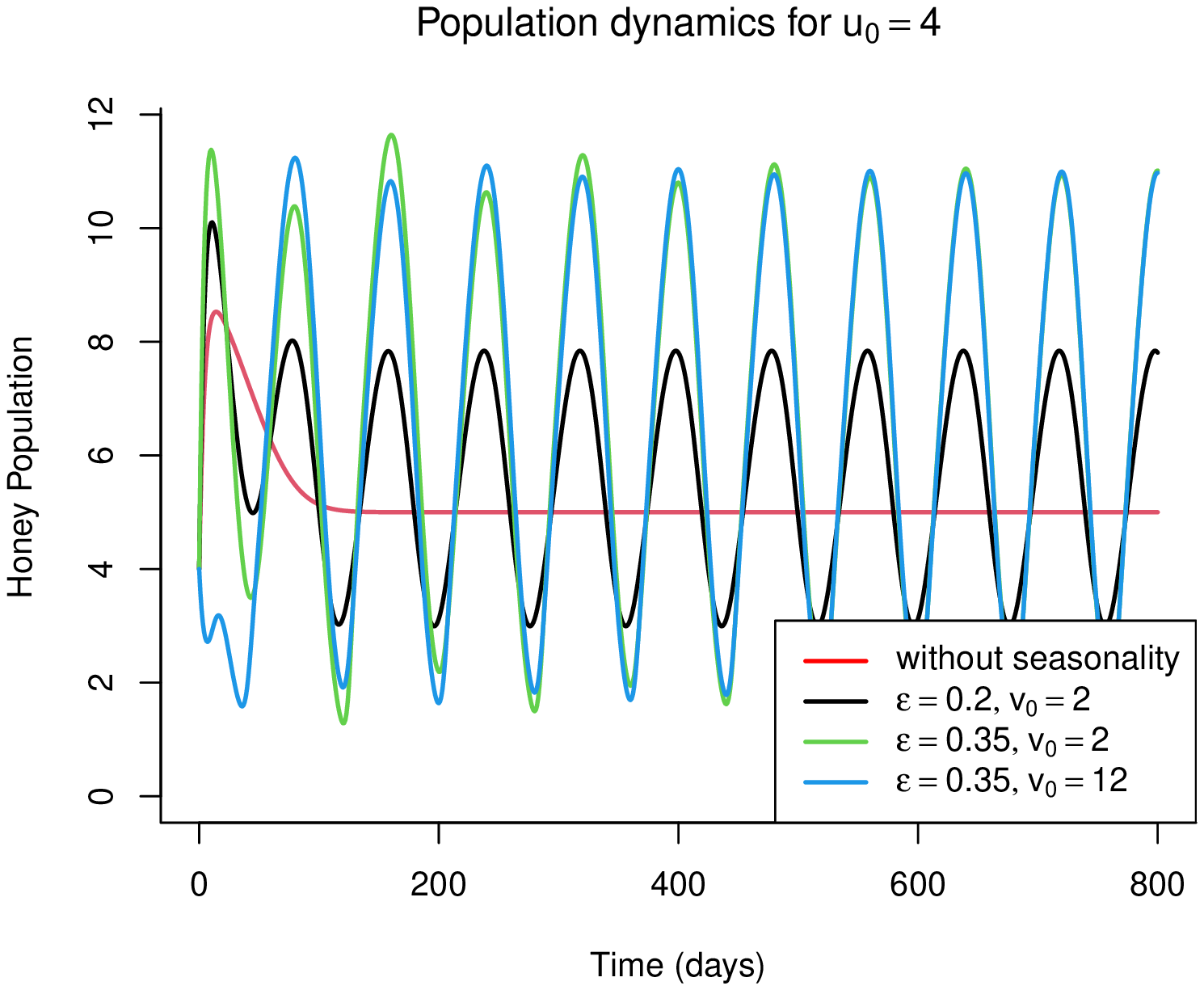}\label{pop-gammaM}
		}
		
			
\subfigure[$\gamma=80, \epsilon=0.2$]{
			\includegraphics[width=5cm]{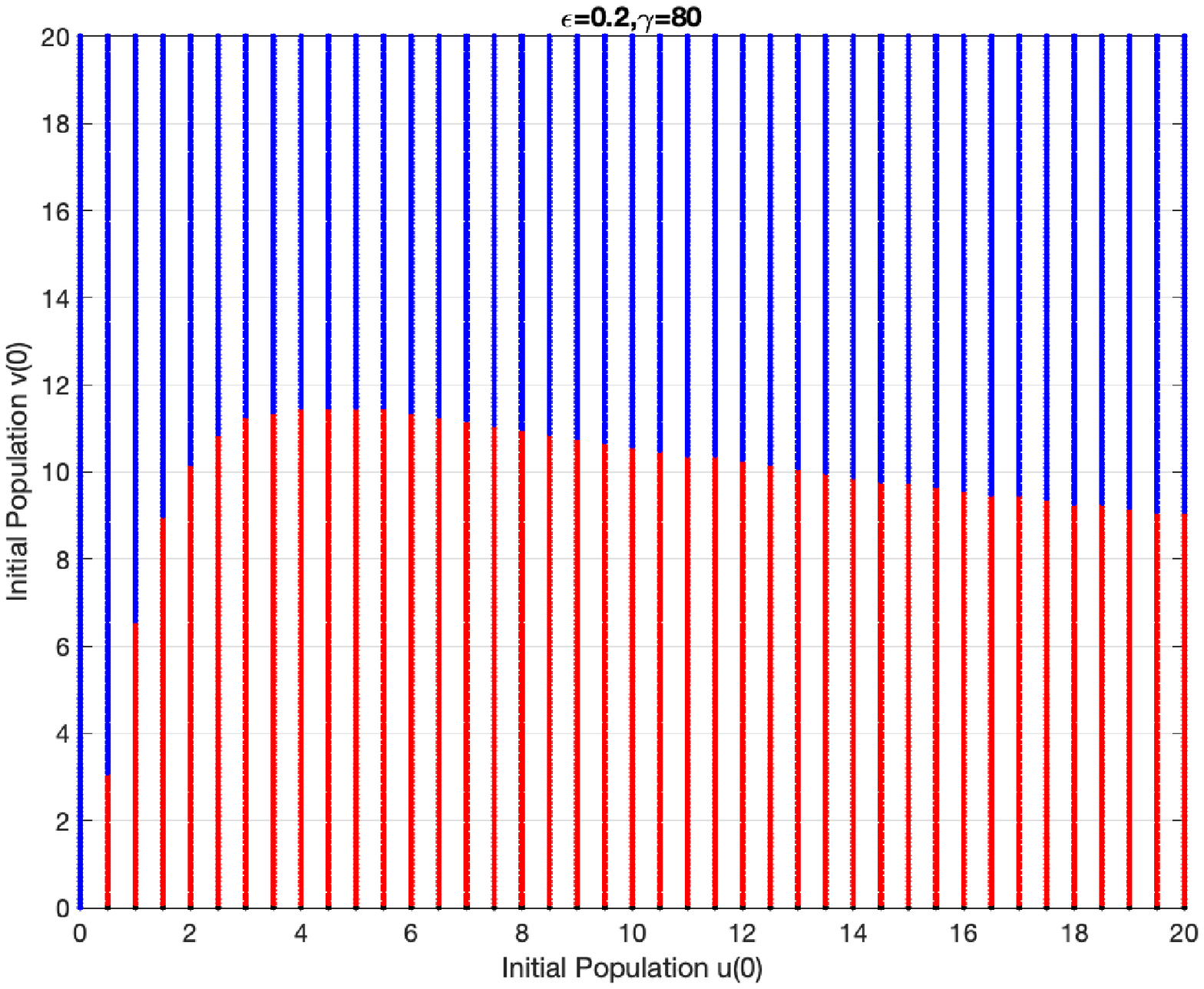}
			\label{before_802}
		}
		\subfigure[$\gamma=80, \epsilon=0.35$]{
			\includegraphics[width=5cm]{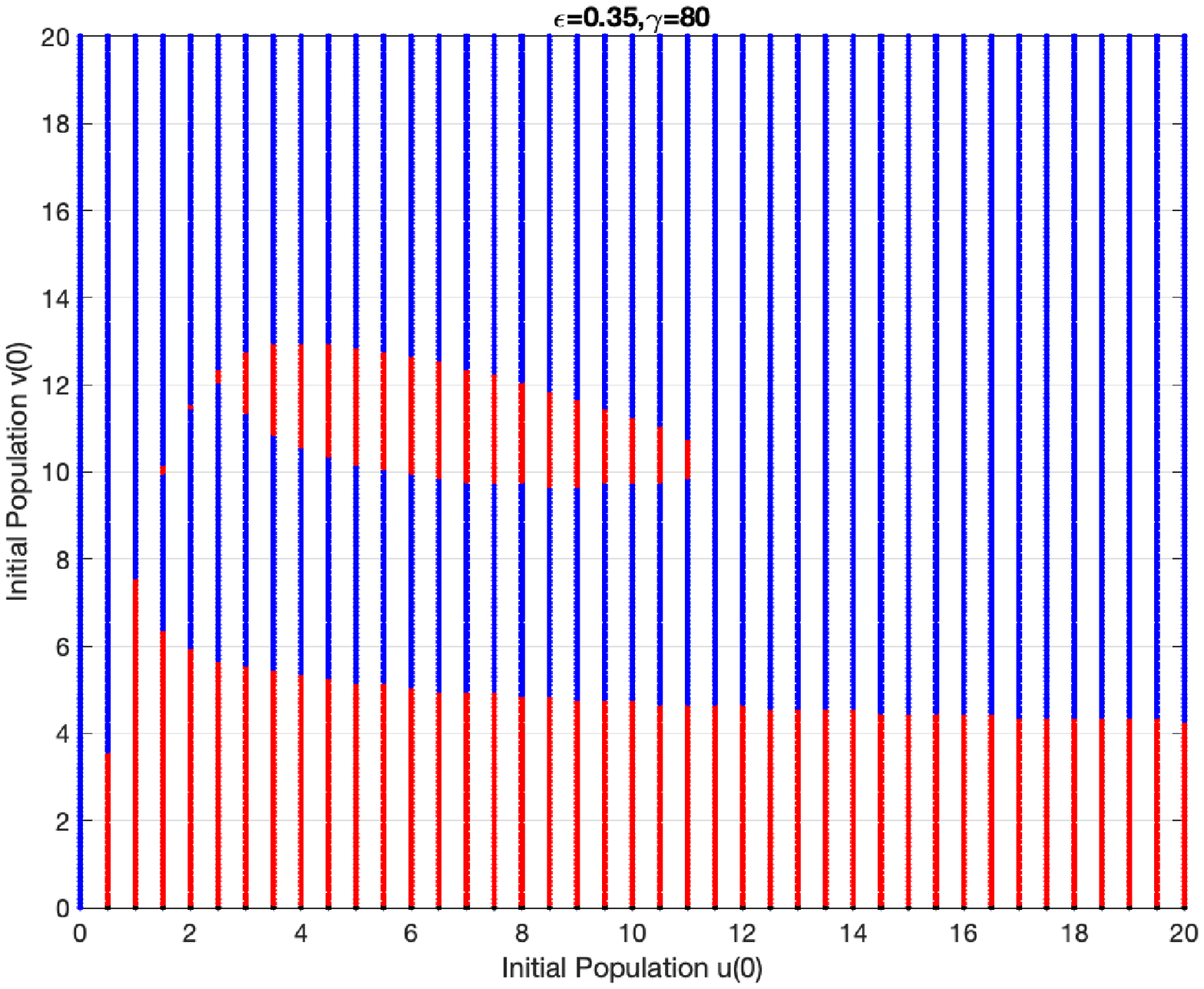}\label{before_8035}
		}
		\caption{Impacts of seasonality on the honey bee colony survival when the period of seasonality $\gamma$ is intermediate; and
		$\bar{r}_0=2.86$, $\bar{d}_h=\bar{d}_m=0.25$, $\omega=0.3$ and $\hat{K}=2.04$ and $\psi=0$. Initial population is $u_0 \in [0, 20]$, and  $v_0 \in [0, 20]$. The blue area is the basin attraction that leads to colony collapse, while the red area is the basin attraction the colony can survive.}
			\label{fig:2D_gammaM}
\end{figure}

\item When the period of seasonality $\gamma$ is large, e.g., $\gamma=100, 250$, comparisons of the basin attractions for the colony survival suggest that the small strength  seasonality $\epsilon$ may not impact the basin attractions of the colony survival while its large value may cause the colony collapsing (see Figure \ref{after_1002} \& \ref{after_1007}). In some cases, the large strength of seasonality $\epsilon$ may have a positive influence on the colony survival by increasing the area of basin attractions of the colony survival (see the comparison of Figure \ref{before_no season} \& \ref{after_2507}). From the population dynamics point of view, Figure \ref{pop-gammaL} suggests that the population has a larger amplitude when $\epsilon$ is larger.\\

\begin{figure}[ht]
		\centering
	\subfigure[no seasonality]{
			\includegraphics[width=5cm]{Figure/scale_no_season.eps}\label{before_no season}
		}
		\subfigure[Honey bee Population for those three cases]{
			\includegraphics[width=5cm]{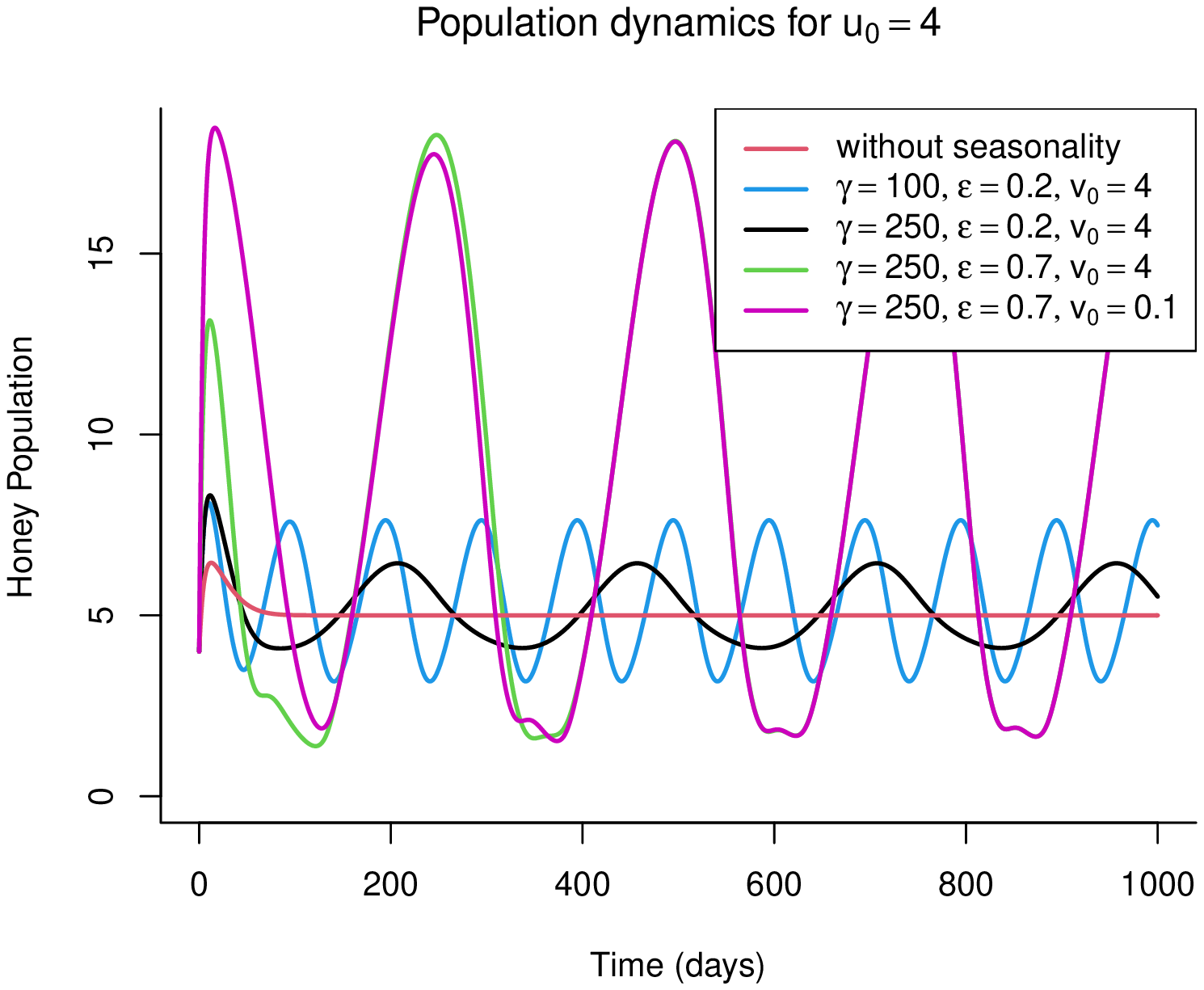}\label{pop-gammaL}
		}
			\subfigure[$\gamma=100, \epsilon=0.2$]{
			\includegraphics[width=5cm]{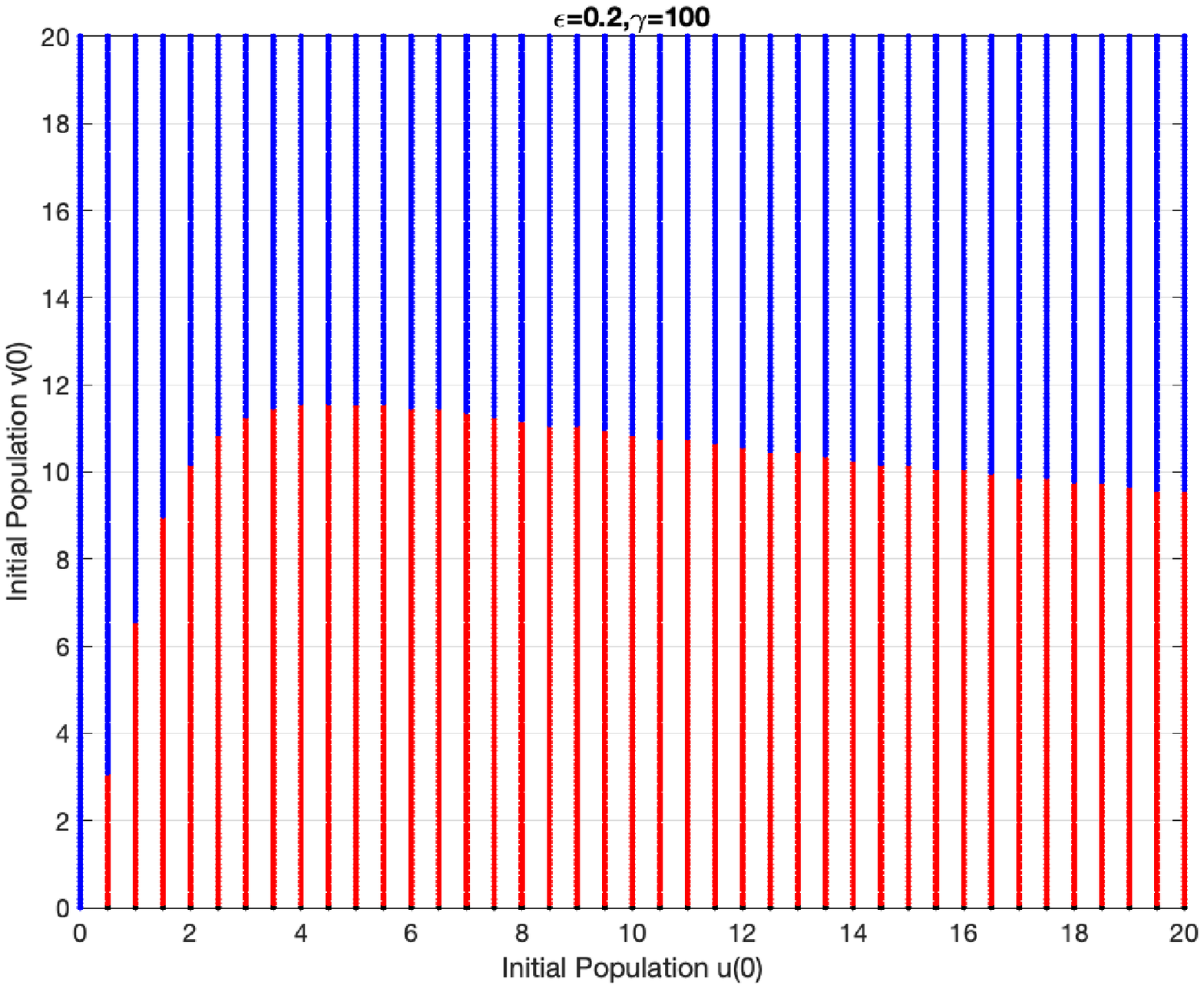}
			\label{after_1002}
		}
		\subfigure[$\gamma=100, \epsilon=0.7$]{
			\includegraphics[width=5cm]{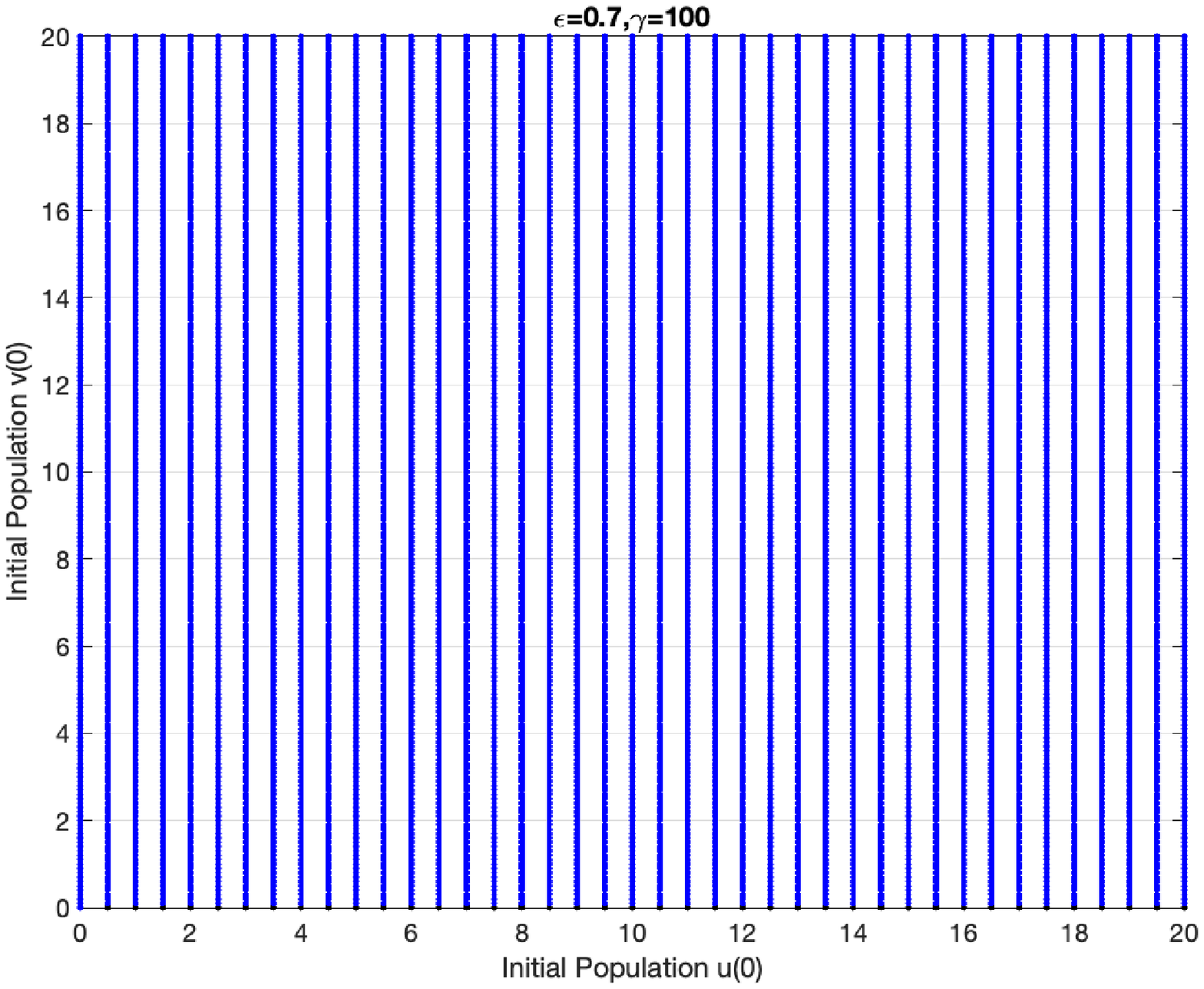}
			\label{after_1007}
		}
	\subfigure[$\gamma=250, \epsilon=0.2$]{
			\includegraphics[width=5cm]{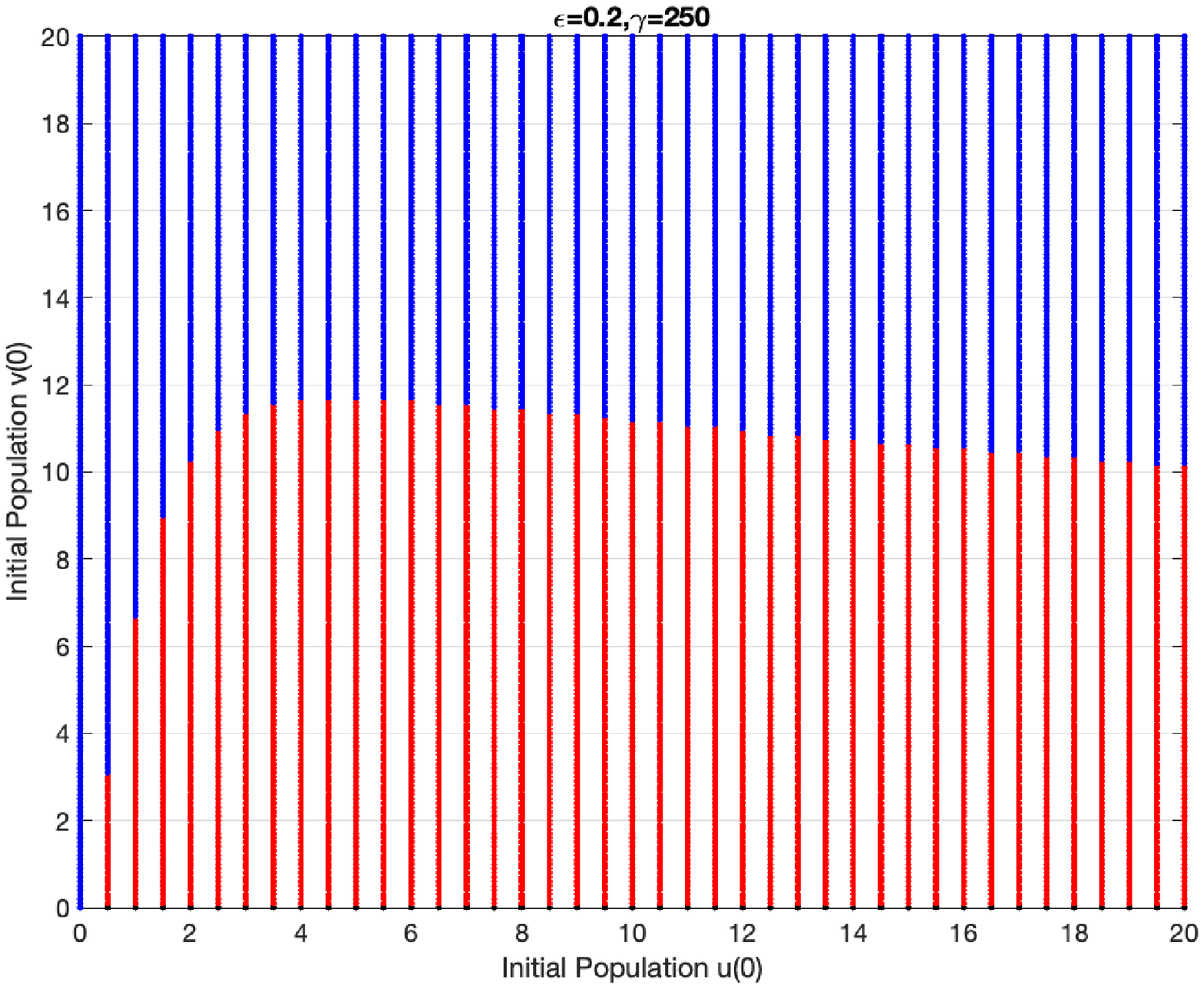}
			\label{after_2502}
		}
		\subfigure[$\gamma=250, \epsilon=0.7$]{
			\includegraphics[width=5cm]{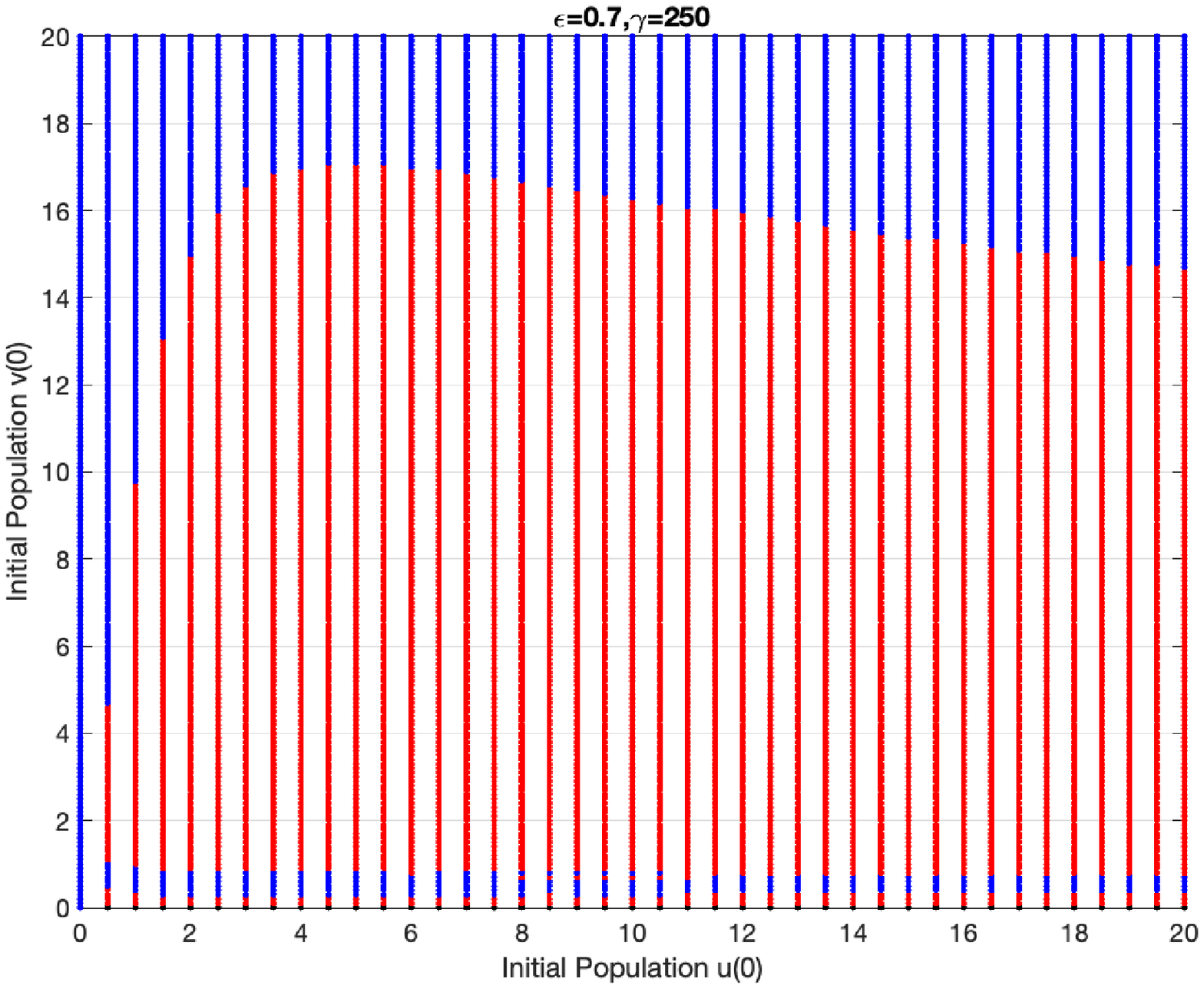}
			\label{after_2507}
		}
		\caption{Impacts of seasonality on the honey bee colony survival when the period of seasonality $\gamma$ is small; and
		$\bar{r}_0=2.86$, $\bar{d}_h=\bar{d}_m=0.25$, $\omega=0.3$ and $\hat{K}=2.04$ and $\psi=0$. Initial population is $u_0 \in [0, 20]$, and  $v_0 \in [0, 20]$. The blue area is the basin attraction that leads to colony collapse, while the red area is the basin attraction the colony can survive.}
			\label{fig:2D_gammaL}
		\end{figure}
		
\end{enumerate}

\begin{figure}[htp]
		\centering
  \subfigure[$\epsilon=0.2, \psi=0$]{
			\includegraphics[width=3cm]{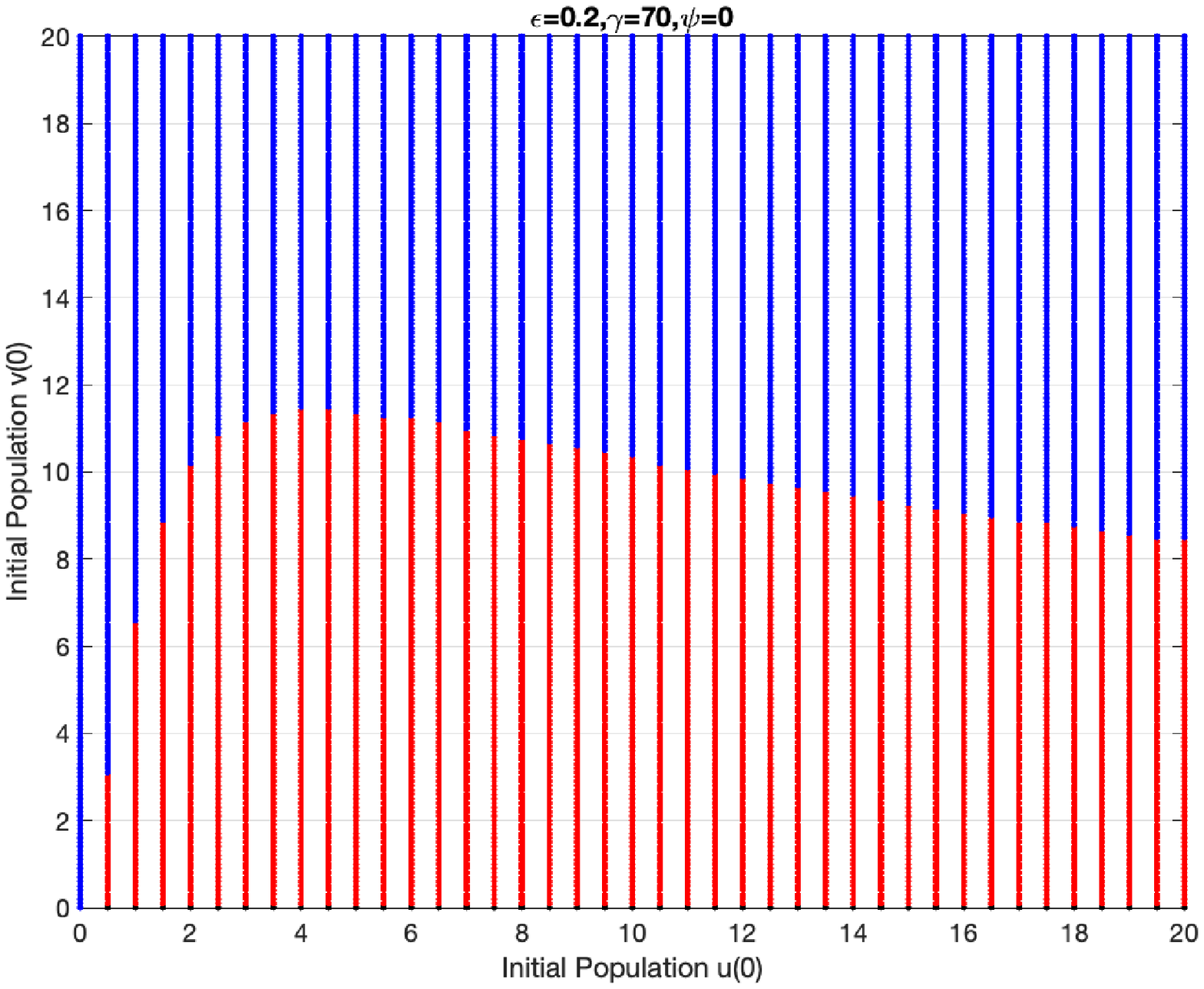}
			\label{70-psi0_1}
		}	
		\subfigure[$\epsilon=0.2, \psi=10$]{
			\includegraphics[width=3cm]{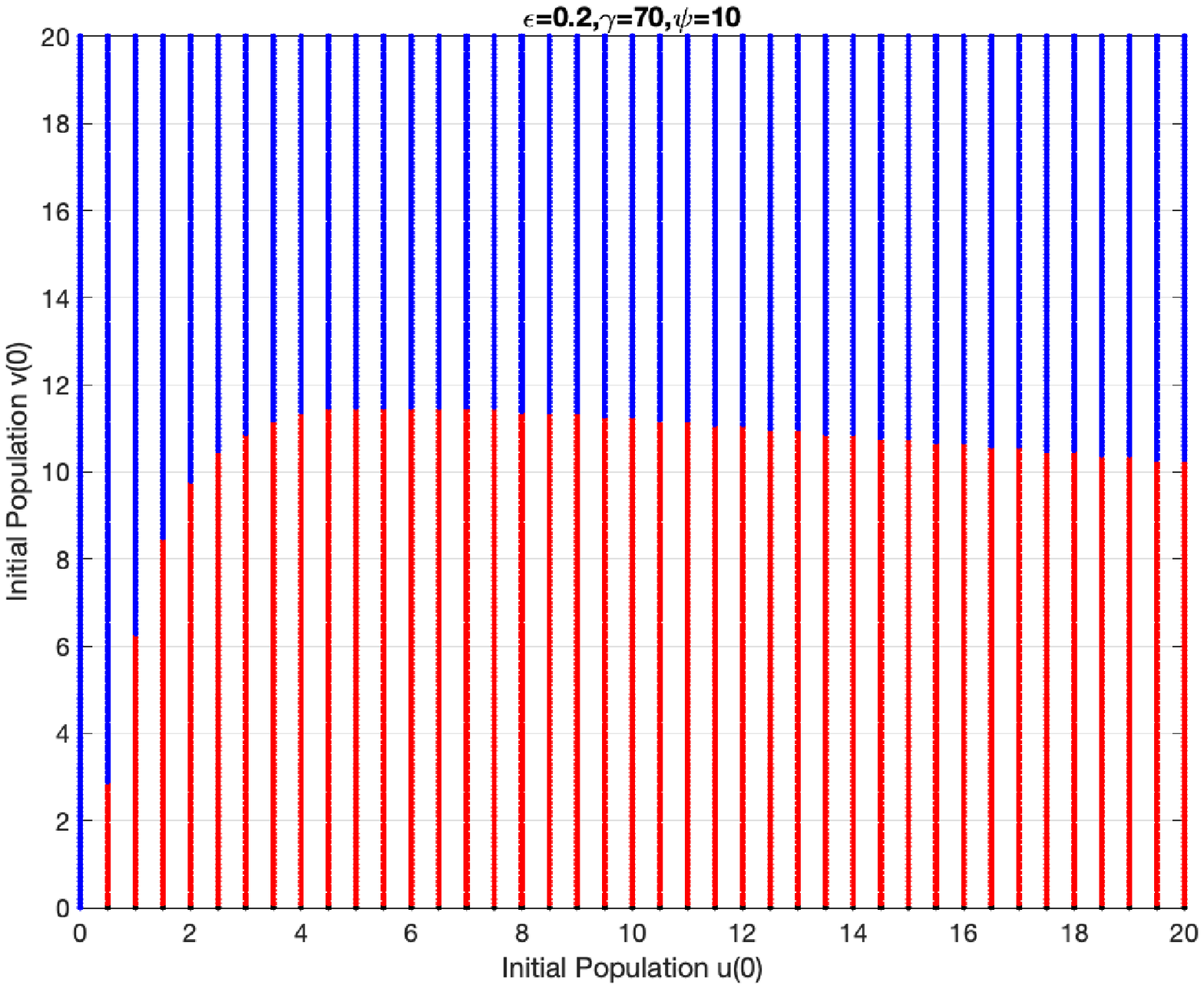}
			\label{70-psi10_1}
		}
  \subfigure[$\epsilon=0.2, \psi=30$]{
			\includegraphics[width=3cm]{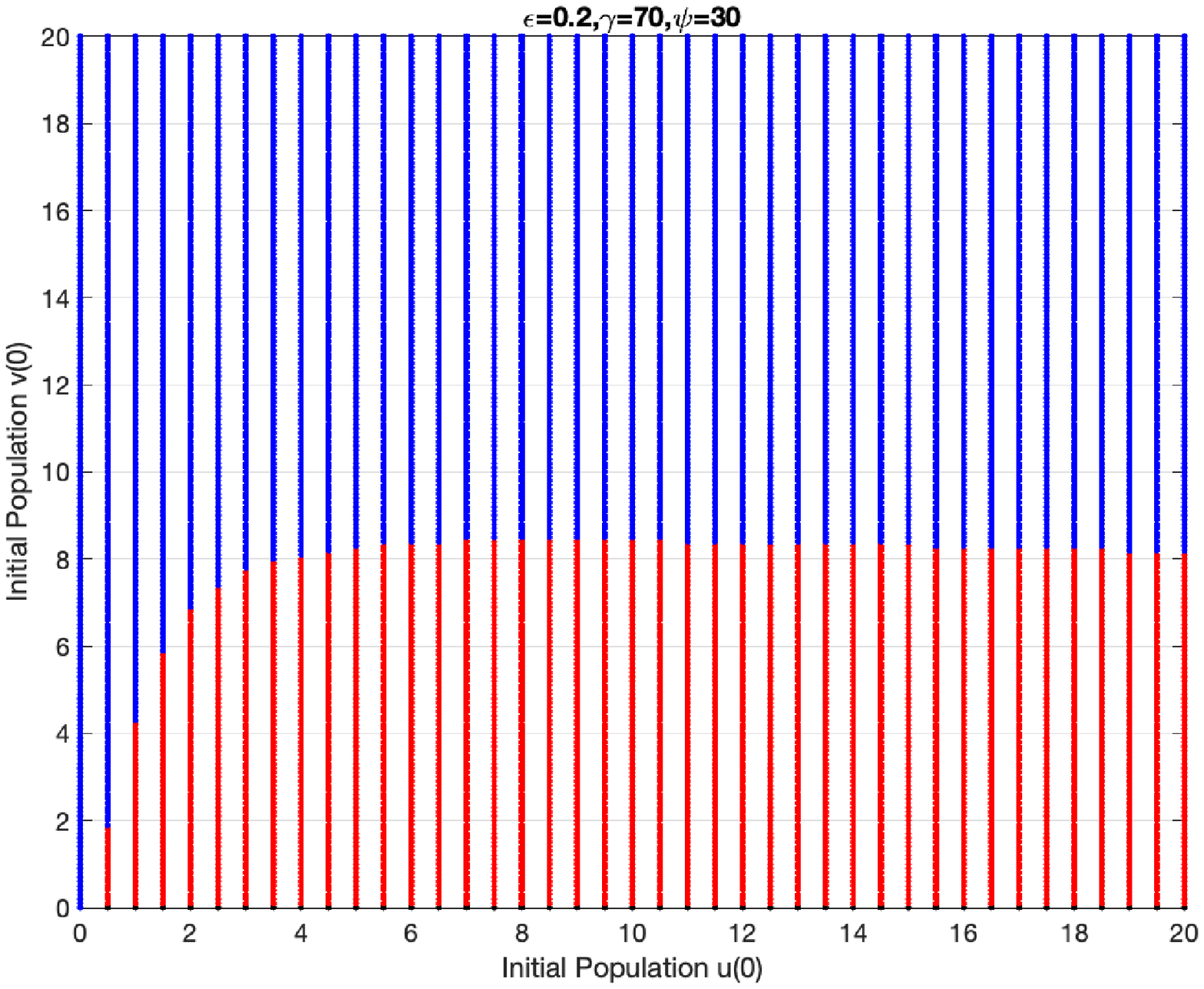}
			\label{70-psi30_1}
		}
  \subfigure[$\epsilon=0.2, \psi=35$]{
			\includegraphics[width=3cm]{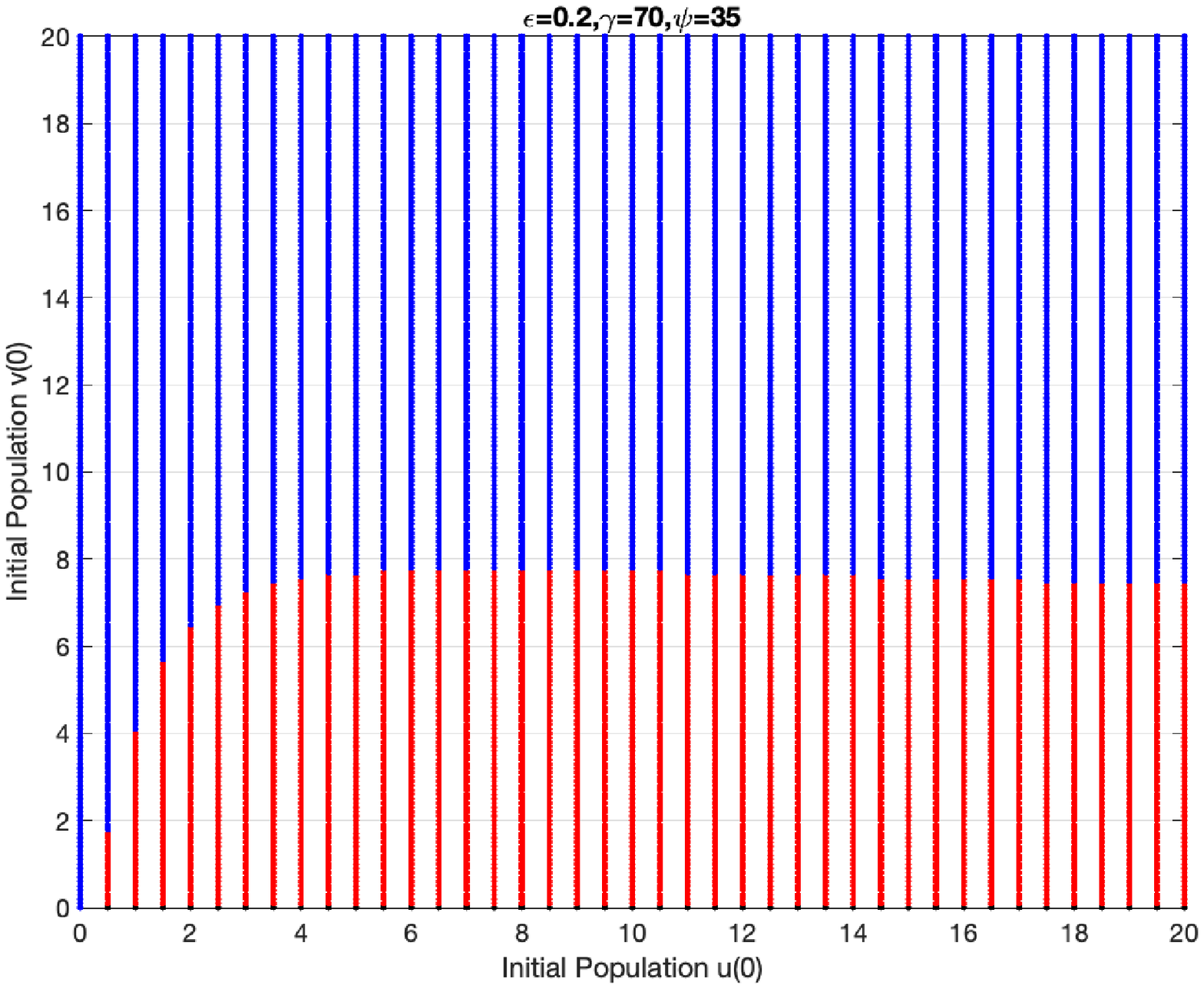}
			\label{70-psi35_1}
		}
    \subfigure[$\epsilon=0.2, \psi=40$]{
			\includegraphics[width=3cm]{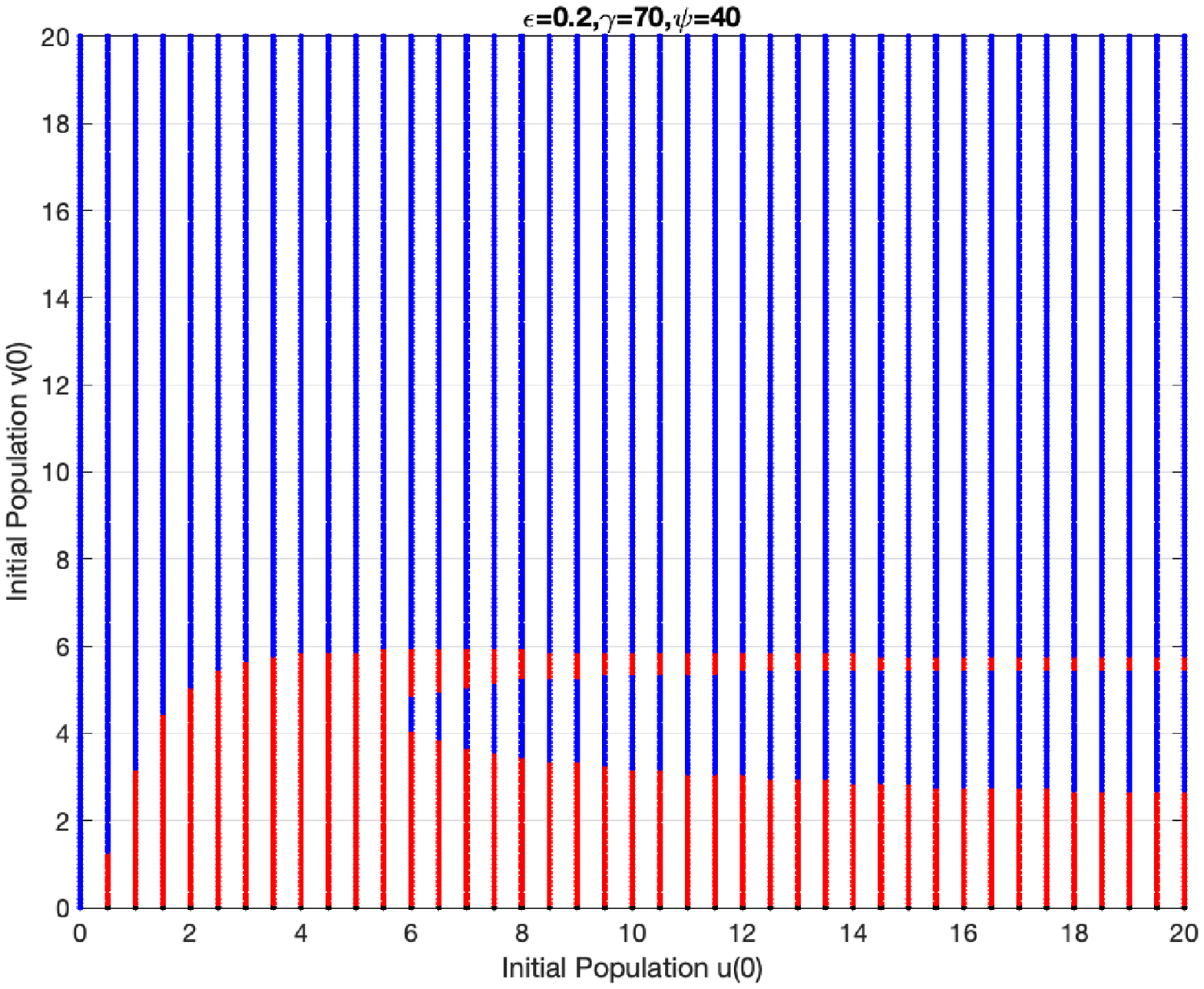}
			\label{70-psi40_1}
		}
		\subfigure[$\epsilon=0.2, \psi=60$]{
			\includegraphics[width=3cm]{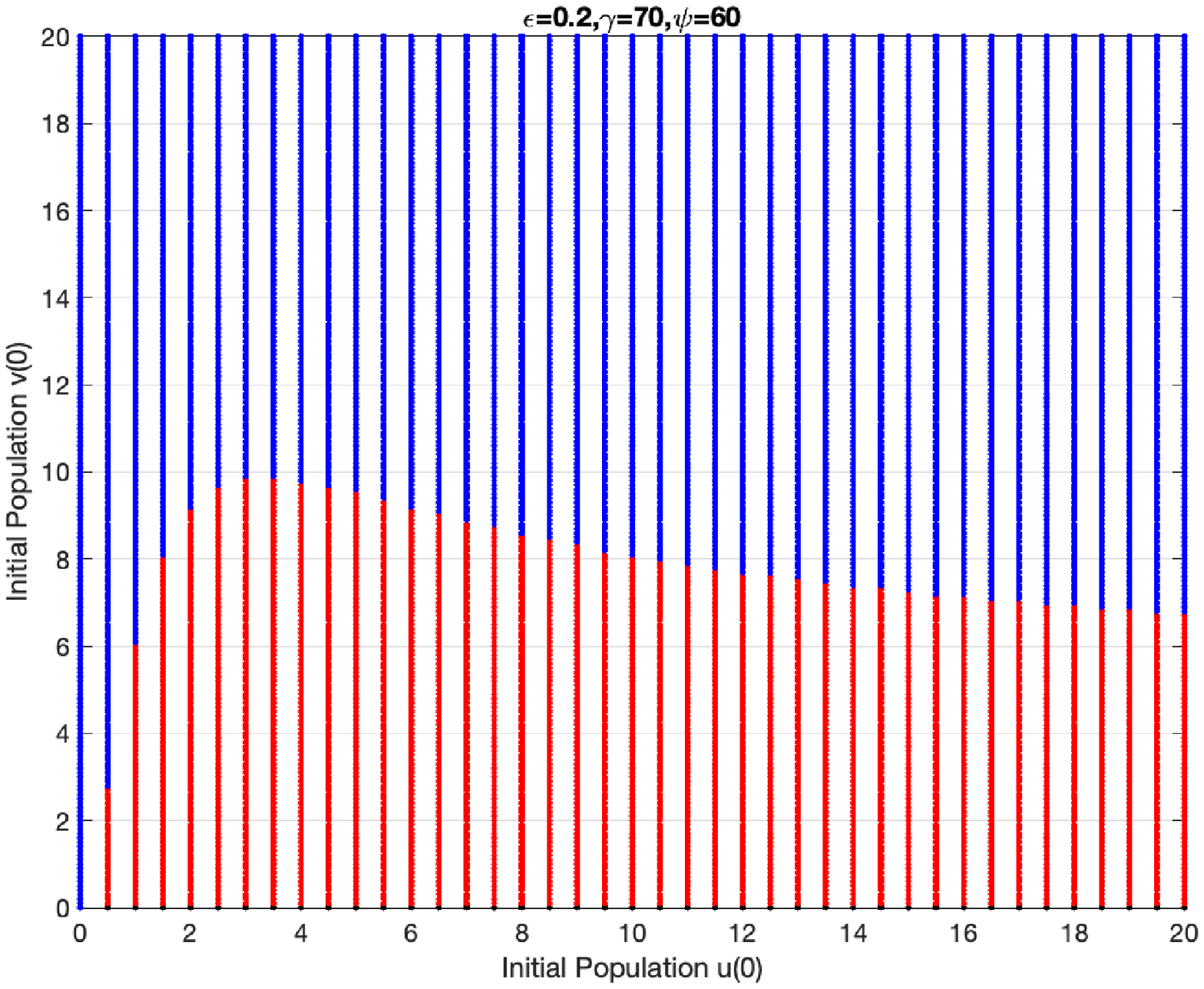}
			\label{70-psi60_1}
		}
\subfigure[$\epsilon=0.35, \psi=0$]{
			\includegraphics[width=3cm]{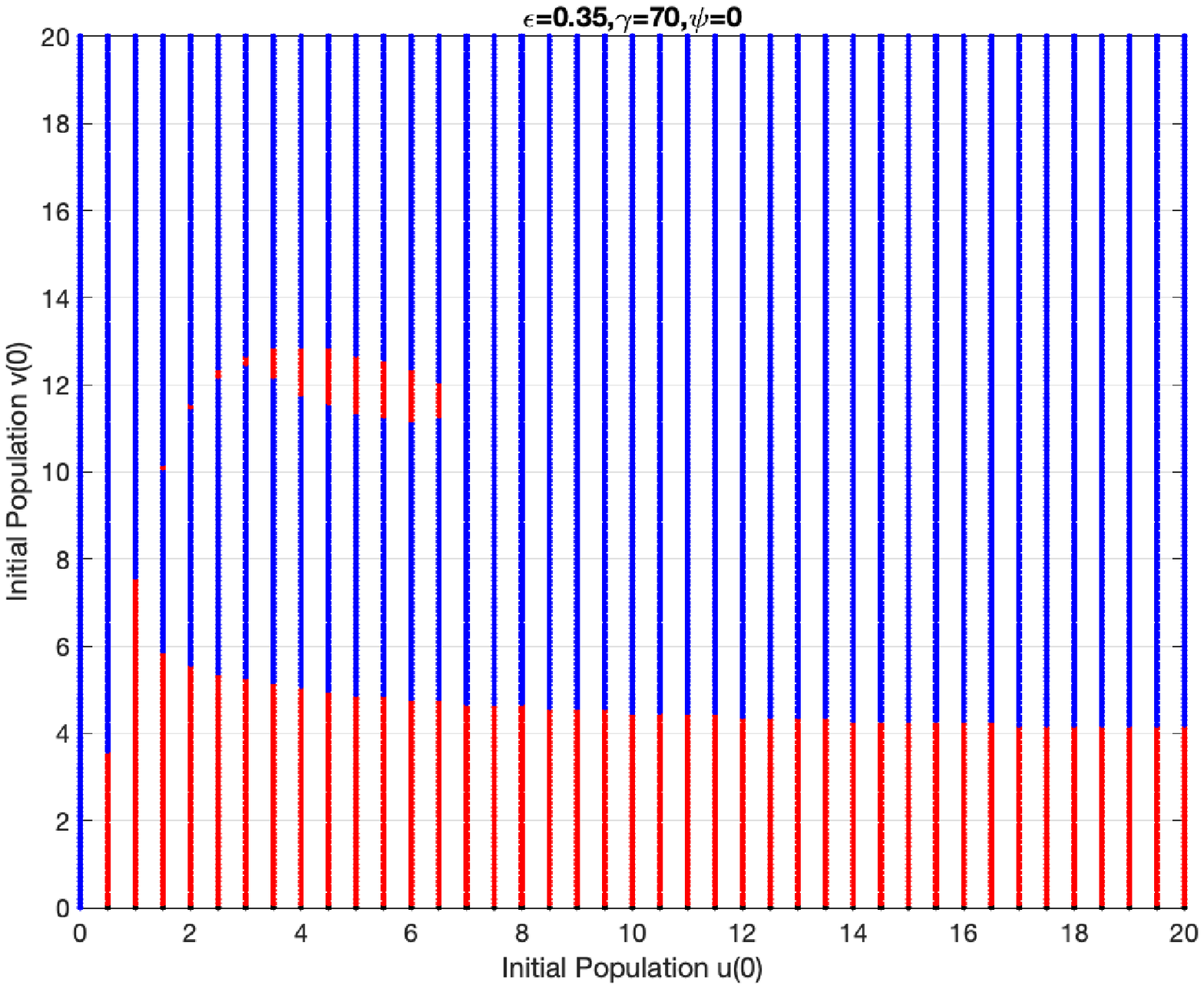}
			\label{70-psi0}
		}	
		\subfigure[$\epsilon=0.35, \psi=10$]{
			\includegraphics[width=3cm]{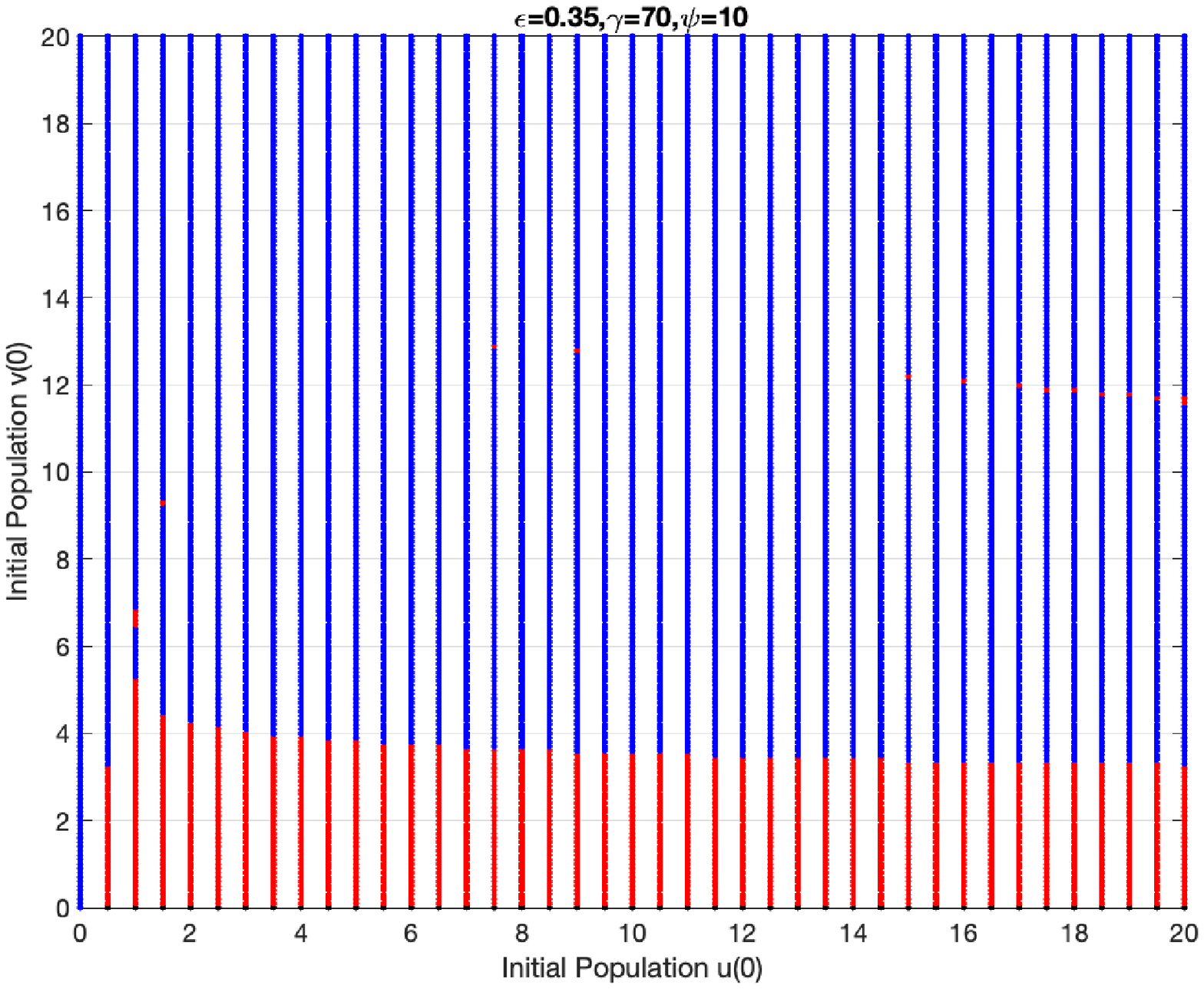}
			\label{70-psi10}
		}
  \subfigure[$\epsilon=0.35, \psi=30$]{
			\includegraphics[width=3cm]{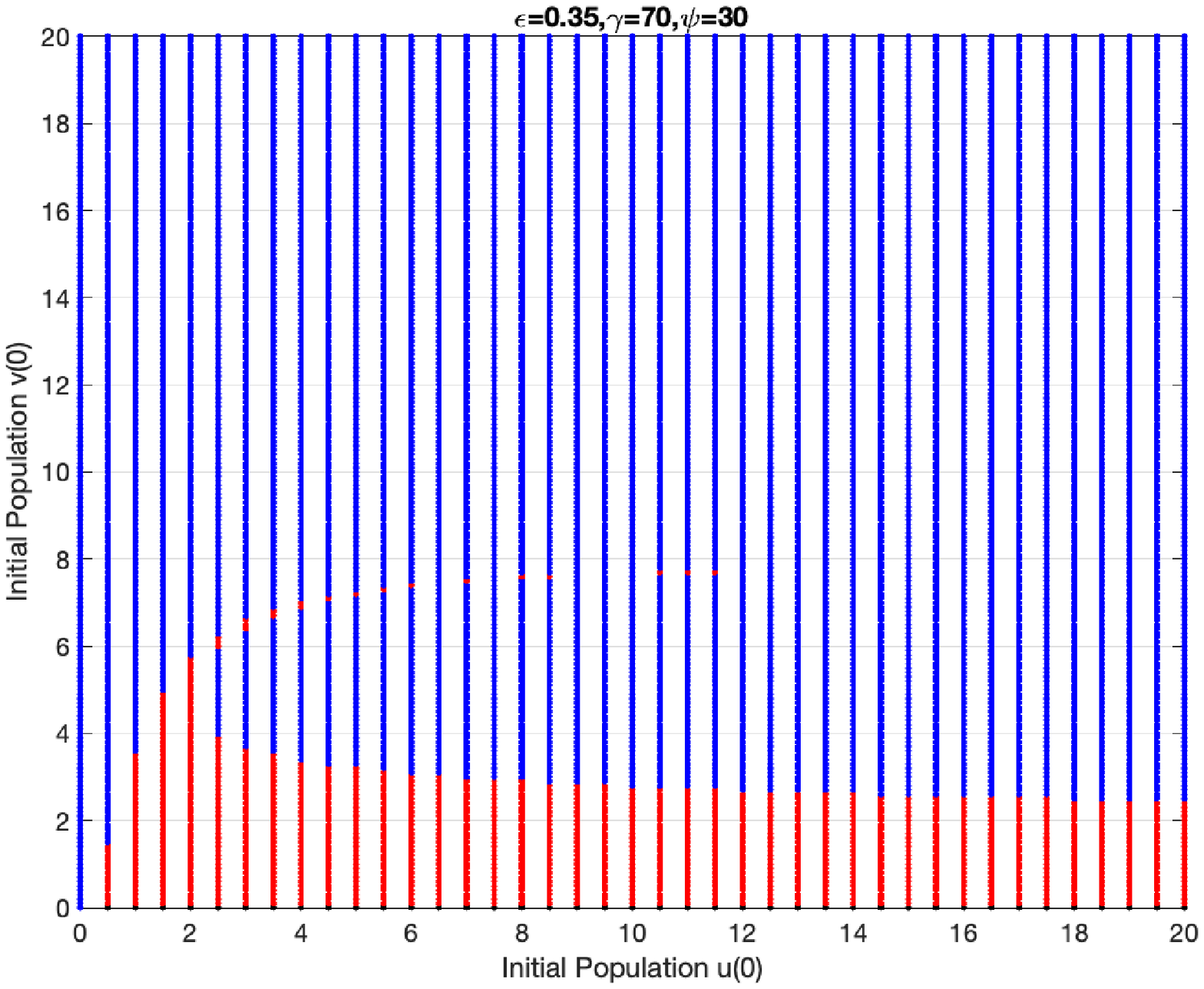}
			\label{70-psi30}
		}
  \subfigure[$\epsilon=0.35, \psi=35$]{
			\includegraphics[width=3cm]{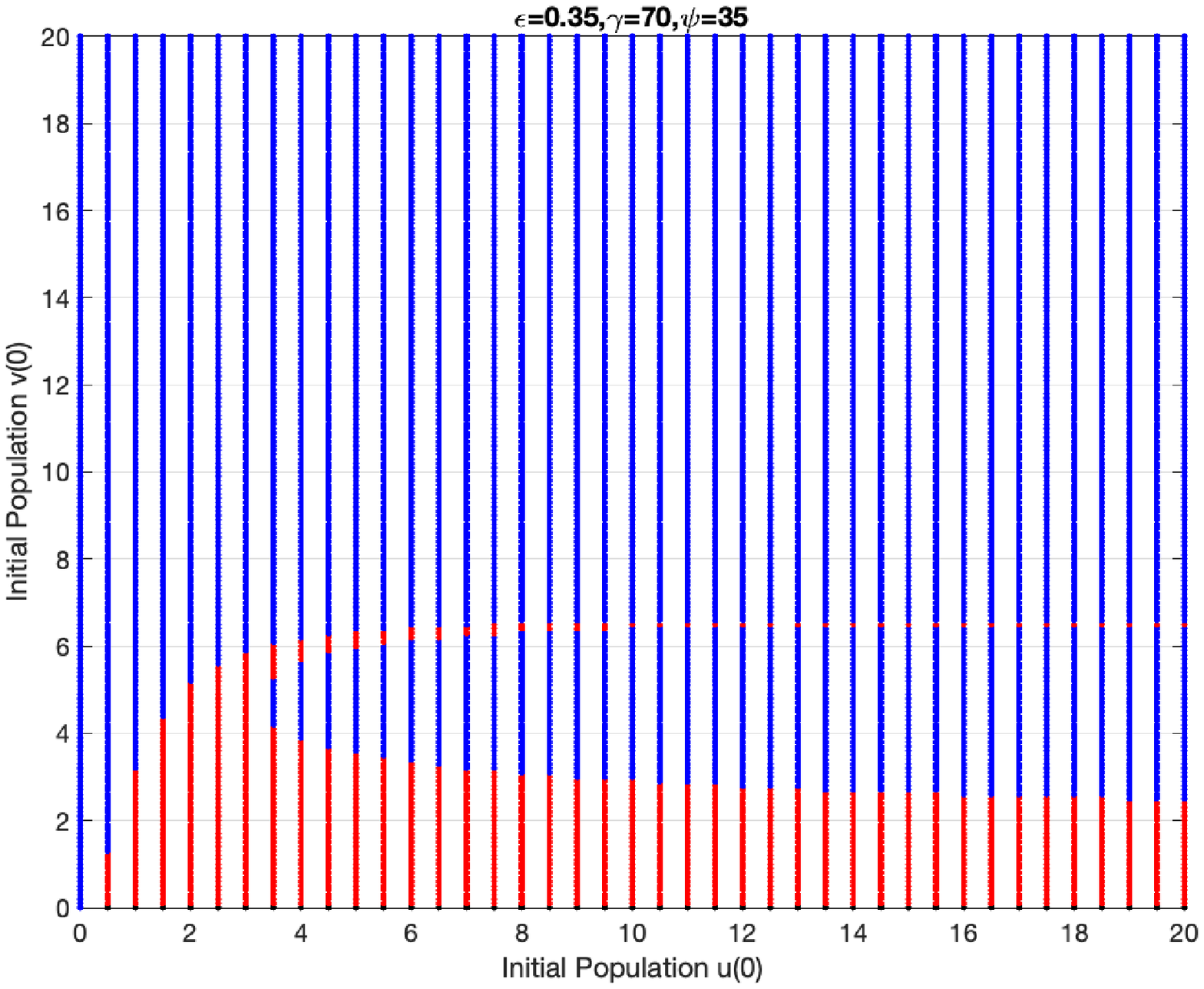}
			\label{70-psi35}
		}
    \subfigure[$\epsilon=0.35, \psi=40$]{
			\includegraphics[width=3cm]{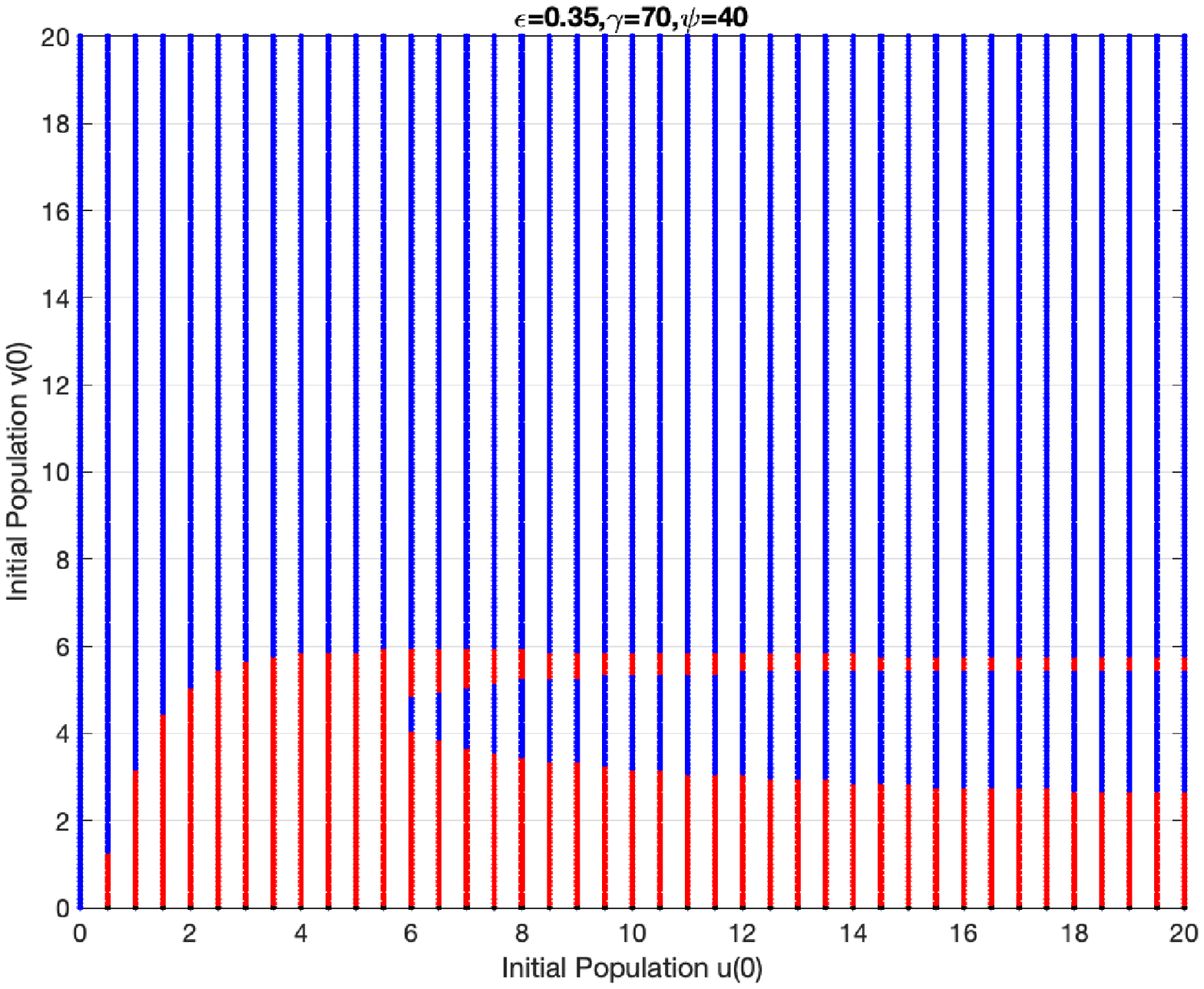}
			\label{70-psi40}
		}
		\subfigure[$\epsilon=0.35, \psi=60$]{
			\includegraphics[width=3cm]{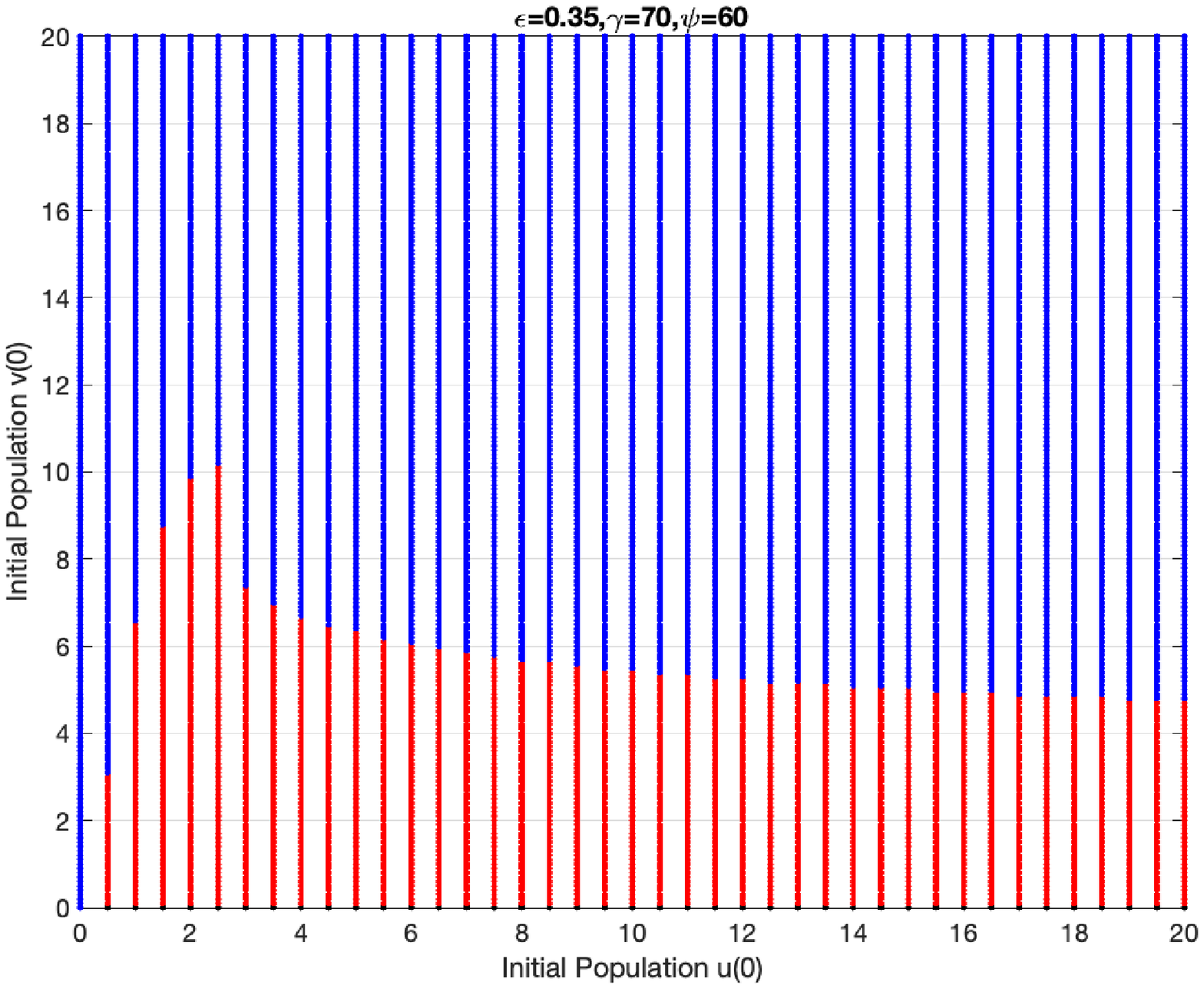}
			\label{70-psi60}
		}
  \subfigure[Honey bee population dynamics for $\epsilon=0.35$]{
			\includegraphics[width=3cm]{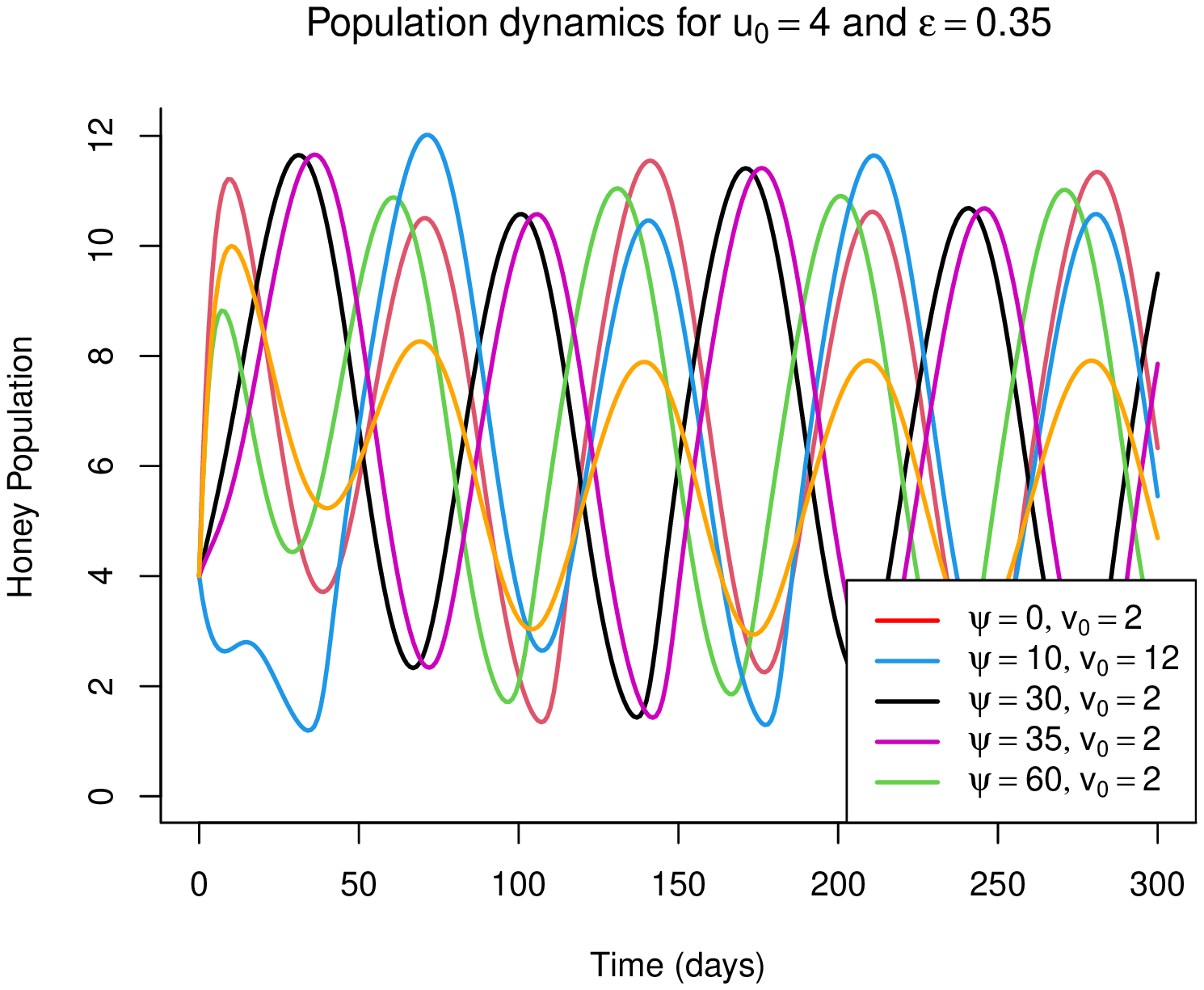}
			\label{psi-dynamic}
		}
  \subfigure[Honey bee population dynamics for $\epsilon=0.2$]{
			\includegraphics[width=3cm]{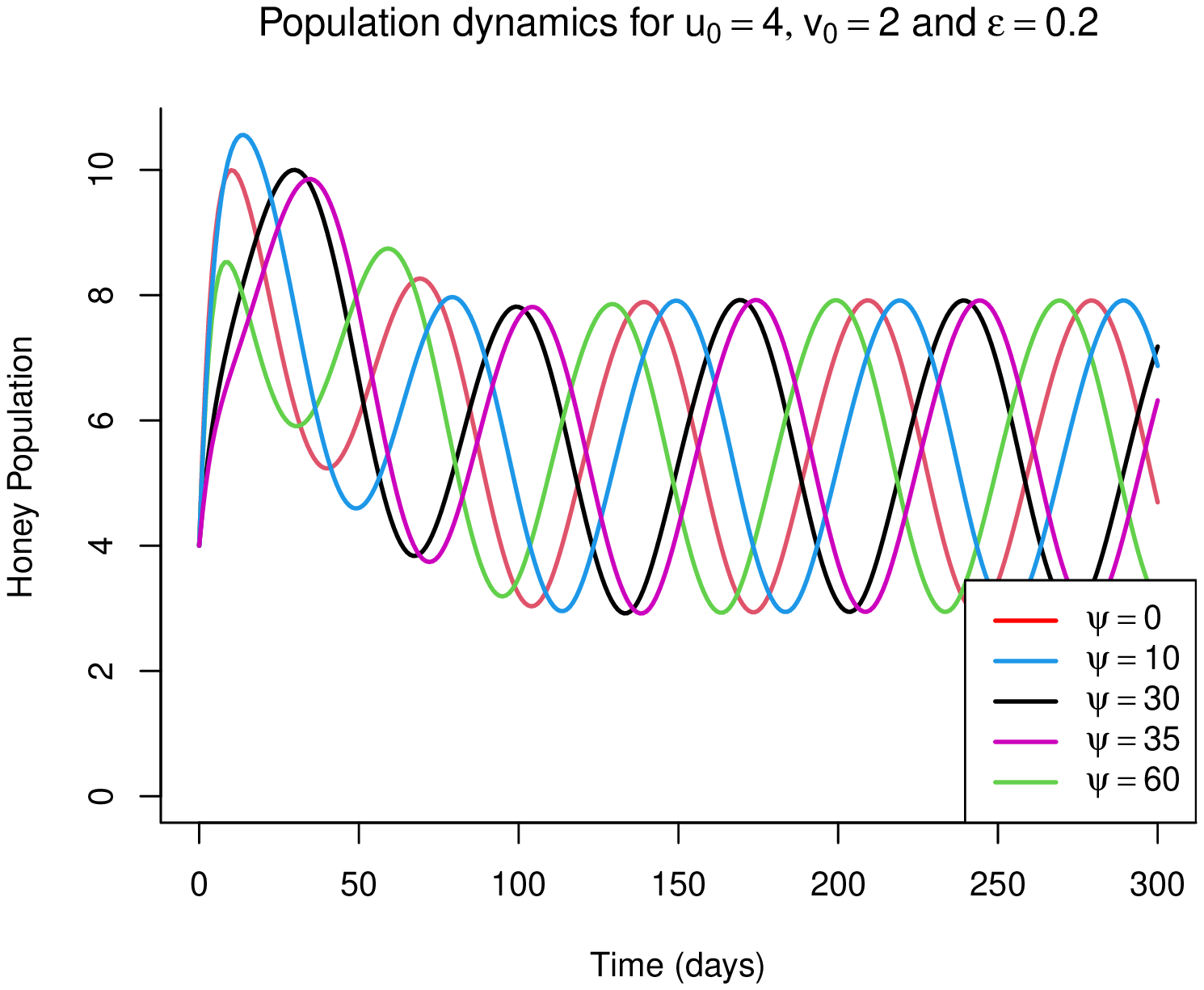}
			\label{psi-dynamic_2}
		}
\caption{Impacts of the timing of the maximum egg-laying rate ($\psi$). The blue area is colony collapse, and the red area is colony coexistence. $\bar{r}_0=2.86$, $\bar{d}_h=\bar{d}_m=0.25$, $\omega=0.3$,  $\hat{K}=2.04$, and $\gamma=70, \epsilon=0.2 \& 0.35$  Honey bee initial population is $u_0 \in [0, 20]$, and mite initial population is $v_0 \in [0, 20]$}
			\label{fig:psi-2d}
\end{figure}

Next, we explore the impacts of the timing of the maximum egg-laying rate ($\psi$) on colony survival and population dynamics in Figure \ref{fig:psi-2d} by fixing 
$$\bar{r}_0=2.86, \bar{d}_h=\bar{d}_m=0.25, \omega=0.3, \hat{K}=2.04, \gamma=70, \epsilon \in \{ 0.2, 0.35\}.$$

$\gamma=70:$ 1) $\epsilon=0.2$ (the order from large to small): $\psi=10$ is largest, then $\psi=0$, $ \psi=60=30$ these two cases have same survival area, $\psi=35$, and $\psi=40$ is the smallest.

2) $\epsilon=0.35$ (the order from large to small): $\psi=60$ is largest, then $\psi=0$, $\psi=40$, $\psi=10$, $\psi=35$, and $\psi=30$ is the smallest.

Notice that the seasonality period is $\gamma=70$ and $\epsilon=0.35$. We choose the timing of the maximum egg-laying rate $\psi\in \{0, 10, 30, 35, 40$ and $60\}$ and observe that the red area of the basin attractions for the colony survival is largest when the timing of the maximum laying rate ($\psi$) is $\psi=60$ (Figure \ref{70-psi60}), then the second largest in the case when $\psi=0$ (Figure \ref{70-psi0}), the smallest one is $\psi=30$ (Figure \ref{70-psi30}), and the second smallest in the case when $\psi=35$ (Figure \ref{70-psi35}). These observations from Figure \ref{fig:psi-2d} regarding the impacts of the maximum laying rate ($\psi$) of our honeybee-parasite model \ref{Honeybee-mite-scaled} seem to show similar trends of our honey bee-only model \ref{honeybee} (see Figure \ref{fig:1D-psi}): as the $\psi$ increases, the seasonality can suppress the survival of the colony; and after the minimum survival area, the $\psi$ can promote the survival of the colony. But the significant difference with the bee-only model is the smallest area is not $\psi=\frac{\gamma}{2}$. Figure \ref{psi-dynamic} and \ref{psi-dynamic_2} show how different timing of the maximum egg-laying rate $\psi$ can lead to different colony dynamics.\\

To further understand the impacts of the timing of the maximum laying rate ($\psi$) of our honeybee-parasite model \ref{Honeybee-mite-scaled}, we set the strength of the seasonality being $\epsilon=0.2$, and choose the timing of the maximum egg-laying rate $\psi\in \{0, 10, 30, 35, 40$ and $60\}$, respectively. The basin attractions for the colony survival is largest when the timing of the maximum laying rate ($\psi$) is $\psi=10$(Figure \ref{70-psi10_1}), the second largest in the case when $\psi=0$ (Figure \ref{70-psi0_1}), the smallest one is $\psi=40$ (Figure \ref{70-psi40_1}), and the second smallest in the case when $\psi=35$ (Figure \ref{70-psi35_1}). These observations are different than the case of $\epsilon =0.35$ shown in Figure \ref{fig:psi-2d} and the case of the honey bee only model \ref{honeybee} (see Figure \ref{fig:1D-psi}). 
The significant difference is that $\psi$ can promote the survival of the colony at the very beginning of $\psi$ growth ($\psi=10$ in our simulation). These comparisons and our further simulations suggest that the impacts of the timing of the maximum laying rate ($\psi$) on the honey bee colony survival in the presence of parasitism are very complicated. The area of the basin attractions for the colony survival may be increasing or decreasing with respect to the value of $\psi$ and $\epsilon$ without clear patterns.\\

 \begin{figure}[ht]
		\centering
	\subfigure[honey bee population]{
			\includegraphics[width=5cm]{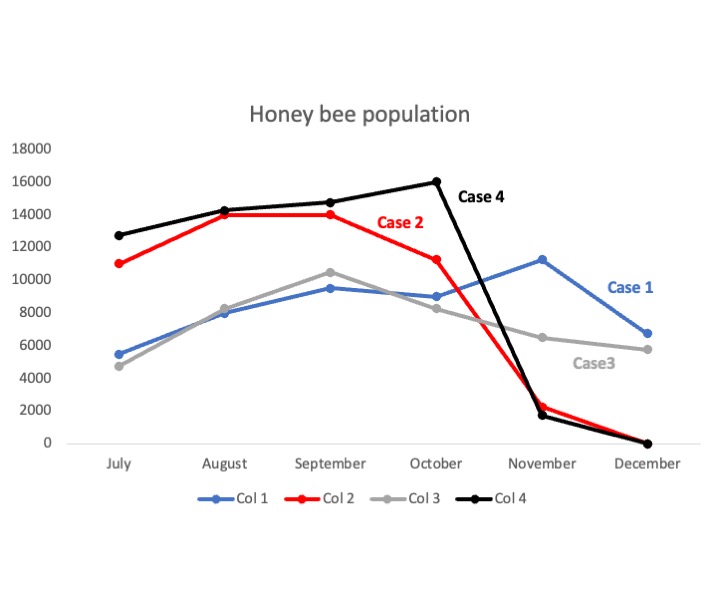}\label{fig:bee}
		}
		\subfigure[Population Dynamics]{
			\includegraphics[width=5cm]{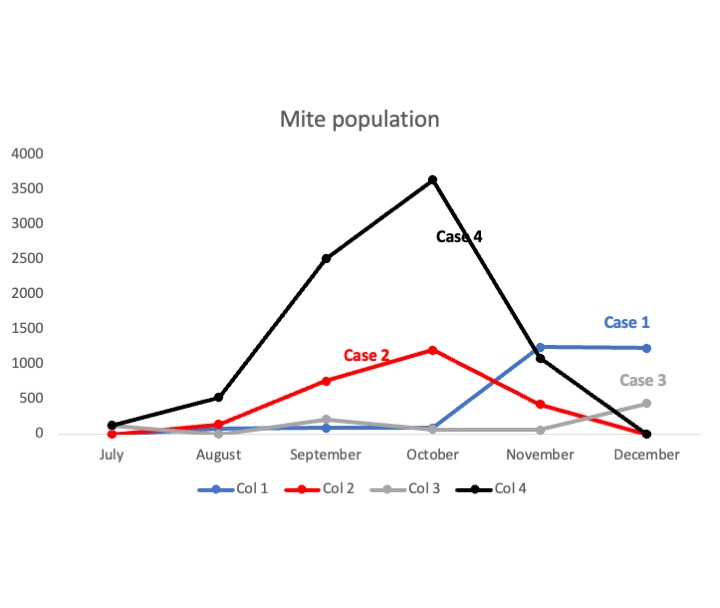}\label{fig:mite}
		}
  
    \subfigure[no seasonality]{
			\includegraphics[width=5cm]{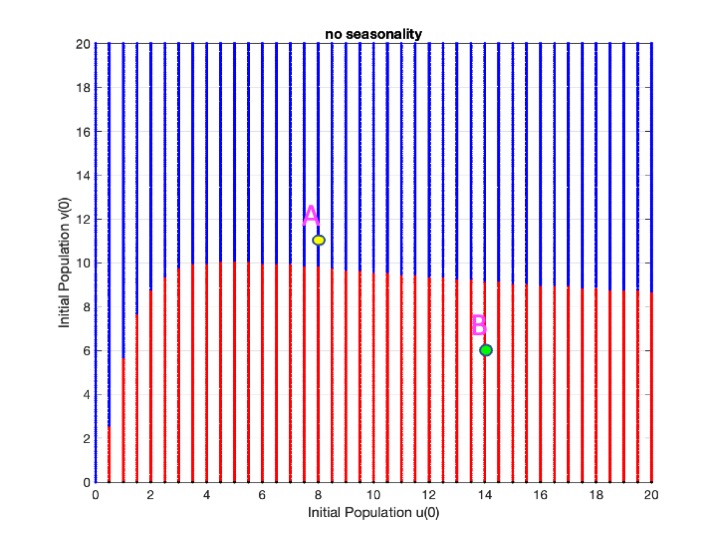}\label{fig:fig7a}}  
  \subfigure[$\gamma=80, \epsilon=0.35$]{
			\includegraphics[width=5cm]{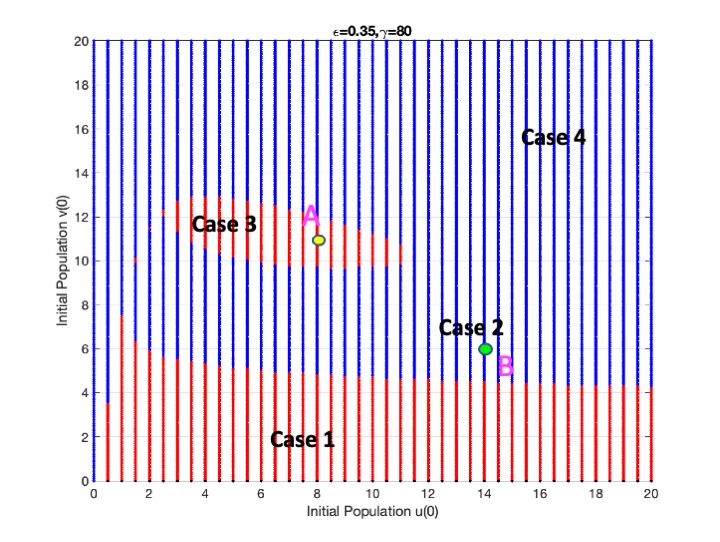}\label{fig:fig7d}
		}
  
 \caption{Figure \ref{fig:bee}: the total bee population of four colonies from July to December. Colonies 1 (Case 1, blue) and 3 (Case 3, gray) survive, and Colonies 2 (Case 2, red) and 4 (Case 4, black) collapse. Figure \ref{fig:mite}: the total mite population in four colonies from July to December. Colonies had different initial populations. Figure \ref{fig:fig7a}: the simulation result from Figure \ref{before_no season2} with two signed points A and B. Figure \ref{fig:fig7d}: the simulation result from Figure \ref{before_8035} with two signed points A and B and cases. These four cases correspond to Figure \ref{fig:bee} \& \ref{fig:mite} colonies.}
			\label{fig:pop_data}
\end{figure}

By comparing the basins of attractions of the honeybee-mite system without seasonality in Figure \ref{before_no season2} to the honeybee-mite system with seasonality in Figure \ref{before_8035}, \ref{70-psi40_1} \& \ref{70-psi0}, we observe that seasonality can split the basins of attractions into disconnected regions. This may lead to two scenarios after adding seasonality: (1) the colony may survive from collapsing (see Point A in Figure \ref{fig:fig7a} versus Figure \ref{fig:fig7d}), and (2) the colony may be prone to collapsing (see Point B in Figure \ref{fig:fig7a} versus Figure \ref{fig:fig7d}). This suggests that seasonality may generate varied outcomes depending on initial conditions. For instance, while an initial rise in the parasite population is generally perceived as detrimental, it can enhance colony survival under specific circumstances, particularly when considering seasonal factors (compare points A and B in Figure \ref{fig:fig7d}). This phenomenon has been observed in experimental data \cite{degrandi2020can}. 
 To illustrate those observations, we use Figure \ref{before_8035} as an example where we list four cases. Among them, the initial bees population of Colony 1 (Case 1, blue) and Colony 3 (Case 3, gray) are similar, and Colony 2 (Case 2, red) and Colony 4 (Case 4, black) are close. While, the initial mite population is increasing in the order of Case 1, Case 2, Case 3, and Case 4. Figure \ref{fig:bee} shows the colony of Case 1 and Case 3 survived while Case 2 and Case 4 collapsed, especially Colony 3 has fewer bees and more mites than Colony 2, but survives.\\
 
\begin{figure}[ht]
		\centering
\subfigure[Point A bee population]{
			\includegraphics[width=5cm]{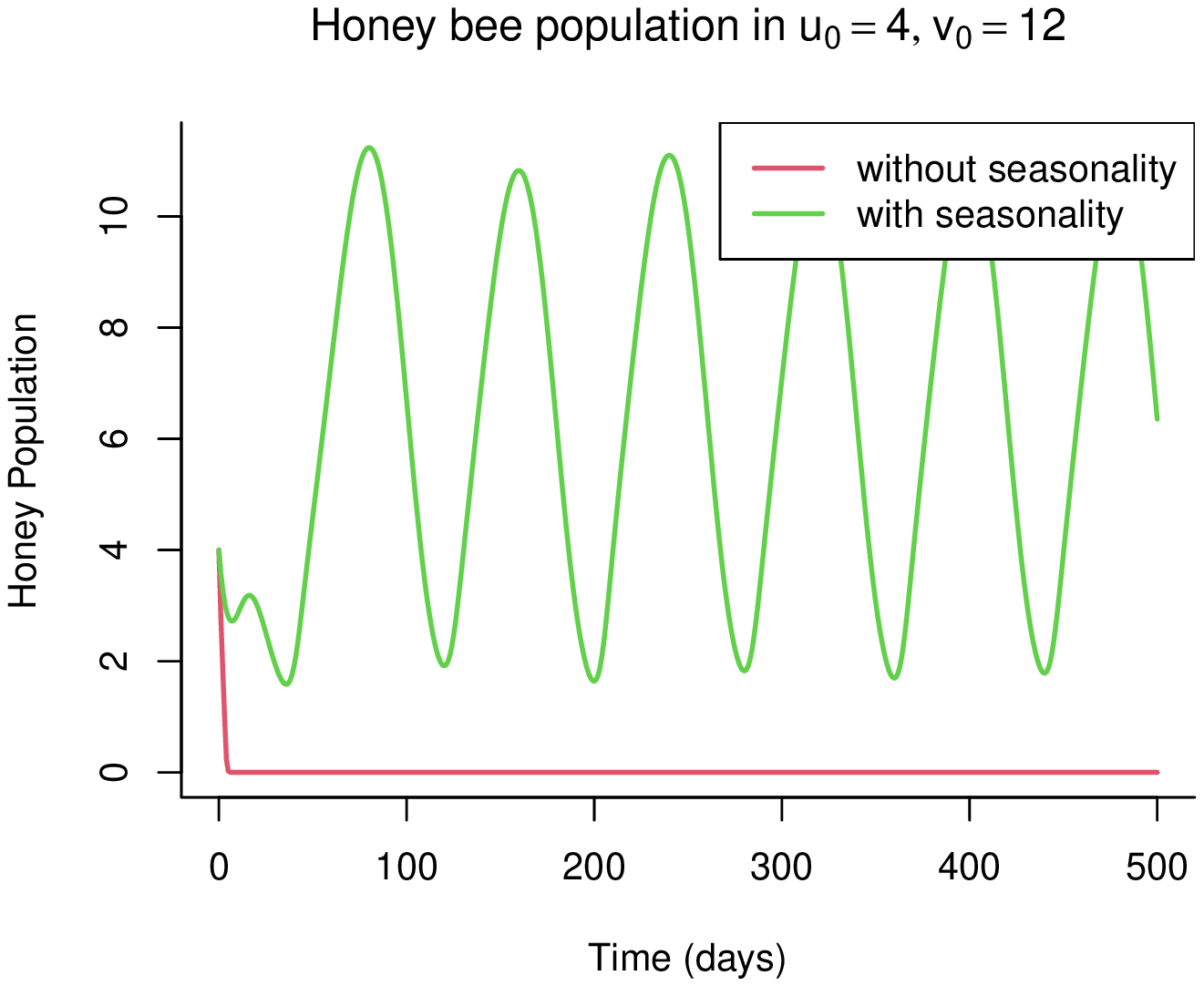}\label{fig:A-bee}
		}
  \subfigure[Point A mite population]{
			\includegraphics[width=5cm]{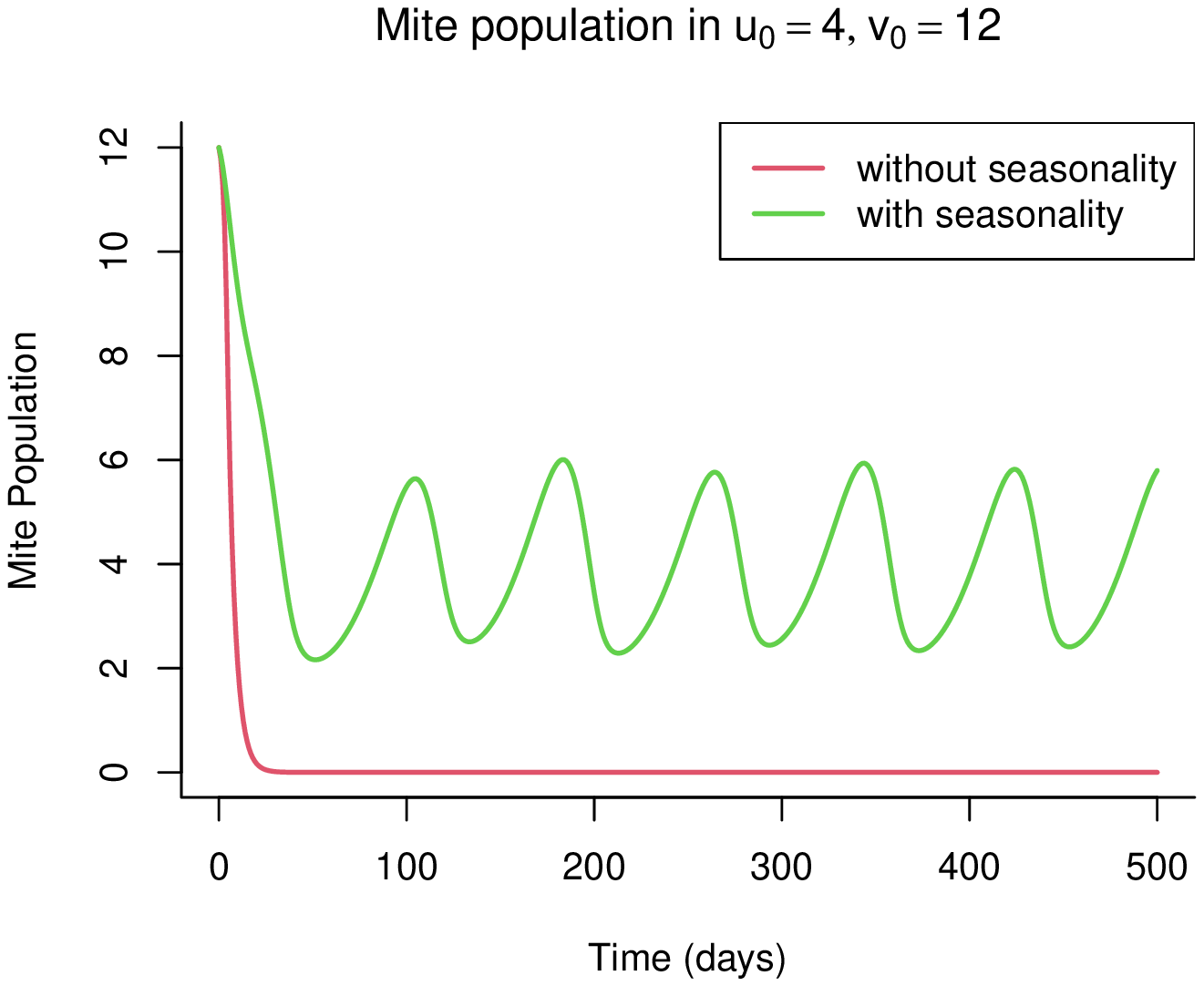}\label{fig:A-mite}
		}
  
   \subfigure[Point B bee population]{
			\includegraphics[width=5cm]{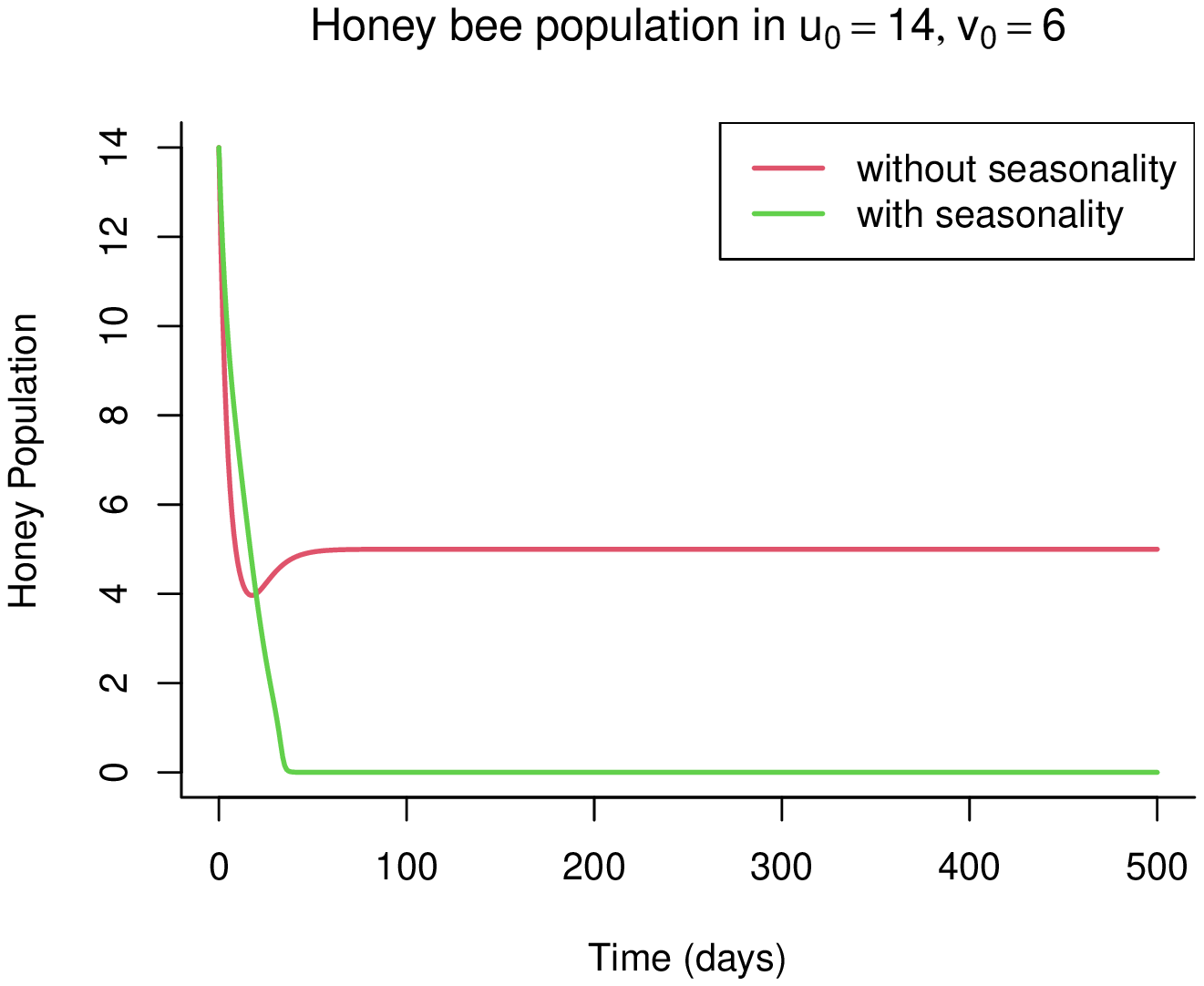}\label{fig:B-bee}}  
   \subfigure[Point B mite population]{
			\includegraphics[width=5cm]{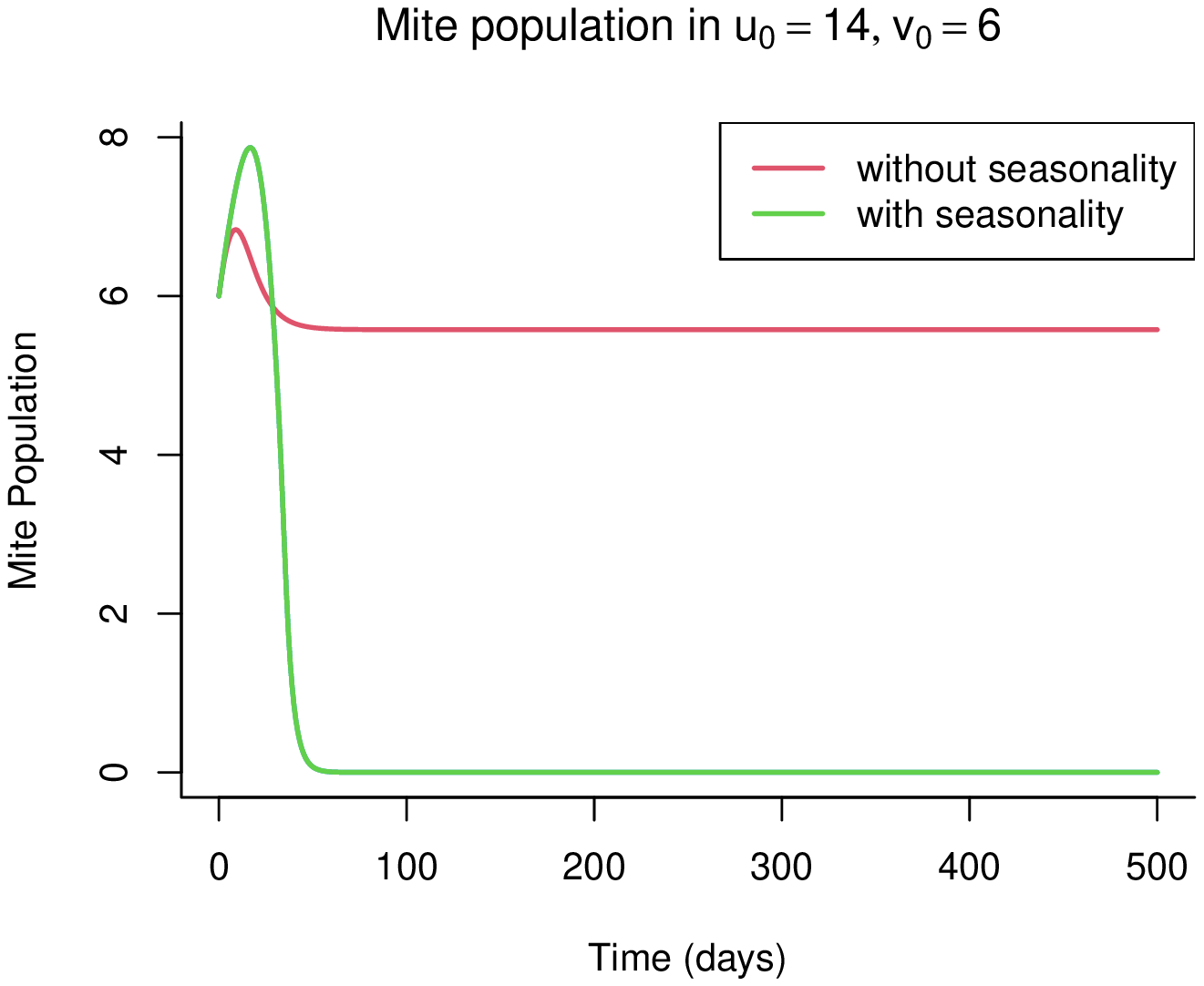}\label{fig:B-mite}
		}
  
  \caption{Colony dynamics with time series. These point A and point B correspond with Figure \ref{fig:fig7a} \& \ref{fig:fig7d}. Point A: seasonality leads the colony from collapsing to survive. Point B: seasonality leads the colony from survival to collapse.}
			\label{fig:AB}
\end{figure}

The potential biological explanation for this phenomenon lies in the heart of seasonality impacts on the egg laying rate incorporated in the model, and mite population is impacted through bee population. For the mite population to grow, colonies must have enough bees. If the system doesn't consider the impacts of seasonality (Model \ref{honeybee-mite-constant}), then a higher parasite level ($v_0$: Point A $>$ Point B) leads to colony collapse (see red curves in Figure \ref{fig:A-mite} \& \ref{fig:B-mite}), because parasitism reduces colony population growth by shortening the lifespan of adult workers, then the population of bees is reduced and so will Varroa population growth. However, the system with seasonality (Model \ref{Honeybee-mite-scaled}) leads to a switch in the outcomes of these two colonies, which is survival colony goes to collapse because of seasonality, whereas the collapsing colony becomes survival. The point is that the egg-laying rate of bees is periodic due to seasonal effects, then the number of bees will increase at some time intervals (the green curve in Figure \ref{fig:A-bee}). At a higher parasite level, fewer bees will bring the mite population down ($\frac{\omega u}{1+u} v$) to a manageable level, and the seasonality egg-laying rate helps the colony grow up periodically (seasonality in Point A). At a lower parasite level, seasonality also leads mites to grow up more than without seasonality effects (Figure \ref{fig:B-mite}). Seasonality and high numbers of bees may lead to excessive mite growth beyond the colony's sustainable threshold and colony collapse. This principle is similar to one method of controlling Varroa mites: removing the brood from the hive and interrupting the brood reproductive cycle. With no brood present, mites are compelled to feed on adult bees, which can limit the mites' ability to reproduce, helping to control their populations \cite{jack2021integrated}. Nevertheless, this method will be affected by seasonality. Removing lots of broods in the fall may have strong negative impacts on overwintering survival \cite{jack2020evaluating}.\\

\begin{figure}[ht]
		\centering
	
		\subfigure[$\omega=0.18,\epsilon=0.2$]{
			\includegraphics[width=3cm]{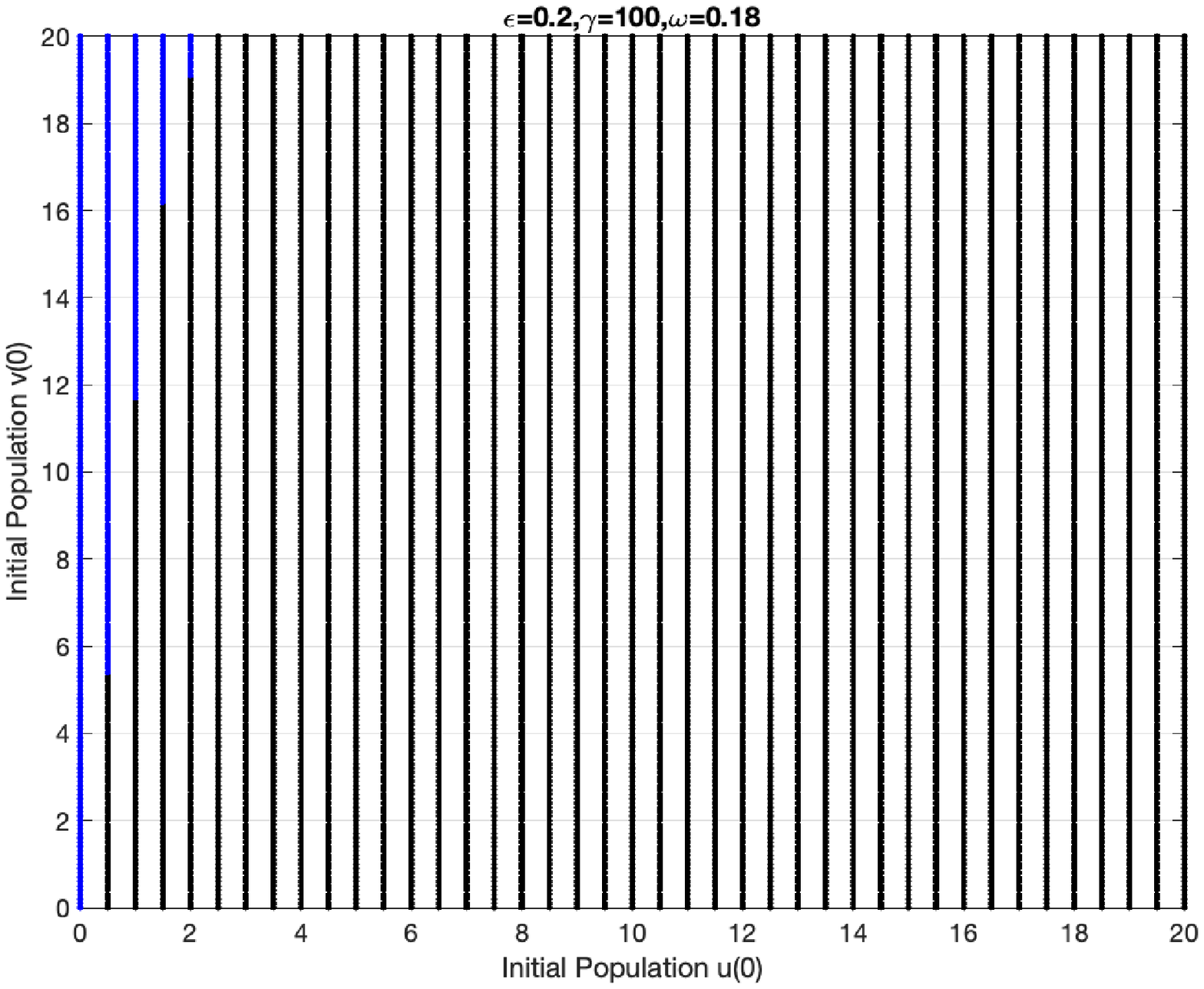}
			\label{mite-0.18}
		}
		\subfigure[$\omega=0.3,\epsilon=0.2$]{
			\includegraphics[width=3cm]{Figure/gamma1002.eps}
			\label{mite-0.3}
		}
		\subfigure[$\omega=0.5,\epsilon=0.2$]{
			\includegraphics[width=3cm]{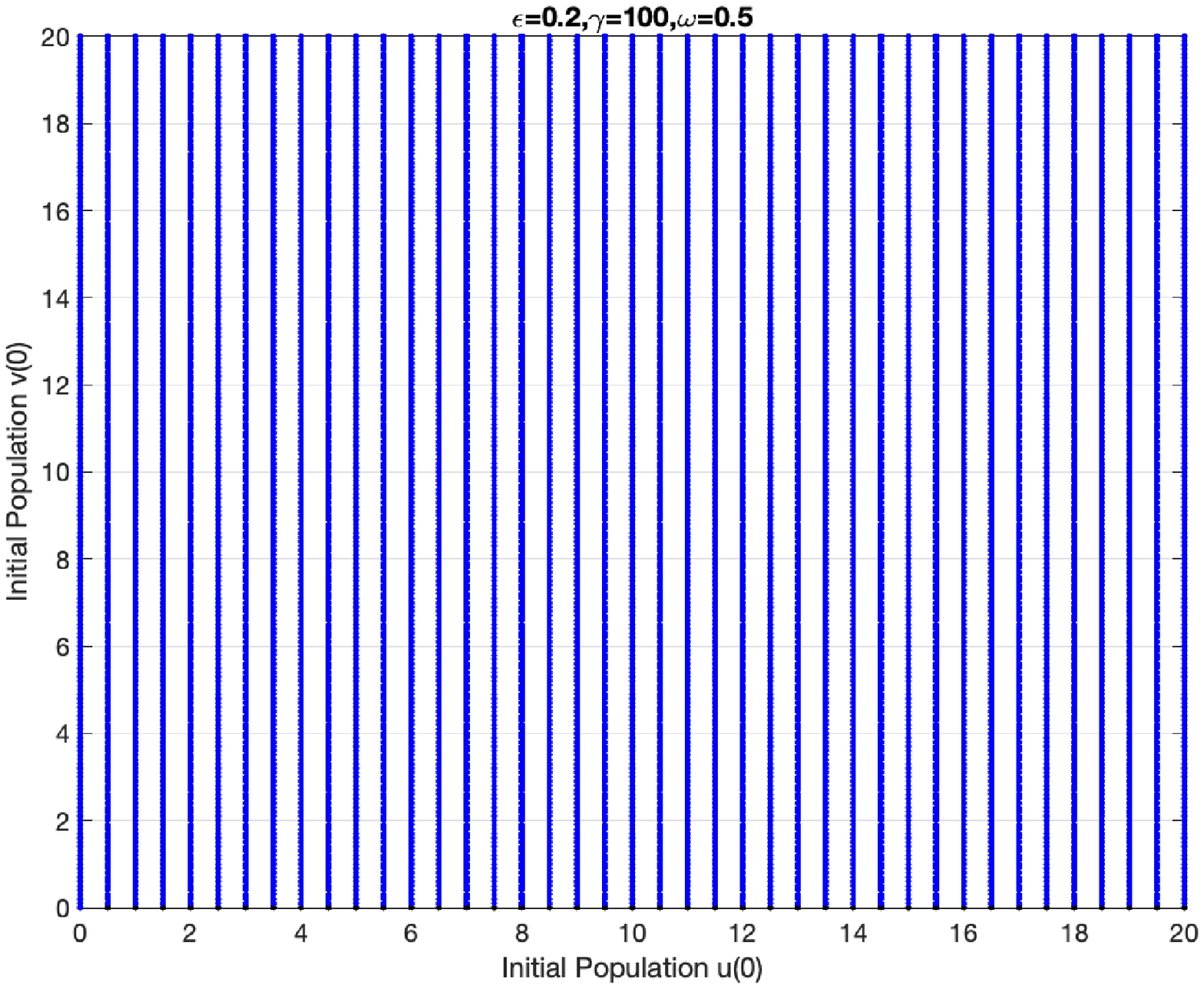}
			\label{mite-0.5}
		}
		\subfigure[$\omega=0.18,\epsilon=0$]{
			\includegraphics[width=5cm]{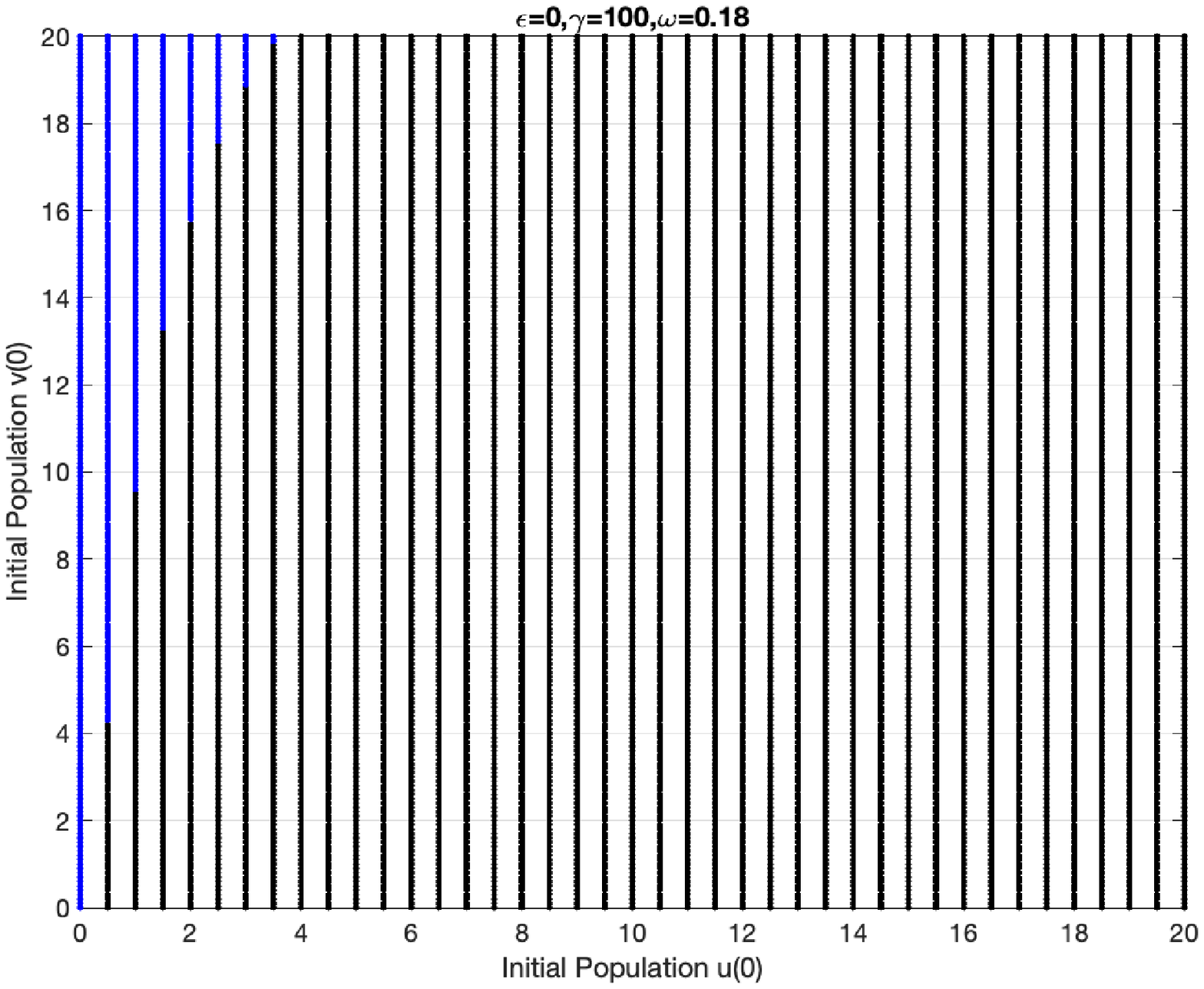}
			\label{mite-0.18-without}
		}
  \subfigure[Population Dynamics]{
			\includegraphics[width=5cm]{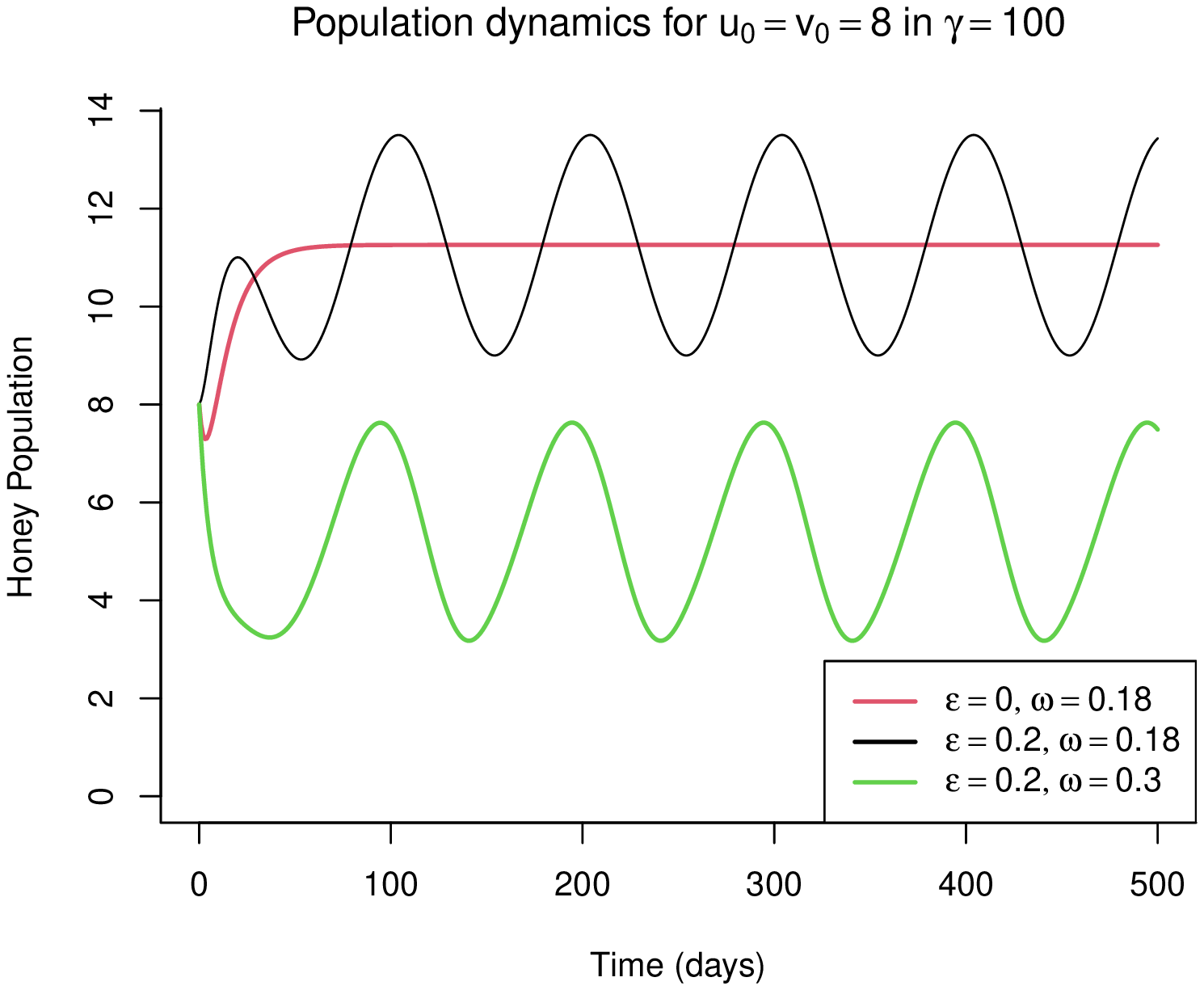}
			\label{omega_dynamic}
		}
			\caption{Impacts of parasitism ($\omega$) on the colony dynamics of honeybee-mite model \eqref{Honeybee-mite}. The blue area is colony collapse, the red area is colony coexistence, and the black area is only bee survive with $\bar{r}_0=2.86$, $\bar{d}_h=\bar{d}_m=0.25$, $\gamma=100$, $\psi=0$, $\epsilon=0.2$    and $\hat{K}=2.04$. Honey bee initial population is $u_0 \in [0, 20]$, and mite initial population is $v_0 \in [0, 20]$ }
			\label{fig:parasitism}
\end{figure}

Now we explore the impacts of parasitism $\omega$ on honey bee population dynamics and its colony's survival in Figure \ref{fig:parasitism}. Comparison of black areas (which is the basins of attractions of only honey bee survival) in Figure \ref{mite-0.18-without} \& \ref{mite-0.18} suggest that small parasitism (e.g., $\omega=0.18$) with seasonality is more likely to lead to the colony survival than the case without seasonality. When parasitism is not small (e.g., $\omega=0.30$) (see Figure \ref{mite-0.3}), seasonality can destabilize the system and decrease the average population of the honey bee.  When $\omega$ is large (e.g., $\omega=0.5$), parasitism has negative impacts on the honey bee colony that lead the colony to collapse (see all blue areas in Figure \ref{mite-0.5}). Figure \ref{omega_dynamic} shows increasing parasitism, colonies may still survive but the average population of honey bees decreases (see black and green curves). These observations are in line with our theorem \ref{th:Ee} for the case without seasonality.\\


\subsection{Impacts of seasonality on stable limit cycle coexistence}

\begin{figure}[ht]
		\centering
		\subfigure[$\gamma=4, \epsilon=0.2$]{
			\includegraphics[width=5cm]{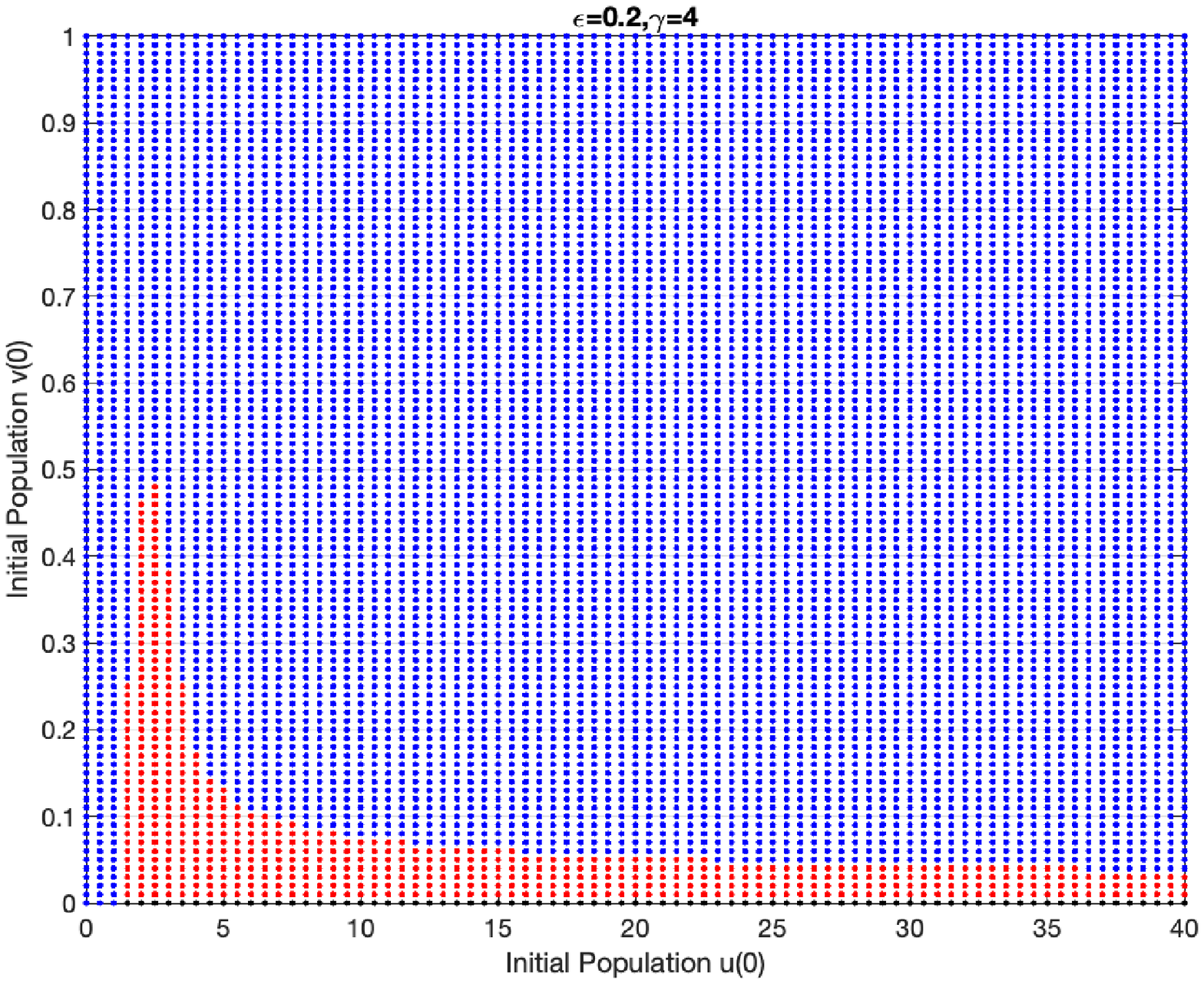}
			\label{cycle_42}
		}
  \subfigure[$\gamma=4, \epsilon=0.5$]{
			\includegraphics[width=5cm]{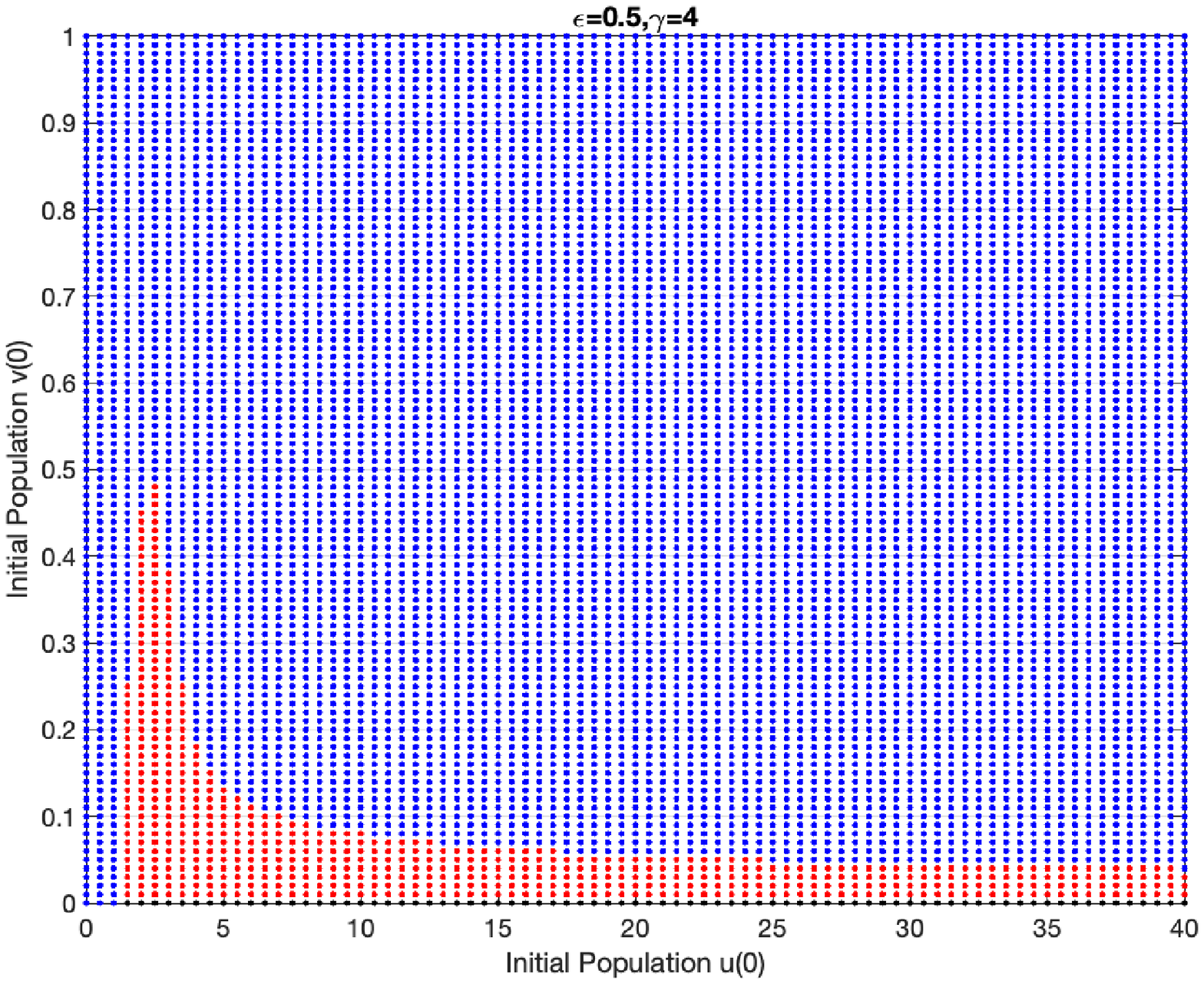}
			\label{cycle_45}
		}
\subfigure[$\gamma=40, \epsilon=0.2$]{
			\includegraphics[width=5cm]{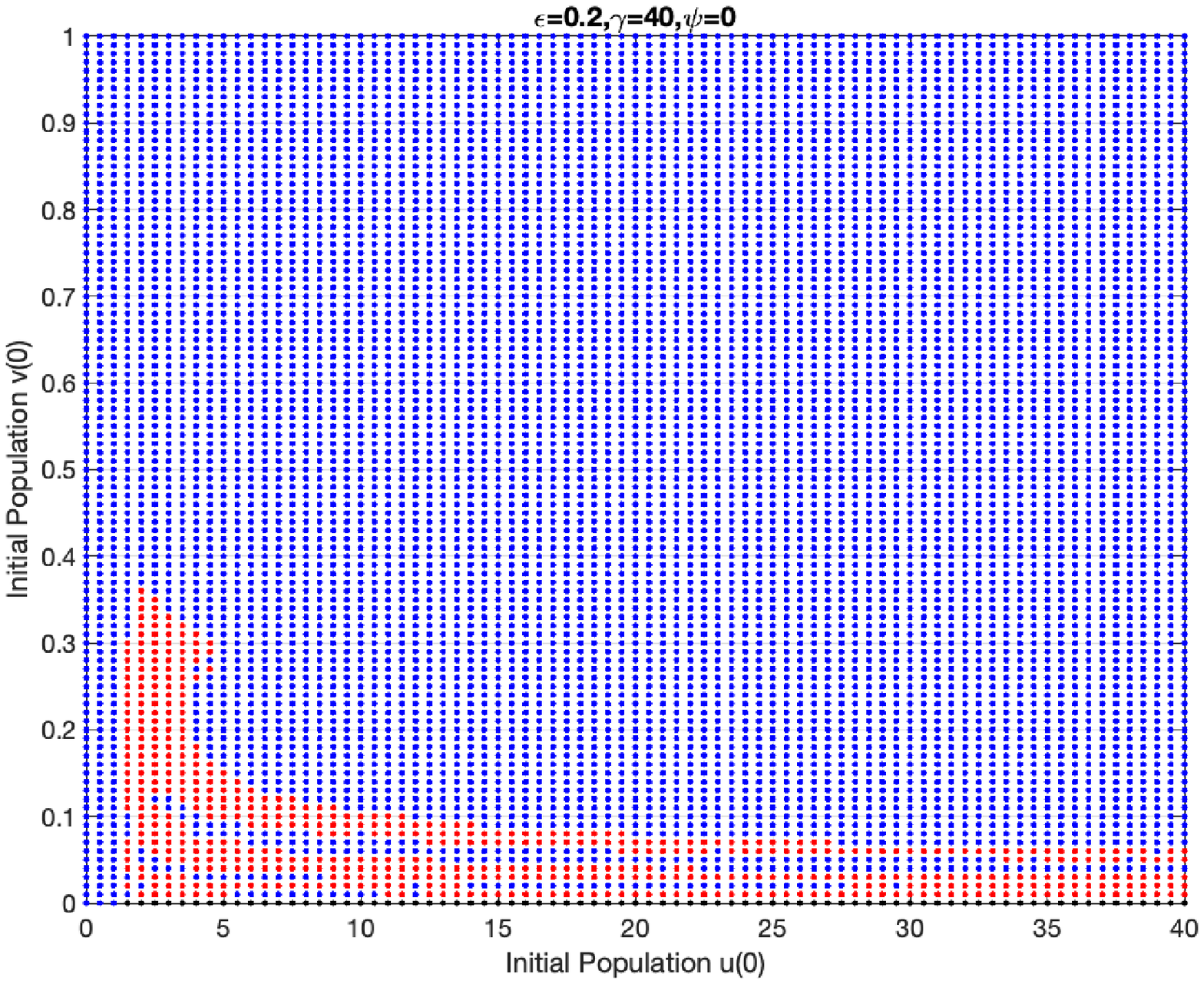}
			\label{cycle_402}
		}
		\subfigure[$\gamma=40, \epsilon=0.5$]{
			\includegraphics[width=5cm]{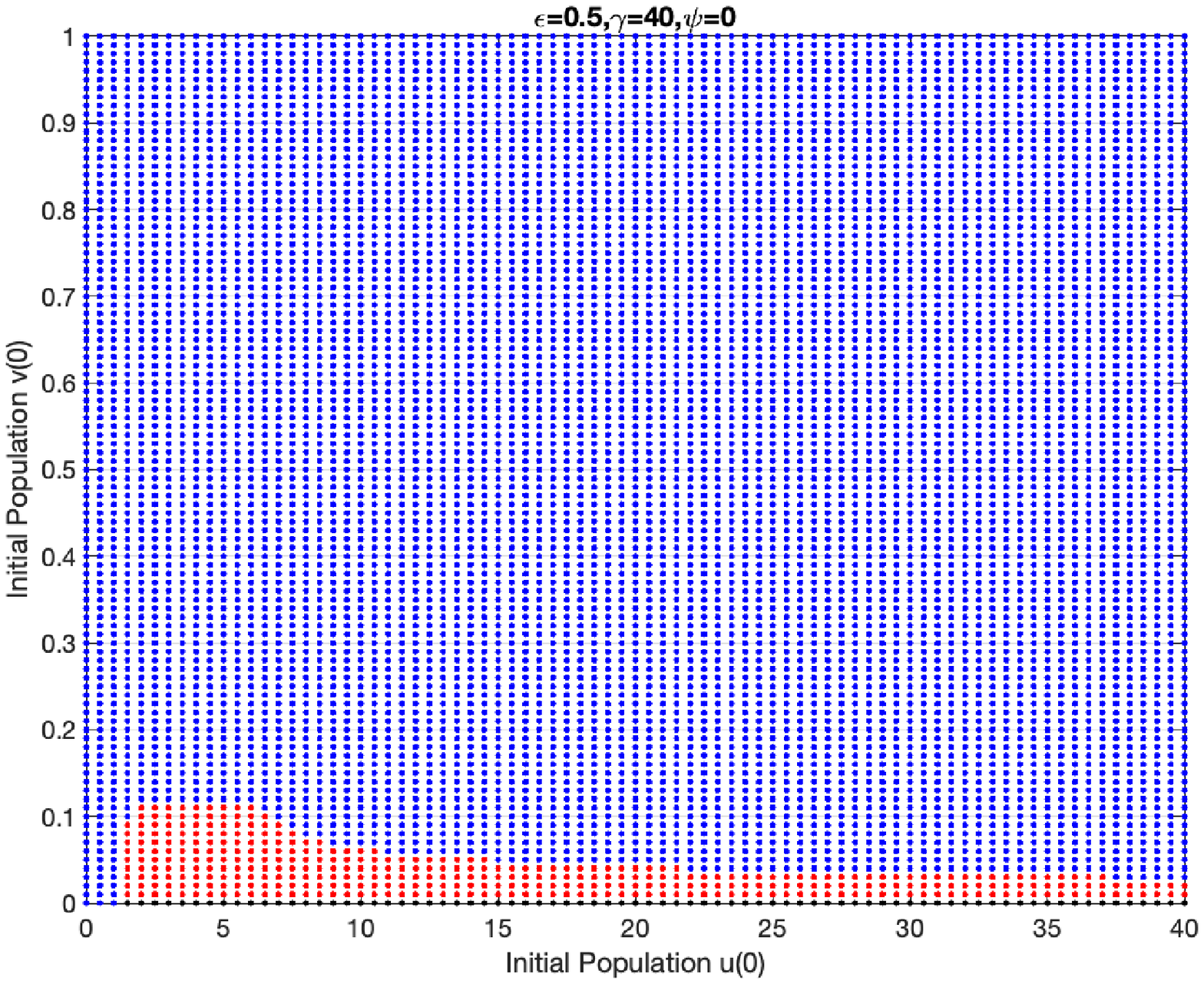}
			\label{cycle_405}
		}
			\subfigure[no seasonality]{
			\includegraphics[width=5cm]{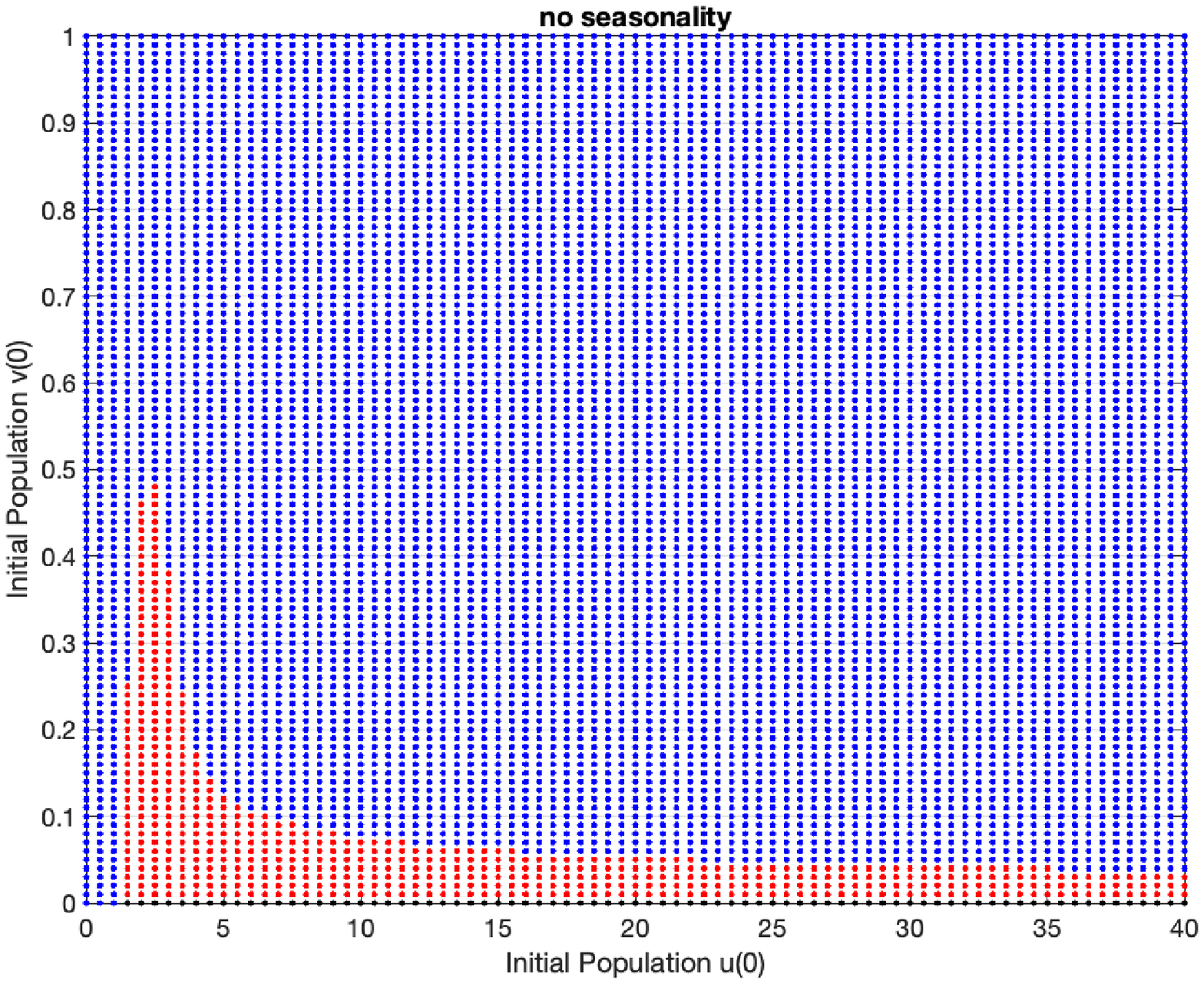}
			\label{cycle_no}
		}
  \subfigure[Honey bee population dynamics]{
			\includegraphics[width=5cm]{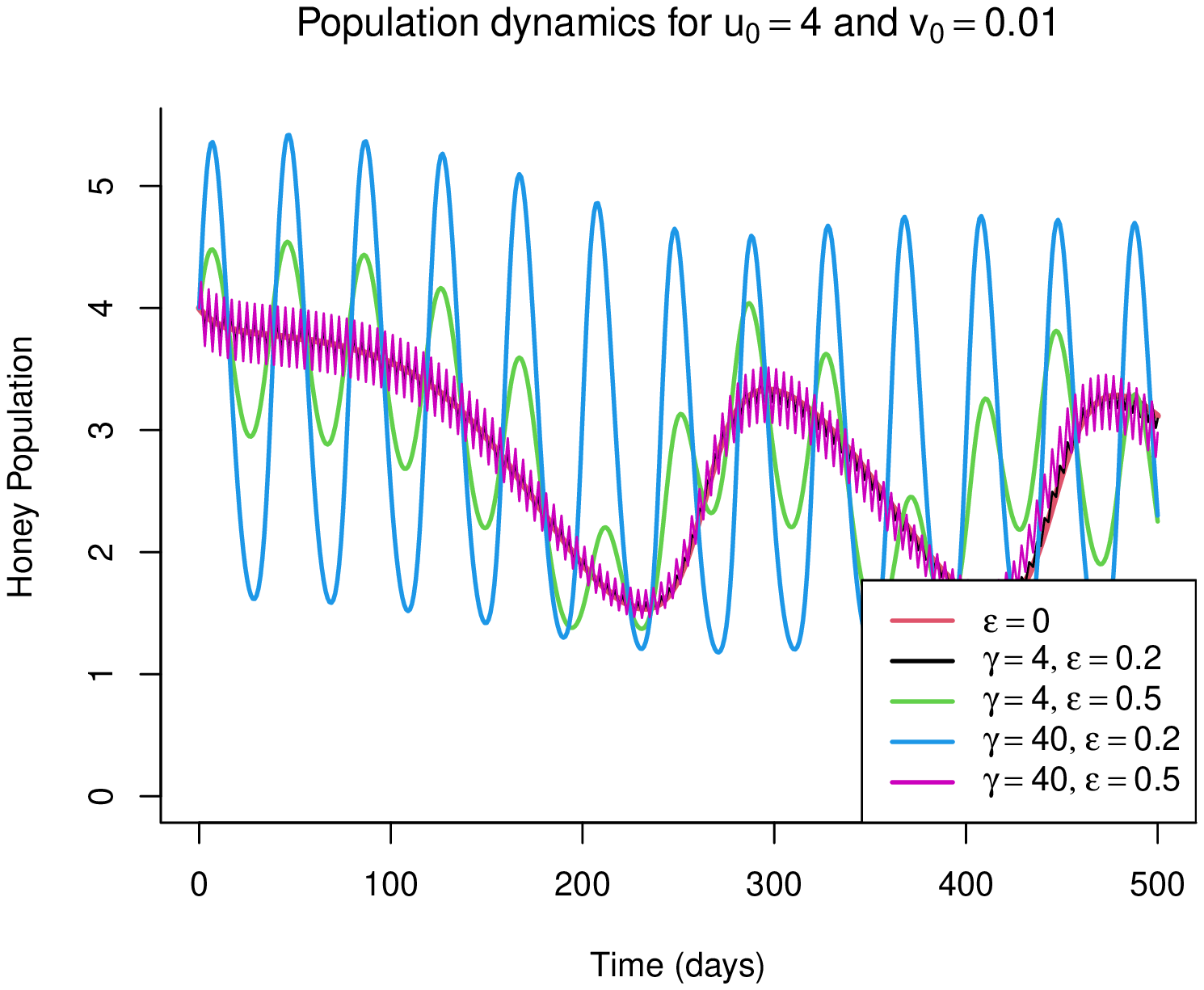}
			\label{cycle_dynamic}
		}
 			\caption{Impacts of seasonality on the stable limit cycle: the strength of seasonality $\epsilon$ and the period of seasonality $\gamma$ when $\bar{r}_0=1$, $\bar{d}_h=0.2$, $\bar{d}_m=0.21$, $\omega=0.3$, $\psi=0$, and $\hat{K}=4.49$.  Honey bee initial population is $u_0 \in [0, 40]$, and mite initial population is $v_0 \in [0, 1]$.  The blue area is colony collapse, and the red area is colony coexistence.}
			\label{fig:limit_cycle1}
\end{figure}

We choose a stable limit cycle example of our honeybee-parasite interaction model \ref{Honeybee-mite-scaled} by setting 
$$\bar{r}_0=1,\,\bar{d}_h=0.2,\, \bar{d}_m=0.21,\,\omega=0.3,\,\hat{K}=4.49,\,\psi=0$$
which has a stable collapsing state $(0,0)$ for the colony, and a stable limit cycle around the source interior equilibrium $(2.33, 0.3875)$ whose basins of attractions are red area shown in Figure \ref{cycle_no}.\\

We explore the impacts of the seasonality strength $\epsilon$, the period of seasonality $\gamma$, the queen laying her maximum number of eggs at time $\psi$, and the parasitism effects $\omega$ on the colony survival and population dynamics. We perform the following simulation in Figure \ref{fig:limit_cycle1} on the basin’s attractions of our honeybee-parasite model \ref{Honeybee-mite-scaled}. We set the queen laying her maximum number of eggs at time $\psi=0$ to observe the impacts of $\gamma$ and $\epsilon$. 

\begin{enumerate}
    \item When the period of the seasonality is small, i.e., $\gamma=4$, comparisons of areas of basin attractions for the colony survival among Figure \ref{cycle_no} (no seasonality), \ref{cycle_42} (the seasonality strength $\epsilon=0.2$) suggest that small seasonality strength $\epsilon$ may not significantly impact the survival of the colony much but larger seasonality strength $\epsilon$ can generate larger population amplitude (see Figure \ref{cycle_dynamic}).
    \item When the period of the seasonality is in the intermediate range, e.g., $\gamma=40$, comparisons of areas of basin attractions for the colony survival among Figure \ref{cycle_no} (no seasonality), \ref{cycle_402} (the seasonality strength $\epsilon=0.2$) and \ref{cycle_405} (the seasonality strength $\epsilon=0.5$) suggest that seasonality strength $\epsilon$ seems to suppress the survival of the colony. 
    \item When the the seasonality strength $\epsilon$ is fixed, increasing the period of the seasonality $\gamma$ seems to suppress the survival of the colony (See Figures \ref{cycle_42} \& \ref{cycle_402} and Figures \ref{cycle_45} \& \ref{cycle_405}).
    \item The large $\gamma$ and $\epsilon$ would lead to the colony collapsing as we observe that the colony collapses when $\gamma > 60$.
\end{enumerate}

\begin{figure}[ht]
		\centering
	\subfigure[$\gamma=40, \psi=0$]{
			\includegraphics[width=5cm]{Figure/e=0.2_gamma=40.eps}
			\label{cycle_402_2}
		}
\subfigure[$\gamma=40,  \psi=15$]{
			\includegraphics[width=5cm]{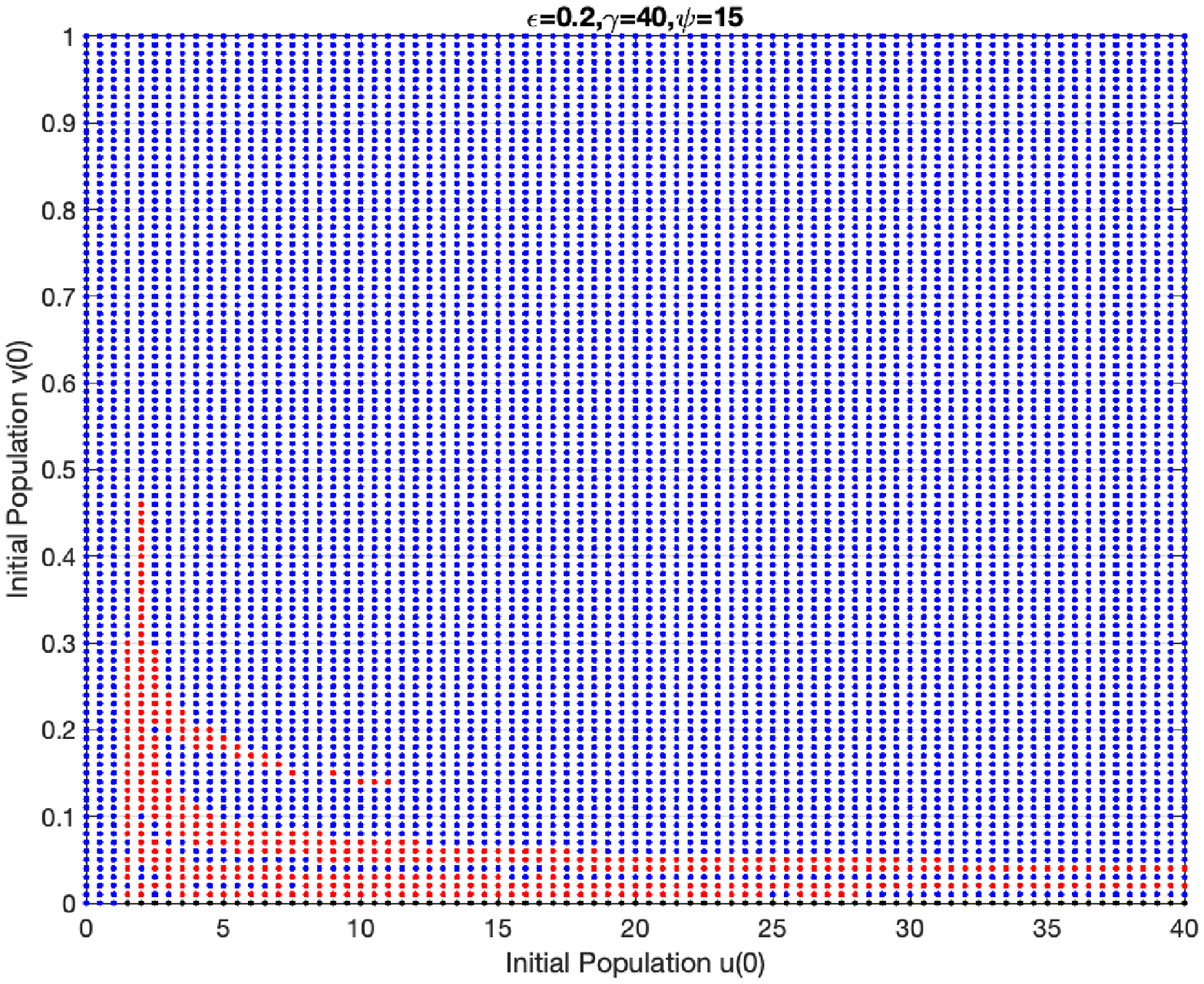}
			\label{cycle_psi15}
		}
		\subfigure[$\gamma=40, \psi=35$]{
			\includegraphics[width=5cm]{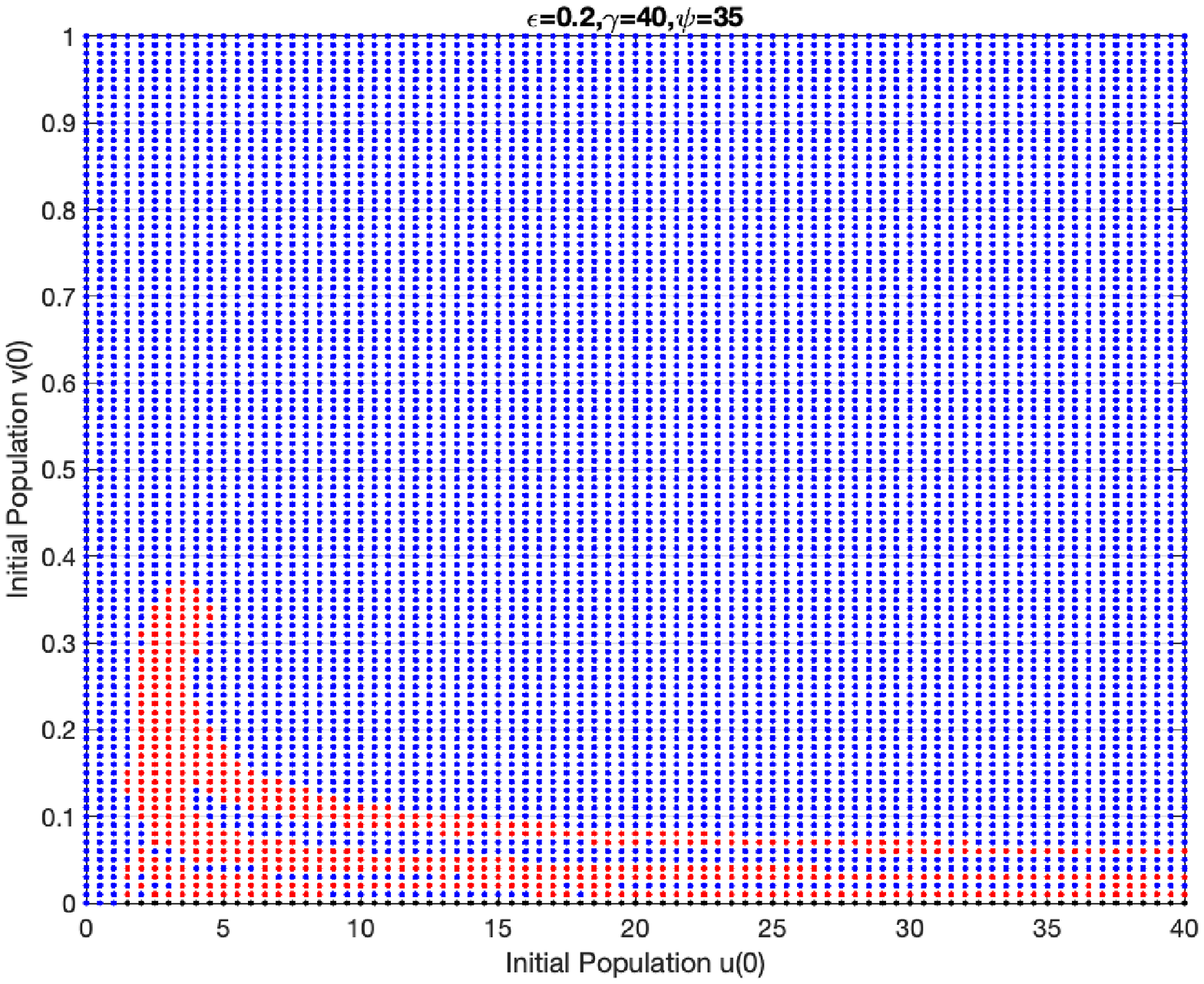}
			\label{cycle_psi35}
		}
  \subfigure[Honey bee population dynamics]{
			\includegraphics[width=5cm]{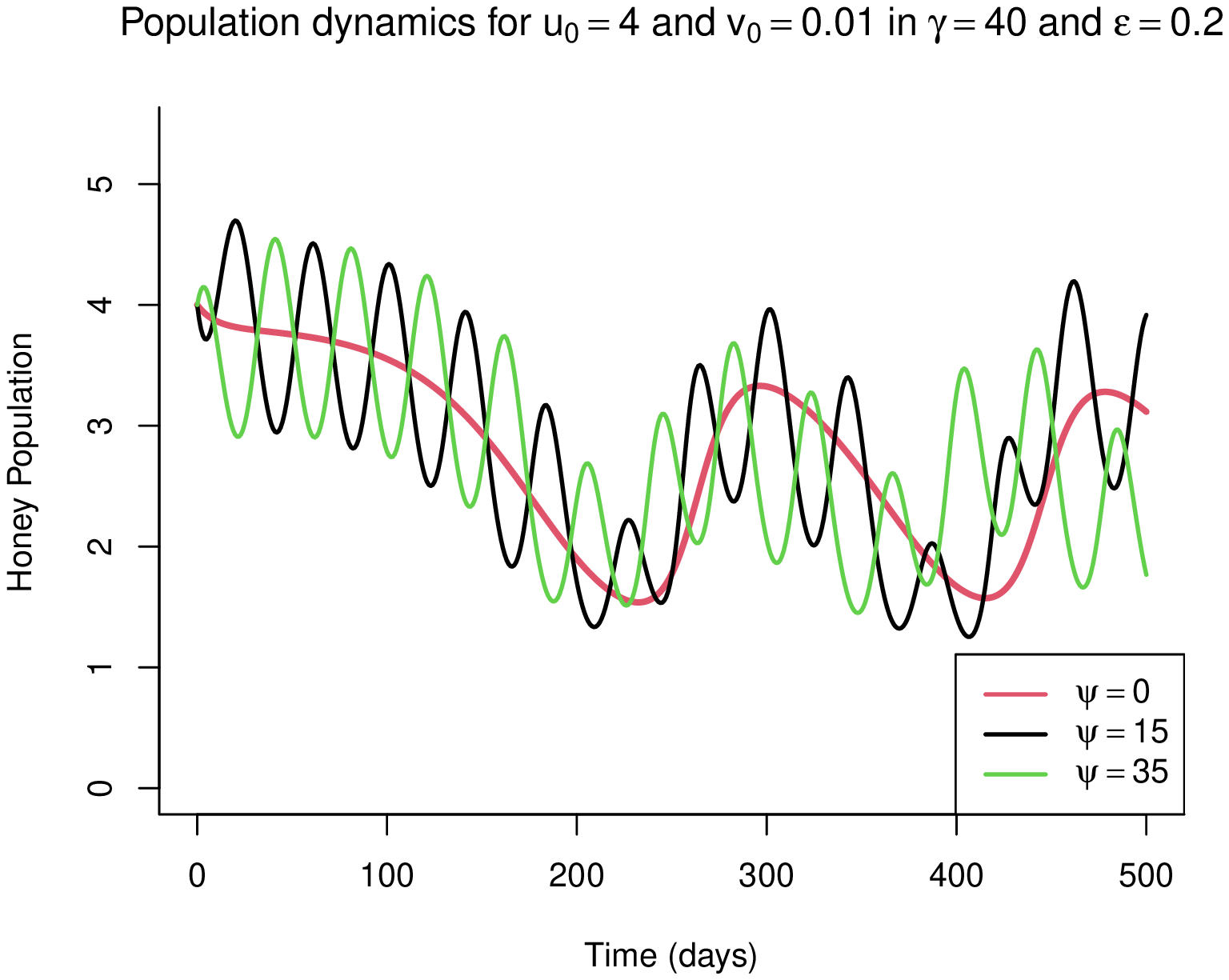}
			\label{cycle_dynamic2}
		}
 			\caption{Impacts of seasonality on the stable limit cycle: the timing of the maximum egg-laying rate $\psi$ when $\bar{r}_0=1$, $\bar{d}_h=0.2$, $\bar{d}_m=0.21$, $\omega=0.3$, $\gamma=40$, $\epsilon=0.2$, and $\hat{K}=4.49$.  Honey bee initial population is $u_0 \in [0, 40]$, and mite initial population is $v_0 \in [0, 1]$.  The blue area is colony collapse, the red area is colony coexistence.}
			\label{fig:limit_cycle-psi}
\end{figure}

Let the period of the seasonality be $\gamma = 40$ and the strength of the seasonality be $\epsilon=0.2$. We explore the impacts of the timing of the maximum egg-laying rate $\psi$ by varying $\psi$=0, $15 (<\frac{\gamma}{2}=20)$, $35(>\frac{\gamma}{2}=20)$ in Figure \ref{fig:limit_cycle-psi}.  We observe that the basin attractions for the colony survival seem to have similar shapes:  the largest area is $\psi=0$ (Figure \ref{cycle_402_2}), the second largest being $\psi=35$ (Figure \ref{cycle_psi35}), and the smallest one is $\psi=15$ (Figure \ref{cycle_psi15}). The observation under this particular parameter set regarding the impacts of $\psi$ of our honeybee-parasite model \ref{Honeybee-mite-scaled} seems to show similar trends as our honey bee only model \ref{honeybee} (see Figure\ref{fig:1D-psi}). Figure \ref{cycle_dynamic2} provides some visual insights on how may $\gamma$ and $\epsilon$ impact population dynamics.\\

\begin{figure}[ht!]
		\centering
				\subfigure[$\gamma=40,  \omega=0.1$]{
			\includegraphics[width=5cm]{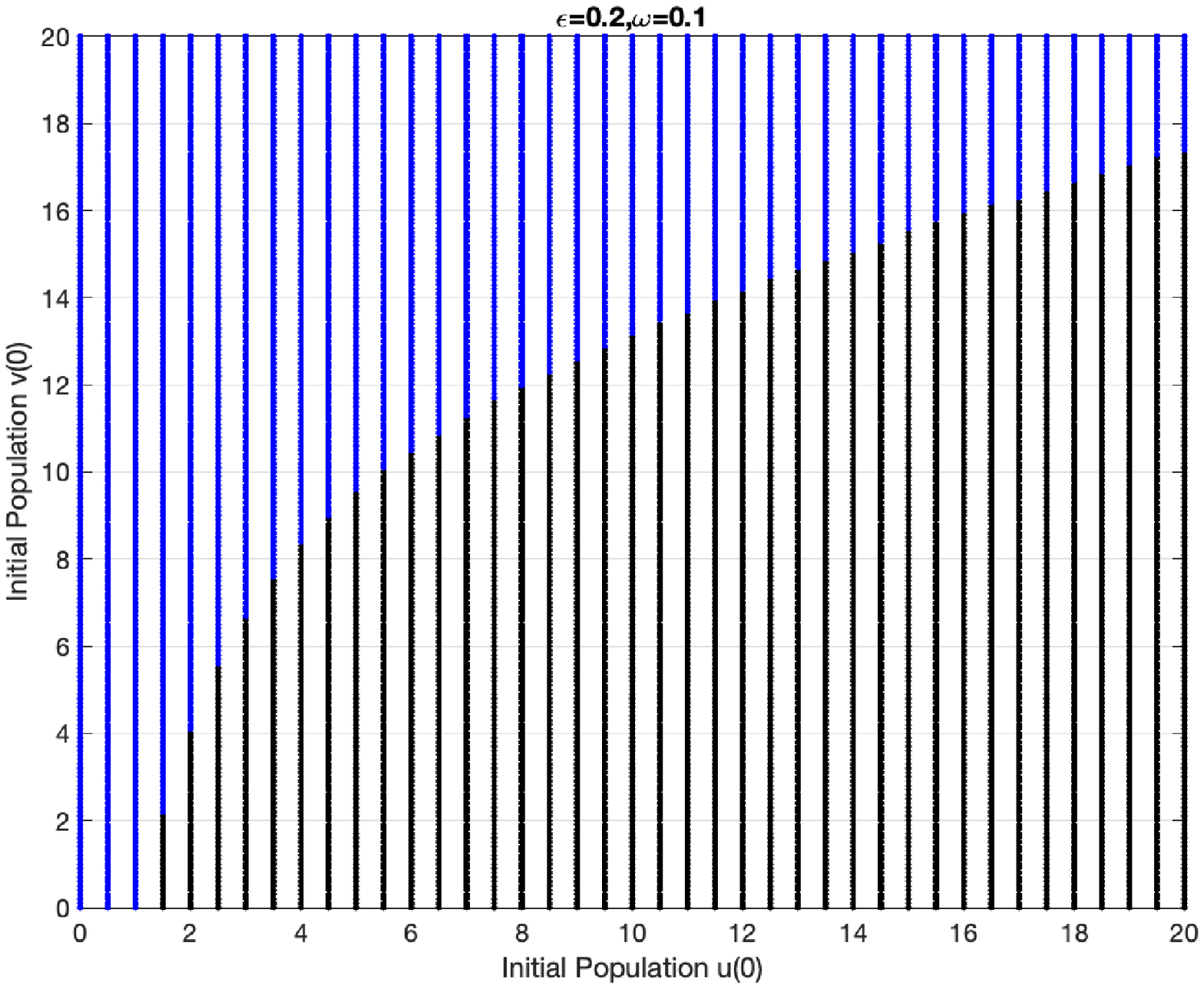}
			\label{cycle_omega}
		} 
 \subfigure[$\gamma=40,  \omega=0.292$]{
			\includegraphics[width=5cm]{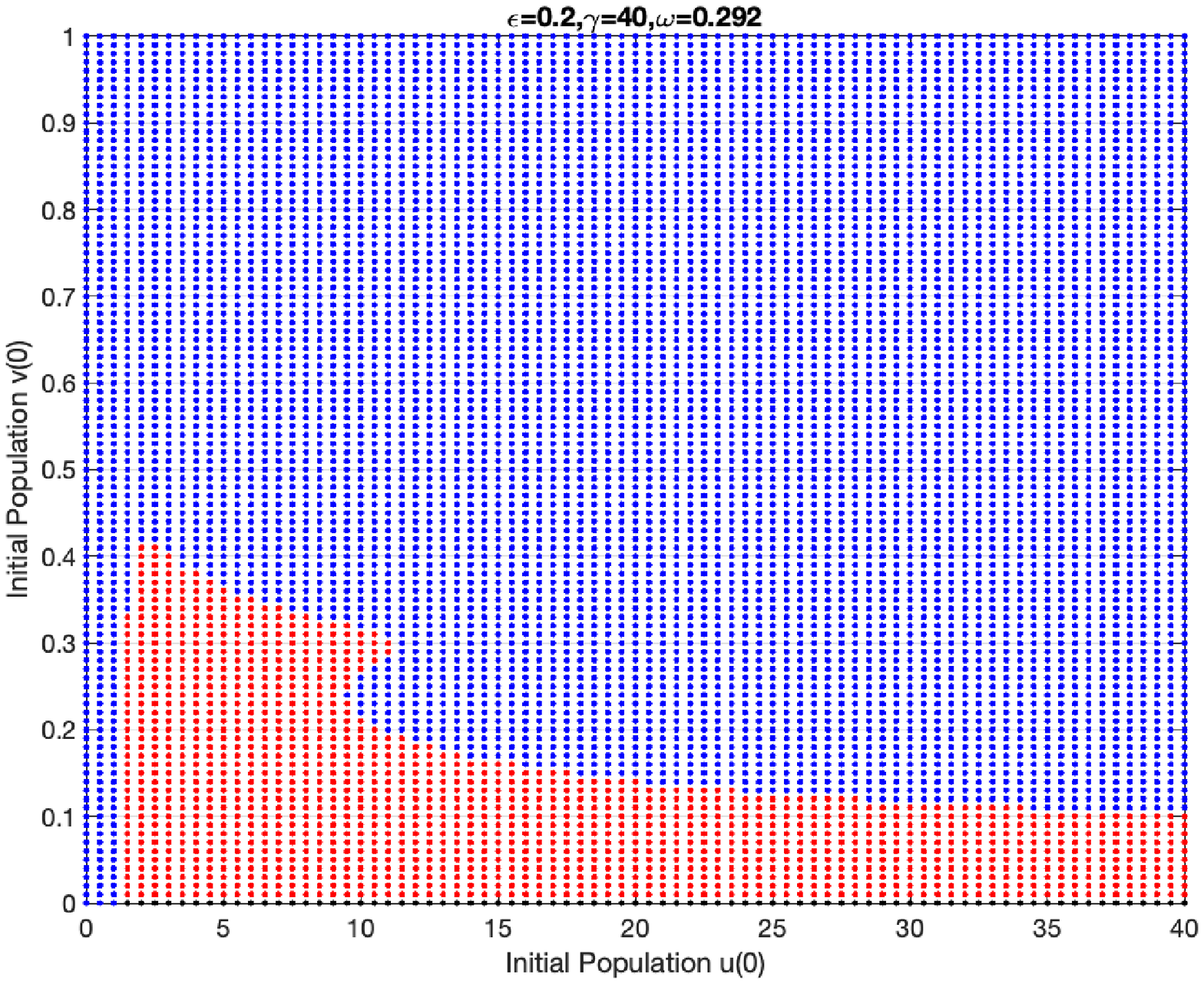}
			\label{cycle_omega0.292}
		}
\subfigure[$\gamma=40,  \omega=0.3$]{
			\includegraphics[width=5cm]{Figure/e=0.2_gamma=40.eps}
			\label{cycle_402_3}
		}
			\subfigure[no seasonality with $\omega=0.3$]{
			\includegraphics[width=5cm]{Figure/No_seasonality.eps}
			\label{cycle_no_2}
		}
  \subfigure[$\gamma=40$, $u_0=5$,$v_0=0.04$]{
			\includegraphics[width=5cm]{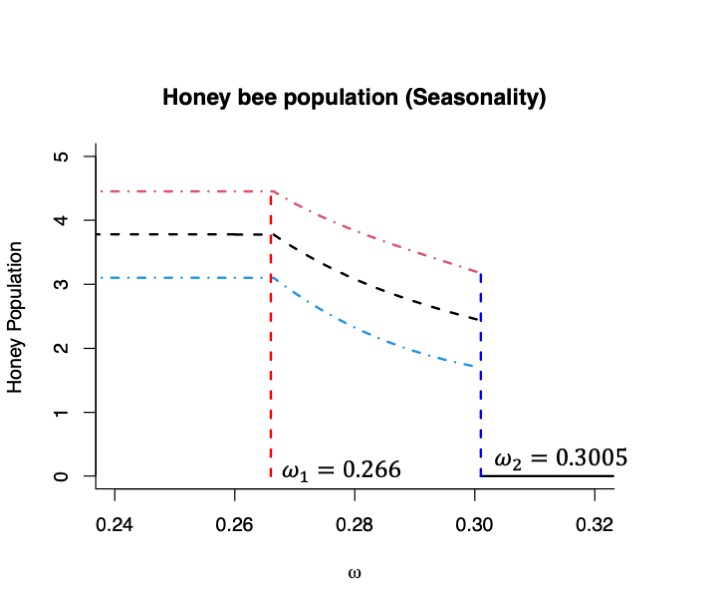}
			\label{cycle_omega_bif}
		}
  \subfigure[no seasonality, $u_0=5$,$v_0=0.04$]{
			\includegraphics[width=4.5cm]{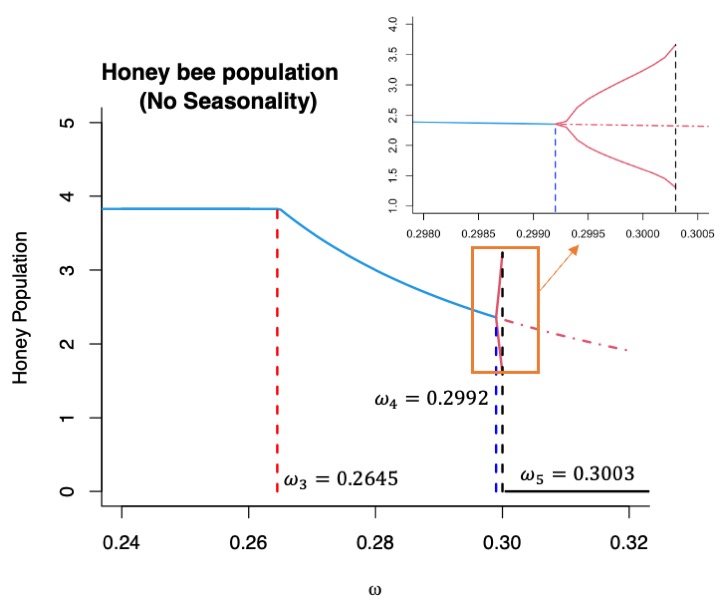}
			\label{cycle_omega_bif_no}
		}
 			\caption{Impacts of seasonality and parasitism on the stable limit cycle when $\bar{r}_0=1$, $\bar{d}_h=0.2$, $\bar{d}_m=0.21$, $\gamma=40$, $\psi=0$, $\epsilon=0.2$ and $\hat{K}=4.49$.  Honey bee initial population is $u_0 \in [0, 40]$, and mite initial population is $v_0 \in [0, 1]$. The blue area is colony collapse, the red area is colony coexistence, and the black area is only bee survival. Figure \ref{cycle_omega_bif}: Max and min honey bee population with seasonality. The red dot-dashed curve indicates the maximum bee population of the period, and the blue dot-dashed curve indicates the minimum bee population of the period. The black dashed curve shows the average of the max and min population. Figure \ref{cycle_omega_bif_no}: Max and min honey bee population without seasonality. The blue solid curve indicates the locally stable equilibrium, the red solid curves indicate the stable limit cycle of the Hopf-bifurcation, and the red dot-dashed curve indicates the source equilibrium. The black dashed line indicates the critical of $\omega$ which makes the colony survive to collapse. The orange square zooms in the Hopf-bifurcation details. Both figures: The black lines indicate collapse. The red and blue indicate critical values of $\omega$.}
			\label{fig:limit_cycle-omega}
\end{figure}

Let the period of the seasonality be $\gamma = 40$ and the strength of the seasonality be $\epsilon=0.2$. We explore the impacts of the parasitism by varying $\omega \in [0.1, 0.35]$ in Figure  \ref{fig:limit_cycle-omega}.  We observe follows:
\begin{enumerate}
    \item  When the parasitism $\omega$ is small (e.g., $\omega=0.1$),  the honey bee can survive while the parasite dies out (see Figure \ref {cycle_omega}). 
    \item When parasitism is increased to $\omega=0.292$,  colonies can survive with parasitism, but the area of basins of attractions for survival decreases as parasitism increases (see Figure \ref{cycle_omega}, \ref{cycle_omega0.292} \& \ref{cycle_402_3}). Thus parasitism has a negative influence on the colonies' survival. 
    \item When the parasitism is large (e.g., $\omega>0.3$ when $u_0=5, v_0=0.04$),  colonies collapse.
\end{enumerate}
We observe that (1) If the colony can survive, increasing the parasitism attack degree can decrease the average population of honey bees. (2) Large parasitism can lead to a colony collapsing. In general, Seasonality with parasitism can have negative impacts in terms of either decreasing the average honey bee population or the colony collapsing.\\

Without seasonality, the value of parasitism rate $\omega$ can lead to destability through hopf-bifurcation (see Theorem \ref{th:hopf}). To further explore how parasitism may impact the honeybee population dynamics with or without seasonality, we perform bifurcation on the impacts of parasitism $\omega=[0.2, 0.33]$ with (see Figure \ref{cycle_omega_bif}) or without (see Figure \ref{cycle_omega_bif_no}) seasonality by setting 
$\bar{r}_0=1$, $\bar{d}_h=0.2$, $\bar{d}_m=0.21$, $\psi=0$, $\gamma=40$, $\epsilon=0.2$, $\hat{K}=4.49$, $u(0)=5$ and $v(0)=0.04$.\\ 


In the absence of seasonality (Model \ref{Honeybee-mite-scaled}), the $\omega_3$ is the bifurcation value where  the mite-free equilibrium ($(\bar{N}^*_h,0)$) changes from being locally stable to unstable (see Theorem \ref{th:Ee} item (2)), and the coexistence of bee and mite population emerges as the locally stable interior equilibrium, and the interior equilibrium become unstable (see Theorem \ref{th:Ee} item (3)) through supercritical Hopf-bifurcation at $\omega_4$, where exists the stable limit cycle (see Theorem \ref{th:hopf}). After the value of $\omega_5$, the colony collapses. Therefore, the bifurcation diagram in Figure \ref{cycle_omega_bif_no}) suggests that: (1) when the severity of parasitism ($\omega$) is small, the colony survives with non-parasites; (2) when the value of $\omega$ rise, bees and parasites coexist in the colony and gradually decreases the population of bees; (3) under the conditions of supercritical Hopf-bifurcation, bees and parasites coexist in a periodic state; (4) when $\omega$ is large enough, the parasites leads to the colony collapse.\\

In the seasonality model (Model \ref{honeybee} and see Figure \ref{cycle_omega_bif}), before the value of $\omega_1$, the system is locally stable around mite-free solutions (see Theorem \ref{th:non-2d}); after this bifurcation point, bees and parasites coexist as the periodic interior solutions. The $\omega_2$ is the critical value when colony collapses. \\

We observe that seasonality can delay the impact of parasitism in two bifurcation points: (1) $\omega_1>\omega_3$: parasite needs larger attacking rates to survive in the periodic environment. And (2) $\omega_2>\omega_5>\omega_4$: colony can still survive with the larger attacking rates from parasites in the periodic environment.\\


\section{Conclusion}
\indent Studies \cite{chen2021review,ullah2021viral,vanbergen2021cocktail,vercelli2021qualitative} suggest that pollinators like honey bees are facing a crisis of dwindling numbers, due to combinations of stressors. In this paper, we proposed and study a non-autonomous, nonlinear differential equations model that describes the interactions between honey bees' and parasite' while including seasonality in the queen's egg-laying rate. The seasonality logistics are adopted from the literature \cite{chen2020model,messan2021population,messan2018effects}.  The proposed model with related theoretical and bifurcation analysis aims to address how 1) seasonality can influence honey bee colony dynamicsies? 2) parasitism impacts honey bee colonies? and 3) seasonality and parasitism jointly influence honey bee colonies?\\

We first explored the seasonality impacts on the honey bee colony. Our theoretical results (Theorem \ref{th:honeybee-rt}) imply that the egg-laying rate plays an important role in determining the colony's survival. If the egg-laying rate is low, the colony is expected to die. When egg laying is not low, the colony’s fate depends on the initial population size in varied seasonal conditions. Our mathematical analysis of the honeybee-parasite model \eqref{Honeybee-mite-scaled} in a constant environment shows that parasitism most likely has negative impacts on honeybee population dynamics and the survival of the colony. Our theoretical work on Model \eqref{Honeybee-mite-scaled} indicates that parasites decrease the honeybee population (Theorem \ref{th:Ee}) and destabilize the dynamics through subcritical or supercritical Hopf-bifurcation (see Theorem \ref{th:hopf}). The Hopf-bifurcation is determined by the queen egg-laying rate $r_0$, death rates of both honeybee $d_h$ and parasite $d_m$, and parasitism $\omega$. More specifically, the colony collapses through supercritical Hopf-bifurcation, and the colony has fluctuating population dynamics through supercritical Hopf- bifurcation.\\

Seasonality in this paper is defined by its strength of seasonality $\epsilon\in[0,1]$, period $\gamma$, and timing of the maximum queen egg-laying rate $\psi$. These three factors are intertwined and generate complicated impacts on honeybee population dynamics with or without parasitism. Our study shows that seasonality can have both negative and positive influences on honeybee colony survival depending on conditions. The colony is more likely to collapse when the period of seasonality ($\gamma$) is limited and the strength of seasonality ($\epsilon$) is large (see Figure \ref{fig:1D-gamma}). In the absence of parasitism, the colony may benefit from the seasonality when the timing of the maximum egg-laying ($\psi$) is larger than half of the period of seasonality ($\gamma$), i.e. $\psi>\frac{\gamma}{2}$ (see Figure \ref{fig:1D-psi}). In the presence of parasites, the impacts of the timing of the maximum egg-laying ($\psi$) are much more complicated. 
Depending on other parameters' values, in some cases, the smaller timing of the maximum egg-laying ($\psi$) or closer to the $\gamma$ may benefit the colony survival (see Figure \ref{fig:psi-2d}\& \ref{fig:limit_cycle-psi}). There are also situations that are beneficial to the colony when growing $\psi$ in the beginning ($\epsilon=0.2$ in Figure \ref{fig:psi-2d} ).\\


As shown by our model and results, seasonality plays a significant role in honey bee colony dynamics. Seasonality can affect bees' behavior and resources. Bees tend to visit flowers more frequently and forage more actively in warm and favorable weather rather than in cold and harsh weather \cite{tuell2010weather}. Ogilvie and Forrest (2017) \cite{ogilvie2017interactions} have also highlighted the crucial role of floral resources in determining bee community growth rates and foraging decisions, suggesting that periodic seasonal changes can help bee communities recover. However, because of climate change, there are seasonality changes, such as a longer period of low flowering abundance in mid-summer, which negatively affects bees \cite{aldridge2011emergence}. Moreover, studies have shown that Africanized bees are better adapted to low-shade habitats than native bees in Mexico, indicating that hotter or longer summers because of seasonality or climate change is unfriendly for native bees \cite{jha2009contrasting}. Bees can adapt to seasonal changes by altering their brood production and lifespan throughout the year \cite{jha2009contrasting,feliciano2020honey}. These phenomena all reflect that the impact of seasonality on bee populations is complex and tied to factors both within the colony and in the environment.\\ 

Seasonality also affects the reproduction and spread of parasites. Jack et al. (2023) \cite{jack2023seasonal} pointed out that reducing the Varroa mites’ population in the spring is important for long-term mite control. Winter also can be an effective time for treating Varroa because there is no  brood and all mites are feeding on adult bees and therefore exposed to the miticide. However, interrupting brood rearing in the fall may not be an effective strategy for mite control by  \cite{jack2020evaluating}, as mite populations increase after treatment \cite{jack2023seasonal}. These findings are consistent with the conclusion of our model, which underscores the complex impacts of seasonality on bee-parasite dynamics. At present, seasonal temperatures are rising due to climate change, and will affect resource availability, bee abundance, and varroa parasitism especially in the fall \cite{smolinski2021raised}. Our model is can predict the different fates of bee colonies by changing the seasonal parameters of egg-laying rate. Such research underscores the importance of studying the effects of seasonality and our research further highlights the need for investigation to quantify these impacts mathematically.\\

Bifurcations and simulations (see Figure \ref{pop-gammaL}) suggest that larger strength of seasonality $\epsilon$ leads to a larger amplitude in population oscillating dynamics. Large strength of seasonality $\epsilon$ alone can cause colony collapse, especially when the colony exhibits oscillations due to parasitism (see Figure \ref{fig:limit_cycle1}). Both our theoretical and bifurcation (see Figures \ref{cycle_omega_bif} \& \ref{cycle_omega_bif_no}) results show that parasitism with or without seasonality can lead to the colony collapsing and decrease the average population dynamics of honey bees.\\

As bee numbers continue to decline, it is crucial to understand the factors that can help honeybees face these threats and/or help them mitigate these ecological disturbances. Strong evidence suggests that climate changes contribute greatly to pollinators' population decline. Seasonality is one aspect of climate change. Our current work and literature \cite{chen2020model,messan2021population,messan2018effects} provide useful insights into how seasonality in the queen egg-laying rate and parasites impact honeybee colonies. Our study suggests that these impacts can be positive or negative depending on the environment. Based on our results, it is possible to develop specific strategies to take advantage of the positive impacts and avoid situations when certain attributes of seasonality lead to colony collapsing or population decreasing. For example, beekeepers may regulate the honeybee population by altering the timing and amount of the egg-laying rate through the amount of food such as sugar and pollen fed to the colonies. Seasonality affects parasite reproduction, maturation, and transmission rates of the viruses they carry. Colony losses might be reduced if the beekeeper can actively respond to the colonies’ needs by observing the colonies’ situation, she/he can help the colony to reduce or even eliminate the impacts of seasonality with well-timed treatments \cite{piot2022honey,vercelli2021qualitative}.\\

Climate change has been considered one of the current significant threats to honey bees and beekeeping \cite{flores2019climate}. As beekeepers have observed in the past ten years, climate impacts on honeybees include scarcity of floral resources and greater spread of disease \cite{vercelli2021qualitative}. Climate change affects the flowering period, directly affecting foraging and resource gathering through weather conditions and extreme heat and shifts in the timing and duration of bloom. Available nectar and pollen affect brood rearing and colony growth impacting both colony survival and pollination services \cite{vercelli2021qualitative,reddy2012potential}, potentially affecting societal risk and prolonging exposure to more extreme events within a season \cite{EPA2021Climate}. Including seasonality in our model is the first step towards studying the impacts of climate changes on honeybee colonies. To better understand how climate change affects the seasonality of bee behavior, including brood rearing, colony growth, and foraging. There is a need for further field studies that provide data to validate our models and direct  our future work.\\

\begin{appendices}

\section{Proofs}\label{secA1}

\subsection*{Proof of Theorem \ref{th:positive}}
\begin{proof}
Let $$f_1(u,v)=\bar{r}(t)\frac{u^2}{\hat{K}+u^2}-
\bar{d}_h u-\frac{\omega u}{1+u}v$$ and $$f_2(u,v)=\frac{\omega u}{1+u}v-\bar{d}_m v.$$ Assume each point $(u_1,v_1) \in \mathbb  X$ in functions $f_1$ and $f_2$ has a neighbour $(u_2,v_2) \in \mathbb  X_0$, and $u_1>u_2$. 
As we know, $r(t)=r_0(1+\epsilon \cos(\frac{2\pi(t -\psi)}{\gamma}))$, the maximum of $r(t)$ is $r_{max}=r_0(1+\epsilon)$, and the minimum of $r(t)$ is $r_{min}=r_0(1-\epsilon)$. Then $\bar{r}_{max}=\frac{r_{max}*c}{R*b}$ and $\bar{r}_{min}=\frac{r_{min}*c}{R*b}$.
Then we can get:

\begin{equation*}
    \begin{split}
     \lvert f_1(u_1,v_1)-f_1(u_2,v_2)\rvert&=\lvert \bar{r}(t)(\frac{u_1^2}{\hat{K}+u_1^2}-\frac{u_2^2}{\hat{K}+u_2^2})+
\bar{d}_h (u_2-u_1)+(\frac{\omega u_2}{1+u_2}v_2-\frac{\omega u_1}{1+u_1}v_1) \rvert\\
&< \lvert \bar{r}_{max}(\frac{u_1^2-u_2^2}{\hat{K}+u_2^2})+\bar{d}_h (u_2-u_1)+\omega(v_2-v_1)\rvert\\
&=\lvert \bar{r}_{max}(\frac{(u_1+u_2)(u_1-u_2)}{\hat{K}+u_2^2})+\bar{d}_h (u_2-u_1)+\omega(v_2-v_1)\rvert\\
&< (\bar{r}_{max}+\bar{d}_h)\lvert u_2-u_1\rvert +\omega\lvert v_2-v_1\rvert
    \end{split}
\end{equation*}

Therefore, there exists two real constants $M_1=\bar{r}_{max}+\bar{d}_h$ and $M_2=\omega$ for $\lvert f_1(u_1,v_1)-f_1(u_2,v_2)\rvert \leq M_1\lvert u_1-u_2\rvert+M_2\lvert v_1-v_2\rvert$. Similarly, function $f_2(u,v)$ has:

\begin{equation*}
    \begin{split}
       \lvert f_2(u_1,v_1)-f_2(u_2,v_2)\rvert&= \lvert \frac{\omega u_1}{1+u_1}v_1-\frac{\omega u_2}{1+u_2}v_2+\bar{d}_m (v_2-v_1)\rvert\\
       &< \omega\lvert v_2-v_1\rvert+\bar{d}_m \lvert v_2-v_1\rvert
    \end{split}
\end{equation*}

Therefore, there exists a real constant $L=\omega+\bar{d}_m$ for $\lvert f_2(u_1,v_1)-f_2(u_2,v_2)\rvert\leq L\lvert v_1-v_2\rvert$. Since eqts.\eqref{Honeybee-mite-scaled} are Lipschitz continuous, following the Lipschitz condition, the system \eqref{Honeybee-mite-scaled} has local existence and uniqueness solution. \\

According to Theorem A.4 (p.423) of Thieme (2003) \cite{thieme2018mathematics}, we can conclude that Model \eqref{Honeybee-mite-scaled} is positive invariant in $\mathbb  X$.
Let $g(u)=\frac{u^2}{\hat{K}+u^2} < 1$ and $h(u)=\frac{\omega u}{1+u}$, then model \eqref{Honeybee-mite-scaled} follows:
$$u'=\bar{r}(t)g(u)-\bar{d}_hu-h(u)v$$ and $$v'=(h(u)-\bar{d}_m)v.$$

From above, 
\begin{equation*}
    \begin{array}{lcl}
       u'&<&\bar{r}-\bar{d}_hu< \bar{r}_{max}-\bar{d}_hu \\ &\Rightarrow& u(t)<\frac{\bar{r}_{max}}{\bar{d}_h}-(\frac{\bar{r}_{max}}{\bar{d}_h}-u_0)e^{-\bar{d}_ht}\\
       &\Rightarrow& u(t) < max\{u_0,\frac{\bar{r}_{max}}{\bar{d}_h} \}.  
    \end{array}
\end{equation*}
Therefore, $u$ is boundedness.

Now, to show the boundedness of $v$, define $H= u+v$, then
\begin{equation*}
    \begin{array}{lcl}
        H'&=& u'+v'=\bar{r}(t) g(u)- \bar{d}_hu-\bar{d}_m v\\
        H'&<&\bar{r}(t)-\max\{\bar{d}_h,\bar{d}_m \} H
    \end{array}
\end{equation*}
Therefore, $H$ is boundedness. Since $u$ is boundedness, $v$ is boundedness. 
\end{proof}

\subsection*{Proof of Proposition \ref{p4honeybee}}
\begin{proof}
Let 
$$f(u)= \frac{r_0 u^2}{\hat{K}+u^2}-\bar{d}_h u=u[\frac{r_0u-d_h(\hat{K}+u^2)}{\hat{K}+u^2}],$$
Then if $r_0>2\bar{d}_h\sqrt{\hat{K}},$ there exists $u_1^*$ and $u_2^*$ such that $f(u_i^*)=0,i=1,2$ and
$$u^*_1=\frac{r_0-\sqrt{r_0^2-4\bar{d}_h^2\hat{K}}}{2\bar{d}_h}\leq u^*_2=\frac{r_0+\sqrt{r_0^2-4\bar{d}_h^2\hat{K}}}{2\bar{d}_h}$$ with 
$$u^*_2=\frac{r_0+\sqrt{r_0^2-4\bar{d}_h^2\hat{K}}}{2\bar{d}_h}>\frac{r_0}{2\bar{d}_h}>\sqrt{\hat{K}}.$$
Notice that
$$f'(u)=\frac{-\bar{d}_h \hat{K}^2-2 \bar{d}_h \hat{K} u^2-\bar{d}_h u^4+2 \hat{K} r_0 u}{\left(\hat{K}+u^2\right)^2}=\frac{-\bar{d}_h\left(\hat{K}+u^2\right)^2+2 \hat{K} r_0 u}{\left(\hat{K}+u^2\right)^2},$$ then we have
 $$f'(0)=-d<0, f'(u_i^*)=-d_h+\frac{2\hat{K} d_h^2}{r_0u_i^*}$$which implies that
  $u^*=0$ is a locally stable equilibrium, and
  $f'(u_1^*)>0$ and   $f'(u_2^*)<0$. Therefore $u^*_2$ is locally stable equilibrium while $u^*_1$ is locally unstable.\\

Note that $$u'=f(u)=u[\frac{r_0u-d_h(\hat{K}+u^2)}{\hat{K}+u^2}]=d_hu[\frac{(u-u_1^*)(u_2^*-u)}{\hat{K}+u^2}].$$ For any initial condition $u(0)\in (u_1^*,u_2^*)$, we have $u'>0$ for all future $t>0$, thus $u(t)$ increases and approaches to $u_2^*$. For any initial condition $u(0)>u_2^*$, we have $u'<0$ for all future $t>0$, thus $u(t)$ decreases and approaches to $u_2^*$.\\
 
If $r_0<2\bar{d}_h\sqrt{\hat{K}}$, then we have
$$u'=f(u)=u[\frac{r_0u-d_h(\hat{K}+u^2)}{\hat{K}+u^2}]=d_h u \left[\frac{-(u-\frac{r_0}{2\bar{d}_h})^2+((r_0/2\bar{d}_h)^2-\hat{K})}{\hat{K}+u^2}\right]<0 .$$
Therefore $u(t)$ converges to 0 if $r_0<2\bar{d}_h\sqrt{\hat{K}}$ holds.

\end{proof}

\subsection*{Proof of Theorem \ref{th:honeybee-rt}}
\begin{proof}
Notice that $u=0$ is an equilibrium of $$u'=\frac{\bar{r}(t)u^2}{\hat{K}+u^2}-\bar{d}_h u=u\left[\frac{\bar{r}(t)u}{\hat{K}+u^2}-\bar{d}_h\right].$$ 

From Theorem \ref{th:positive}, we know that $u \geq 0$ for any initial $u(0)\geq 0$. Define $\mathcal{D}=\{u \in [0, \frac{\bar{d}_h\hat{K}}{r_{M}})\}$. Applying for Lyapunov Stability Theorem \cite{aeyels1995stability} and we define $V(u)=u^2 \geq 0$ $\forall u \in \mathcal{D}$.\\
Notice that $$\dot{V}(t,u)=u'=u\left[\frac{\bar{r}(t)u-\bar{d}_h(\hat{K}+u^2)}{\hat{K}+u^2}\right] < \frac{r_M u}{\hat{K}}-\bar{d}_h.$$
Thus, $$\dot{V}(t,u) \leq 0,  \forall u \in  \mathcal{D} \mbox{and} t \geq 0 $$ which the $\mathcal{D}$ is a neighborhood of the origin, and $t \geq 0$. Thus we can conclude that $u=0$ is locally stable.\\

Define $f(u,t)= \frac{\bar{r}(t)u^2}{\hat{K}+u^2}-\bar{d}_h u$, then we have
$$f(u,t)=u\left[\frac{\bar{r}(t)u-d_h(\hat{K}+u^2)}{\hat{K}+u^2}\right].$$
If $r_{max}=r_M=r_0(1+\epsilon)< 2\bar{d}_h\sqrt{\hat{K}}$, then we have 
$$\bar{r}(t)\leq r_M<2\bar{d}_h\sqrt{\hat{K}}.$$ Thus, 
$$u'=f(u,t)\leq u\left[\frac{r_M u-d_h(\hat{K}+u^2)}{\hat{K}+u^2}\right]=d_h u \left[\frac{-(u-\frac{r_M}{2\bar{d}_h})^2+((r_M/2\bar{d}_h)^2-\hat{K})}{\hat{K}+u^2}\right]<0.$$
This implies that $u=0$ is globally stable when $r_M=r_0(1+\epsilon)< 2\bar{d}_h\sqrt{\hat{K}}.$\\



If $r_{min}=r_m =r_0(1-\epsilon) > 2\bar{d}_h\sqrt{\hat{K}}$ holds, then $r_m\leq \bar{r}(t)\leq r_M=r_0(1+\epsilon)$ and
$$u'=f(u,t)\geq u\left[\frac{r_m u-d_h(\hat{K}+u^2)}{\hat{K}+u^2}\right]=u[\frac{r_m u-d_h(\hat{K}+u^2)}{\hat{K}+u^2}]=d_hu[\frac{(u-u_1^*)(u_2^*-u)}{\hat{K}+u^2}]$$ with
$$u^*_1=\frac{r_m-\sqrt{r_m^2-4\bar{d}_h^2\hat{K}}}{2\bar{d}_h}\leq u^*_2=\frac{r_m+\sqrt{r_m^2-4\bar{d}_h^2\hat{K}}}{2\bar{d}_h}.$$
Similar, we have 
$$u'=f(u,t)\leq u\left[\frac{r_M u-d_h(\hat{K}+u^2)}{\hat{K}+u^2}\right]=u[\frac{r_M u-d_h(\hat{K}+u^2)}{\hat{K}+u^2}]=d_hu[\frac{(u-h_1^*)(h_2^*-u)}{\hat{K}+u^2}]$$ with
$$h^*_1=\frac{r_M-\sqrt{r_M^2-4\bar{d}_h^2\hat{K}}}{2\bar{d}_h}\leq h^*_2=\frac{r_M+\sqrt{r_M^2-4\bar{d}_h^2\hat{K}}}{2\bar{d}_h}.$$
Therefore, we have $u$ being a positive invariant in $[u_1^*, h_2^*]$. Note that for any $u>h_2^*$ we have $u'<0$, thus we have
$$\liminf_{t\rightarrow\infty}u(t)\leq u_1^*\leq \limsup_{t\rightarrow\infty}u(t) <h_2^*$$
if
$r_m\geq 2\bar{d}_h\sqrt{\hat{K}}$ and $u(0)>u_1^*$.


\end{proof}

\subsection*{Proof of Theorem \ref{th:honeybee-stability}}
\begin{proof}
   Let $f(u)=\frac{u^2}{\hat{K}+u^2}$, then Eq. \ref{honeybee} rewrites to
    \begin{equation}\label{eqt.linear_bee}
         u'=L(u)=\bar{r}(t)*f(u)-\bar{d}_hu.
    \end{equation}
    Linearizing Eqt.\ref{eqt.linear_bee} about $u=u^*$ gives,
    $$L(u)\approx L(u^*)+\left[\bar{r}(t)*f'(u^*)-\bar{d}_h\right]*(u-u^*).$$
    Then, this linear equation can be $$h'=\left[\bar{r}(t)*f'(u^*)-\bar{d}_h\right]*h$$ where $h=u-u^*$. After that, we can solve the differential equation by integrating factors:
    $$h(t)=C_0e^{\int_0^t\left[\bar{r}(z)*f'(u^*)-\bar{d}_h\right]dz}=C_0e^\lambda$$
    Therefore, if $\lambda < 0$, the stability of the periodic solution $u=u^*$ is stable; if $\lambda > 0$, then the solution is unstable.
\end{proof}

\subsection*{Proof of Theorem \ref{th:Ee}}
\begin{proof}
\begin{enumerate}
    \item For $E^*_1=(0,0)$, 

$$J_{E^*_1} = \begin{Bmatrix}
-\bar{d}_h & 0 \\
0 & -\bar{d}_m
\end{Bmatrix}.$$

Eigenvalues are $\lambda_1=-\bar{d}_h < 0$ and $\lambda_2=-\bar{d}_m < 0$, therefore $E^*_1$ always stable. \\

    \item For  $E_{b1}=(\frac{\bar{r}-\sqrt{\bar{r}^2-4 \hat{K} \bar{d}_h^2}}{2 \bar{d}_h},0)$, eigenvalues are $$\lambda_1=\frac{\left(\omega-\bar{d}_m\right) \left(-\bar{r}+\sqrt{\bar{r}^2-4 \hat{K} \bar{d}_h^2}\right)+2 \bar{d}_h \bar{d}_m}{2 \bar{d}_h+\sqrt{\bar{r}^2-4 \hat{K} \bar{d}_h^2}+\bar{r}}$$ and $$\lambda_2=\frac{\bar{r} \bar{d}_h \left(\bar{r}-\sqrt{\bar{r}^2-4 \hat{K} \bar{d}_h^2}\right)-4 \hat{K} \bar{d}_h^3}{\bar{r} \left(\sqrt{\bar{r}^2-4 \hat{K} \bar{d}_h^2}-\bar{r}\right)}.$$

Since $\frac{\bar{r}}{2 \sqrt{\hat{K}} \bar{d}_h}>1$, $\lambda_2>0$. If $\bar{d}_m>\omega$, $\lambda_1>0$, then $E_{b1}$ is source. If $\bar{d}_m<\omega$, $u^*=\frac{  \bar{d}_m}{ \omega -\bar{d}_m}>\bar{N}^c_h=\frac{\bar{r}-\sqrt{\bar{r}^2-4 \hat{K} \bar{d}_h^2}}{2 \bar{d}_h}$, then $\frac{2 \bar{d}_h \bar{d}_m}{\omega -\bar{d}_m}>\bar{r}-\sqrt{\bar{r}^2-4 \hat{K} \bar{d}_h^2}$, i.e. $\lambda_1<0$, therefore $E_{b1}$ is saddle.\\

For  $E_{b2}=(\frac{\bar{r}+\sqrt{\bar{r}^2-4 \hat{K} \bar{d}_h^2}}{2 \bar{d}_h},0)$, eigenvalues are $$\lambda_1=\frac{\left(\omega-\bar{d}_m\right) \left(\bar{r}+\sqrt{\bar{r}^2-4 \hat{K} \bar{d}_h^2}\right)-2 \bar{d}_h \bar{d}_m}{2 \bar{d}_h+\sqrt{\bar{r}^2-4 \hat{K} \bar{d}_h^2}+\bar{r}}$$ and $$\lambda_2=\frac{\sqrt{\bar{r}^2-4 \hat{K} \bar{d}_h^2} \left(4 \hat{K} \bar{d}_h^3-2 \bar{r}^2 \bar{d}_h\right)+r \left(8 \hat{K} \bar{d}_h^3-2 \bar{r}^2 \bar{d}_h\right)}{\bar{r} \left(\sqrt{\bar{r}^2-4 \hat{K} \bar{d}_h^2}+\bar{r}\right)^2}.$$

Since $\bar{r}^2>4 \hat{K} \bar{d}_h^2>2 \hat{K} \bar{d}_h^2$, $2\bar{r}^2\bar{d}_h>8 \hat{K} \bar{d}_h^3>4 \hat{K} \bar{d}_h^3$, then $8 \hat{K} \bar{d}_h^3-2\bar{r}^2\bar{d}_h<0$ and $4 \hat{K} \bar{d}_h^3-2\bar{r}^2\bar{d}_h<0$, i.e. $\lambda_2<0$. If $\bar{d}_m>\omega$, $\lambda_1<0$, then $E_{b2}$ is sink. If $\bar{d}_m<\omega$, $\bar{N}^*_h=\frac{\bar{r}+\sqrt{\bar{r}^2-4 \hat{K} \bar{d}_h^2}}{2 \bar{d}_h}>u^*=\frac{  \bar{d}_m}{ \omega -\bar{d}_m}$, then $\left(\omega -\bar{d}_m\right) \left(\sqrt{\bar{r}^2-4 \hat{K} \bar{d}_h^2}+\bar{r}\right)>2 \bar{d}_h \bar{d}_m$, i.e. $\lambda_1>0$. Therefore $E_{b2}$ is saddle.\\

    \item For  $E^*=(\frac{\bar{d}_m}{\omega -\bar{d}_m}, \frac{\left[\bar{r}u^*-\bar{d}_h \left((u^*)^2+\hat{K}\right)\right](u^*+1)}{\omega((u^*)^2+\hat{K})})$, we simplified the matrix J to
$$J_{E^*} = \begin{Bmatrix}
-\frac{u^* \left(d_h \left(\hat{K}+(u^*)^2\right)^2+r \left((u^*)^2-\hat{K} (2 u^*+1)\right)\right)}{(u^*+1) \left(\hat{K}+(u^*)^2\right)^2}
 &-\frac{\omega u^*}{u^*+1} \\
 \frac{\omega  v^*}{(u^*+1)^2} & 0
\end{Bmatrix}.$$

\begin{figure}[htbp]
	\centering
	\includegraphics[width=6cm]{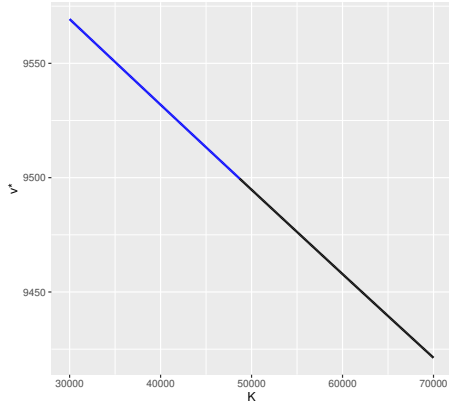}
	\caption{Simulation for the trace ($\lambda_1+\lambda_2$) of $J_{E^*}$ . The black curve indicates the trace is positive, i.e. the stability of $E^*$ is source, and the blue curve indicates the trace is negative, i.e.  the stability of $E^*$ is sink. $\bar{r}=500$, $\omega=0.05$, $\bar{d}_h=0.01$, $\bar{d}_m=0.049969$, and $\hat{K} \in [30000, 70000]$.}
	\label{fig.trace}
\end{figure}

which gives the following two equations:
\begin{equation} \label{eq.trace}
\begin{array}{lcl}
\lambda_1+\lambda_2 &=&  -\frac{u^* \left(d_h \left(\hat{K}+(u^*)^2\right)^2+\bar{r} \left((u^*)^2-\hat{K} (2 u^*+1)\right)\right)}{(u^*+1) \left(\hat{K}+(u^*)^2\right)^2}\\
&=&  \frac{u^*(-\bar{d}_h \hat{K}^2+(\bar{r}  + 2  u^* \bar{r}- 2 \bar{d}_h (u^*)^2)\hat{K}-\bar{r}(u^*)^2-\bar{d}_h (u^*)^4)}{(u^*+1) \left(\hat{K}+(u^*)^2\right)^2}\\
\lambda_1\lambda_2 &=&  \frac{\omega^2  u^* v^*}{(u^*+1)^3} > 0
    \end{array}
\end{equation}

-rEqt. \ref{eq.trace} gets two $\hat{K}_{1,2}$ to make the $\lambda_1+\lambda_2=0$, where $\hat{K}_2=\frac{\bar{r}u^*}{\bar{d}_h}-(u^*)^2+\frac{\bar{r}}{2\bar{d}_h}+\frac{\sqrt{\bar{r}} \sqrt{\bar{r} (2 u^*+1)^2-8 \bar{d}_h  (u^*)^2 (u^*+1)}}{2\bar{d}_h}$ and the condition of $E^*$ is $\hat{K}<\frac{\bar{r}u^*}{\bar{d}_h}-(u^*)^2$, therefore only $\hat{K}_1=\frac{\bar{r}u^*}{\bar{d}_h}-(u^*)^2+\frac{\bar{r}}{2\bar{d}_h}-\frac{\sqrt{\bar{r}} \sqrt{\bar{r} (2 u^*+1)^2-8 \bar{d}_h  (u^*)^2 (u^*+1)}}{2\bar{d}_h}$ exists $E^*$ where is the trace equals 0. Because of $\lambda_1\lambda_2>0$, $\lambda_1+\lambda_2>0$ as $\hat{K}\in(\hat{K}_1, \frac{\bar{r}u^*}{\bar{d}_h}-(u^*)^2)$, i.e. $E^*$ is source, whereas, $\lambda_1+\lambda_2<0$ as $\hat{K}\in(-\infty, \hat{K}_1)$, i.e. $E^*$ is sink. From Fig. \ref{fig.trace}, there exists a $\hat{K}$ that makes interior equilibrium ($E^*$) from sink to source.
\end{enumerate}
\end{proof}

\subsection*{Proof of Theorem \ref{th:hopf}}
\begin{proof}
We re-scaled the system \ref{Honeybee-mite-scaled} to the following model:
\begin{equation}\label{Honeybee-mite-scaled2}
\begin{array}{lcl}
u'&=& g(u)(f(u)-v)\\
v'&=& v(g(u)-\bar{d}_m),
\end{array}
\end{equation}

where $g(u)=\frac{\omega u}{1+u}$ and $f(u)=\frac{\bar{r}}{g(u)}\cdot\frac{u^2}{\hat{K}+u^2}-\frac{\bar{d}_h}{g(u)}\cdot u$.

We would apply Theorem 3.1 in Wei et al. (2011) \cite{wang2011predator} into system \ref{Honeybee-mite-scaled2}, then our system must have:

(a1) $f \in C^1(\bar{\mathbb{R}}), f(a)=f(b)=0$, where $0<a<b$; $f(u)$ is positive for $a<u<b$, and $f(u)$ is negative otherwise; there exists $\bar{\lambda} \in (a,b)$ such that $f'(u)>0$ on $[a,\bar{\lambda})$, $f'(u)<0$ on $(\bar{\lambda},b]$;\\

(a2) $g \in C^1(\bar{\mathbb{R}}), g(0)=0$; $g(u)>0$ for $u>0$ and $g'(u) > 0$ for $u>0$, and there exists $\lambda>0$ such that $g(\lambda)=d$.\\

(a3) $f(u)$ and $g(u)$ are $C^3$ near $\lambda=\bar{\lambda}$ and $f''(\bar{\lambda})<0$.\\

Then the Jacobean matrix of Model \eqref{Honeybee-mite-scaled2} is 
$$J = \begin{Bmatrix}
f'(u)g(u) & -g(u)\\
v g'(u) & 0
\end{Bmatrix}$$

$$g'(u)=\frac{\omega }{(u+1)^2}>0$$
and we set $h(u)=\frac{r u^2}{K+u^2}-u \bar{d}_h$,$f(u)=\frac{h(u)}{g(u)}$ ,then

\begin{equation}\label{eqt.f'}
    \begin{array}{lcl}
   f'(u)&=&\frac{h'(u) g(u) - h(u) g'(u)}{g^2(u)} \\
   &=& \frac{h'(u)}{g(u)}-\frac{f(u)g'(u)}{g(u)}\\
   &=& \frac{\frac{\bar{r} \left(2 \hat{K} u+\hat{K}-u^2\right)}{\left(\hat{K}+u^2\right)^2}-\bar{d}_h }{\omega }
    \end{array}
\end{equation}

(a1) $f \in C^1(\bar{\mathbb{R}}), f(a)=f(b)=0$, where $0<a<b$; $f(u)$ is positive for $a<u<b$, and $f(u)$ is negative otherwise; there exists $\bar{\lambda} \in (a,b)$ such that $f'(u)>0$ on $[a,\bar{\lambda})$, $f'(u)<0$ on $(\bar{\lambda},b]$;\\

$f(u)=\frac{h(u)}{g(u)}$ where $h(u)=\frac{r u^2}{K+u^2}-u\bar{d}_h$ and the solution of $h(u)=0$ being $u_1=\frac{\bar{r}-\sqrt{\bar{r}^2-4 \bar{d}_h ^2 \hat{K}}}{2 \bar{d}_h}$ and $u_2=\frac{\bar{r}+\sqrt{\bar{r}^2-4 \bar{d}_h ^2 \hat{K}}}{2 \bar{d}_h}$. Let $a=u_1$ and $b=u_2$,then you have $f(a)=f(b)=0$.

Then, $$h'(u_1)=-\frac{2 \bar{d}_h \left(\bar{r}^2 \left(\bar{r}-\sqrt{\bar{r}^2-4 \hat{K} \bar{d}_h^2}\right)+2 \hat{K} \bar{d}_h^2 \left(\sqrt{\bar{r}^2-4 \hat{K} \bar{d}_h^2}-2 \bar{r}\right)\right)}{\bar{r} \left(\bar{r}-\sqrt{\bar{r}^2-4 \hat{K} \bar{d}_h^2}\right)^2}$$ and $$h'(u_2)=-\frac{2 \bar{d}_h \left(\bar{r}^2 \left(\sqrt{\bar{r}^2-4 \hat{K} \bar{d}_h^2}+\bar{r}\right)+2 \hat{K} \bar{d}_h^2 \sqrt{\bar{r}^2-4 \hat{K} \bar{d}_h^2}\right)}{\bar{r} \left(\sqrt{\bar{r}^2-4 \hat{K} \bar{d}_h^2}+\bar{r}\right)^2}$$

Since $\bar{r}^2>4 \hat{K} \bar{d}_h^2$, $h'(u_1)>0$ and $h'(u_2)<0$, then we have

$$f'(u_1)=\frac{h'(u_1)}{g(u_1)}>0 \mbox{ and } f'(u_2)=\frac{h'(u_2)}{g(u_2)}<0$$

Therefore, there exists a $\bar{\lambda}\in(a,b)$ make the sign of $f'(u)$ from positive to negative. 

(a2) $g \in C^1(\bar{\mathbb{R}}), g(0)=0$; $g(u)>0$ for $u>0$ and $g'(u) > 0$ for $u>0$, and there exists $\lambda>0$ such that $g(\lambda)=d$.\\

$g(u)=\frac{\omega u}{1+u}$, if $u=0$ then $g(0)=0$, if $u>0$ then $g(u)>0$. $g'(u)=\frac{\omega }{(1+u)^2}>0$. \\
We assume there exist $\lambda >0$ such that $g(\lambda)=\frac{\omega \lambda}{1+\lambda}=d>0$, i.e $\lambda=\frac{d}{\omega-d}$.\\

(a3) $f(u)$ and $g(u)$ are $C^3$ near $\lambda=\bar{\lambda}$ and $f''(\bar{\lambda})<0$.\\

\begin{equation}
\begin{array}{lcl}
f''(\lambda)&=& f''(\bar{\lambda})\\
&=&-\frac{4 \bar{r} \bar{\lambda}^2}{\omega  \left(\hat{K}+\bar{\lambda}^2\right)^2}-\frac{6 \bar{r} (\bar{\lambda}+1) \bar{\lambda}}{\omega  \left(\hat{K}+\bar{\lambda}^2\right)^2}+\frac{2 \bar{r}}{\omega  \left(\hat{K}+\bar{\lambda}^2\right)}+\frac{8 \bar{r} (\bar{\lambda}+1) \bar{\lambda}^3}{\omega  \left(\hat{K}+\bar{\lambda}^2\right)^3}\\
&=& \frac{2 \bar{r} \left(\hat{K}^2-3 \hat{K} (\bar{\lambda}+1) \bar{\lambda}+\bar{\lambda}^3\right)}{\omega \left(\hat{K}+\bar{\lambda}^2\right)^3}\\
\end{array}
\end{equation}

From \eqref{eqt.f'}, since $f'(\bar{\lambda})=0$, $$\frac{\bar{r} \left(2 \hat{K} \bar{\lambda}+\hat{K}-\bar{\lambda}^2\right)}{\left(\hat{K}+\bar{\lambda}^2\right)^2}=\bar{d}_h \Rightarrow \bar{r} (\hat{K}+2 \hat{K} \bar{\lambda}-\bar{\lambda}^2)=\bar{d}_h(\hat{K}+\bar{\lambda}^2)^2$$
Therefore, it must has $\hat{K}+2\hat{K}\bar{\lambda}>\bar{\lambda}^2$.

In addition to this, $f'(u)$ also is following

\begin{equation}
    \begin{array}{lcl}\label{eq.f'_v}
   f'(u)&=&-\frac{\bar{d}_h }{\omega }-\frac{2 \bar{r} (u+1) u^2}{\omega  \left(\hat{K}+u^2\right)^2}+\frac{\bar{r} u}{\omega  \left(\hat{K}+u^2\right)}+\frac{\bar{r} (u+1)}{\omega  \left(\hat{K}+u^2\right)}      \\
   &=& -\frac{2 \bar{r} (u+1) u^2}{\omega  \left(\hat{K}+u^2\right)^2}+\frac{\bar{r} (u+1)}{\omega  \left(\hat{K}+u^2\right)}  +v^*\\
   &=& v^* + \frac{\bar{r}(1+u)}{(\hat{K}+u^2)\omega}(1-\frac{2u^2}{\hat{K}+u^2})
    \end{array}
\end{equation}

From \eqref{eq.f'_v}, $v*>0$ and $f'(u)=0$, if $u$=$\bar{\lambda}$, then $\bar{\lambda}$ and $\hat{K}$ must be $\bar{\lambda}^2>\hat{K}$. Then $$\hat{K}^2-3 \hat{K} \bar{\lambda}^2-3 \hat{K}\bar{\lambda}+\bar{\lambda}^3=(\hat{K}^2-2\bar{\lambda}\hat{K}-\bar{\lambda}^2\hat{K})+(\bar{\lambda}^3- \bar{\lambda}\hat{K}-2 \bar{\lambda}^2\hat{K})<0.$$
In summary, $f''(\bar{\lambda})<0$.\\

\begin{corollary}\label{th.hopf}
(Theorem 3.1 in Wei et al. (2011)) Assume that f,g satisfy (a1)-(a3). Then the system \eqref{Honeybee-mite-scaled2} undergoes a Hopf bifurcation at $(\bar{\lambda}, v_{\lambda})$; the Hopf bifurcation is supercritical and backward (respectively, subcritical and forward) if $a(\bar{\lambda})<0$ ($a(\bar{\lambda})>0$), where $a(\bar{\lambda})$ is defined in \ref{eqt.hopf}.
\end{corollary}

According to the Corollary \eqref{th.hopf}, the direction of the Hopf bifurcation and the stability of bifurcating periodic orbits are determined by the first Lyapunov coefficient

\begin{equation}\label{eqt.hopf}
\begin{array}{lcl}
a(\bar{\lambda
})&=&\frac{f'''(\bar{\lambda}) g(\bar{\lambda}) g'(\bar{\lambda})+2f''(\bar{\lambda})[g'(\bar{\lambda})]^2-f''(\bar{\lambda})g(\bar{\lambda})g''(\bar{\lambda})}{16g'(\bar{\lambda})}\\
&=&\frac{\omega}{16(1+\bar{\lambda})}(2 f''(\bar{\lambda})+\bar{\lambda} f'''(\bar{\lambda}))
\end{array}
\end{equation}

From Eqt.\eqref{eqt.hopf}, since $\bar{\lambda} >0$, $\frac{\omega}{16(1+\bar{\lambda})}>0$, and 

\begin{equation*}
\begin{array}{lcl}
2 f''(\bar{\lambda})+\bar{\lambda} f'''(\bar{\lambda})&=&\frac{2 \bar{r} \left(2 \hat{K}^3-\hat{K}^2 (2 \bar{\lambda} (2 \bar{\lambda}+9)+3)+2 \hat{K} (\bar{\lambda} (4-3 \bar{\lambda})+9) \bar{\lambda}^2+(2 \bar{\lambda}-3) \bar{\lambda}^4\right)}{\omega  \left(\hat{K}+\bar{\lambda}^2\right)^4}
\end{array}
\end{equation*}

\begin{figure}[htbp]
	\centering
	\includegraphics[width=6cm]{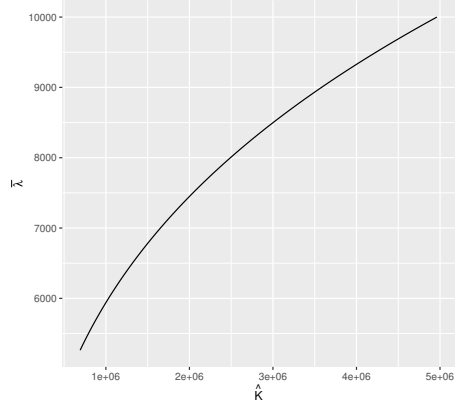}
	\caption{Simulation for the sign of $a(\bar{\lambda})$. All $a(\bar{\lambda})$ is positive. From Theorem \ref{th.hopf}, the bifurcation is subcritical and forward. The black curve indicates positive and red curve indicates negative.  $\bar{r}=100$, $\omega=\in [0.000010001,0.0010002]$, $\bar{d}_h=0.0009$, $\bar{d}_m=0.001$, and $\hat{K} = \hat{K}_1 \in [6.9*10^5, 5.0*10^6]$.}
	\label{fig.a(lambda)}
\end{figure}

From Figure \ref{fig.a(lambda)} and Eqt.\ref{eqt.hopf}, we got $a(\bar{\lambda})>0$. According to Corollary \ref{th.hopf}, the system \ref{Honeybee-mite-scaled} undergoes a Hopf bifurcation at $\hat{K} = \hat{K}_1$; the Hopf bifurcation is subcritical and forward. From Figure \ref{fig.a_lambda} and Eqt.\ref{eqt.hopf}, we can got $a(\bar{\lambda})<0$. According to Corollary \ref{th.hopf}, the system \ref{Honeybee-mite-scaled} undergoes a Hopf bifurcation at $\hat{K} = \hat{K}_1$; the Hopf bifurcation is supercritical and backward .

\begin{figure}[ht]
		\centering
		\subfigure[]{
			\includegraphics[width=5cm]{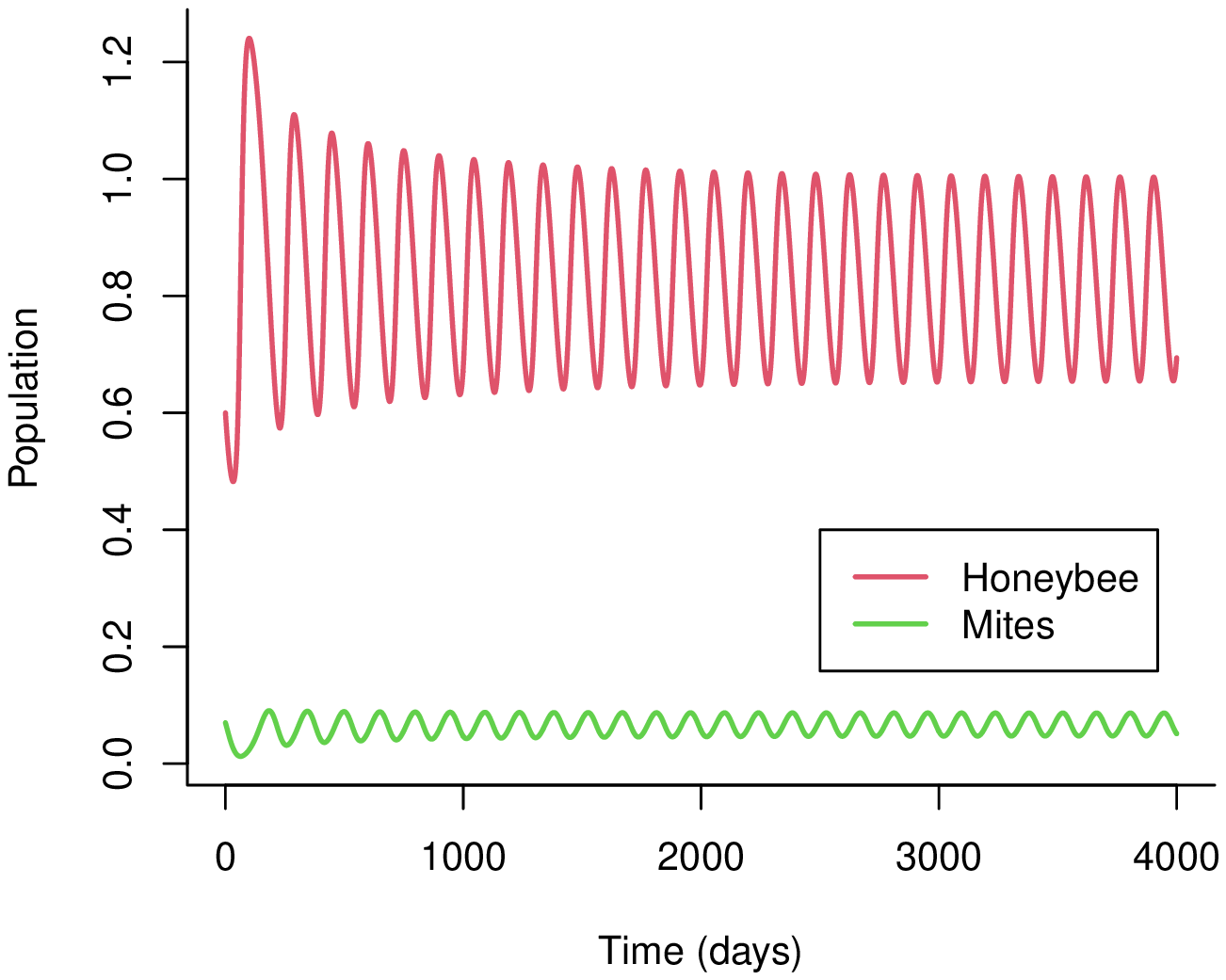}\label{fig.stable}
		}
		\subfigure[]{
			\includegraphics[width=5cm]{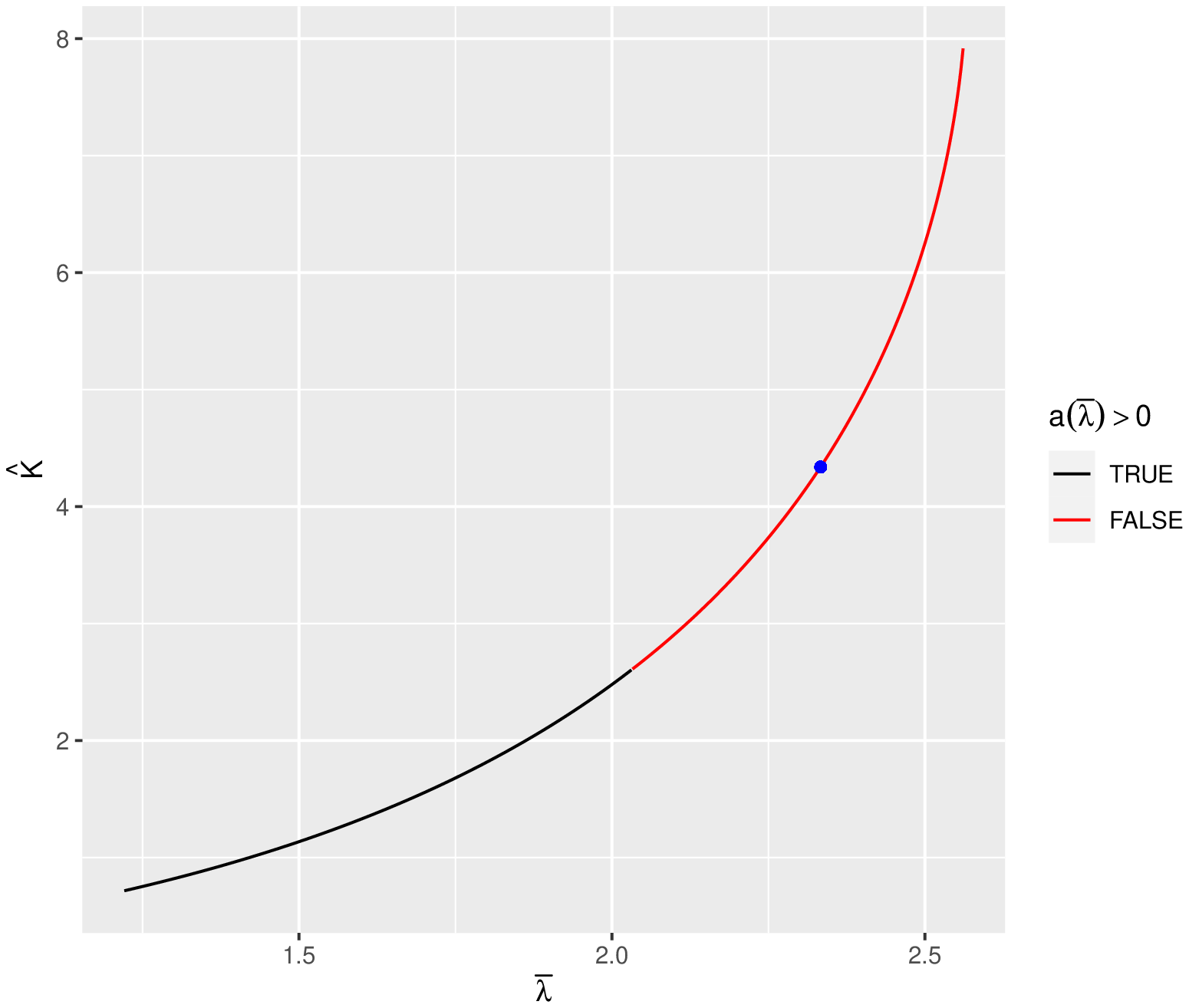}\label{fig.hopf}
		} \label{fig.a_lambda}
			\caption{Simulation for a stable limit cycle around whenever $\hat{K} > \hat{K}_1 $. Figure \ref{fig.hopf}: the conditions for subcritical or supcritical of hopf-bifurcation when $\hat{K}=\hat{K_1}$. The black indicates supcritical i.e. $a(\bar{\lambda})>0$; the blue indicates subcritical i.e. $a(\bar{\lambda})>0$. Choose values at blue dot conditions to get Figure \ref{fig.stable}: $\hat{K}=4.6$, $\hat{K}_1=4.34$, $\bar{r}=1$, $\omega=0.3$, $\bar{d}_h=0.2$, $\bar{d}_m=0.21$, $u^*=2.33$, $\frac{\bar{r}u^*}{\bar{d}_h}-(u^*)^2=6.22$.}
			\label{fig.limit_cycle}
\end{figure}

From Figure \ref{fig.limit_cycle}, $\hat{K} > \hat{K}_1 $, there exists a stable limit cycle.

\end{proof}

\subsection*{Proof of Theorem \ref{th:non-2d}}

\begin{proof}
     Let $f(u)=\frac{u^2}{\hat{K}+u^2}$, then the Jacobian of the system is obtained as:
$$J = \begin{Bmatrix}
-\bar{d}_h+\bar{r}(t)f'(u)-\frac{  \omega v}{(1 +u)^2} & -\frac{  \omega u}{1 +u}\\
\frac{\omega v }{(u+1)^2} & \frac{ \omega u}{u+1}-\bar{d}_m
\end{Bmatrix}.$$
After that, the linearized system at $(u^*,0)$ is
$$\begin{bmatrix}h' \\ g' \end{bmatrix} = \begin{bmatrix}
-\bar{d}_h+\bar{r}(t)f'(u^*) & -\frac{  \omega u^*}{1 +u^*}\\
0 & \frac{ \omega u^*}{u^*+1}-\bar{d}_m
\end{bmatrix}*\begin{bmatrix}
    h \\ g
\end{bmatrix}.$$
Assume the linearly independent set of initial conditions:
$$h_1(0)=1, g_1(0)=0$$
and
$$h_2(0)=0, g_2(0)=1$$
to find linearly independent solutions $(h_1(t),g_1(t))$ and $(h_2(t),g_2(t))$ of linear system.
Then the solutions are:
$$h_1(t)=e^{\int_0^t\left[\bar{r}(z)*f'(u^*)-\bar{d}_h\right]dz} \quad\text{and}\quad g_1(t)=0,$$
$$h_2(t)= e^{\int_0^t\left[\bar{r}(z)*f'(u^*)-\bar{d}_h\right]dz}*\int_0^t\left[-\frac{ \omega u^*}{u^*+1} e^{\int_0^s\left[\frac{  \omega u^*}{1 +u^*}-\bar{d}_m +\bar{d}_h-\bar{r}(s)*f'(u^*)\right]ds}\right]dz$$ and $$g_2(t)=e^{\int_0^t\left[\frac{  \omega u^*}{1 +u^*}-\bar{d}_m\right]dz}.$$
Hence, we can obtain the fundamental matrix $\mathcal{F}(t)$ of the linearized system over the interval $0 \leq t \leq T$ , where T is the period, which is following:
$$\mathcal{F}(T) = \begin{bmatrix}
    h_1(T) & h_2(T) \\
    g_1(T) & g_2(T)
\end{bmatrix},$$
and the eigenvalues of the transition matrix are
$$\lambda_1=e^{\int_0^T\left[\bar{r}(t)*f'(u^*)-\bar{d}_h\right]dt} \quad \text{and} \quad \lambda_2=e^{\int_0^T\left[\frac{  \omega u^*}{1 +u^*}-\bar{d}_m\right]dt}.$$
Therefore, if $\lambda_1 < 0$ and $\lambda_2 < 0$, the $(u^*,0)$ is stable, otherwise, it is unstable.
\end{proof}



\end{appendices}
\section*{Conflict of interest}
The authors declare that they have no conflict of interest.

\bibliographystyle{spmpsci} 
\bibliography{SeasonalityParasitism_accept}
\end{document}